\documentclass[12pt]{article}

\usepackage[utf8]{inputenc}
\usepackage[english]{babel}
\usepackage{amsmath,amssymb,amsfonts,amsthm,hyperref}
\usepackage{graphicx,multirow,caption,bbm,latexsym,epsfig}
\usepackage{cite}
\usepackage{xcolor,colortbl}

\newcommand{\bs}{\begin{subequations}}
\newcommand{\es}{\end{subequations}}
\newcommand{\be}{\begin{equation}}
\newcommand{\ee}{\end{equation}}
\newcommand{\ba}{\begin{eqnarray}}
\newcommand{\ea}{\end{eqnarray}}
\newcommand{\no}{\nonumber \\}

\newcommand{\viz}{\textit{viz.}}
\newcommand{\ie}{\textit{i.e.}}
\newcommand{\vvs}{\textit{versus }}

\def\no{\nonumber\\}
\def\fn{\footnote}

\def\ts{\textsc}
\def\FM{\textsc{FeynMaster}}
\def\FMS{\textsc{FeynMaster} }

\allowdisplaybreaks

\begin{document}

\title{\LARGE The $Z b \bar b$ vertex in a left--right model}

\author{
  Duarte~Fontes,$^{(1)}$\thanks{E-mail:
    \tt dfontes@bnl.gov}
  \
  Darius~Jur\v{c}iukonis,$^{(2)}$\thanks{E-mail:
    \tt darius.jurciukonis@tfai.vu.lt}
  \
  and Lu\'\i s Lavoura$^{(3)}$\thanks{E-mail:
    \tt balio@cftp.tecnico.ulisboa.pt}
  \\*[3mm]
  $^{(1)}\!$
  \small Department of Physics, Brookhaven National Laboratory, \\
  \small Upton, New~York 11973, U.S.A.
  \\*[2mm]
  $^{(2)}\!$
  \small Vilnius University, Institute of Theoretical Physics and Astronomy, \\
  \small Saul\.etekio~ave.~3, Vilnius 10257, Lithuania
  \\*[2mm]
  $^{(3)}\!$
  \small Universidade de Lisboa, Instituto Superior T\'ecnico, CFTP, \\
  \small Av.~Rovisco~Pais~1, 1049-001~Lisboa, Portugal
}

\maketitle

\begin{abstract}
  We consider the one-loop corrections to the $Z b \bar b$ vertex
  in a $CP$-conserving left--right model (LRM),
  \textit{viz.}\ a model with gauge group $SU(2)_L \times SU(2)_R \times U(1)$.
  We allow the gauge coupling constants
  of $SU(2)_L$ and $SU(2)_R$ to be different.
  The spontaneous symmetry breaking is accomplished only
  by doublets and/or singlets of $SU(2)_L$ and $SU(2)_R$.  
  The lightest massive neutral gauge boson of our LRM
is assumed to have the same Yukawa couplings
to bottom-quark pairs as the $Z$ of the Standard Model (SM);
  this assumption has the advantage that, then,
  the infrared divergences
  automatically cancel down in the subtraction
  of the $Z b \bar b$ vertex in the SM from the same vertex in the LRM.
  We effect a proper renormalization of the $Z b \bar b$ vertex
  and check explicitly both its gauge invariance
  and the cancellation of all the ultraviolet divergences.
  We find out that a LRM with the above assumptions
  cannot achieve a better fit to the $Z b \bar b$ vertex
  than a multi-Higgs extension of the SM,
  \viz\ both models can only achieve a decent fit when one admits
  scalar particles with very low masses $\lesssim 50$\,GeV.
  This is true even when we allow for markedly different
  gauge coupling constants of $SU(2)_L$ and $SU(2)_R$.
\end{abstract}

\newpage

\section{Introduction}
\label{introduction}

In the Standard Model (SM),
the $Z b \bar b$ interaction is written
\be
\label{mr003p}
\mathcal{L}_{Zb \bar b} = \frac{g}{c_w}\, Z_\mu\, \bar b\, \gamma^\mu
\left( g_L P_L + g_R P_R \right) b,
\ee
where $c_w$ is the cosine of the weak mixing angle,
$P_L$ and $P_R$ are the projection operators of chirality,
and $g_L$ and $g_R$ are the $Z b \bar b$ couplings.
At tree level $g_L$ and $g_R$ have the values
\be
\label{tree}
g_L^\mathrm{tree,SM} = \frac{s_w^2}{3} - \frac{1}{2},
\qquad
g_R^\mathrm{tree,SM} = \frac{s_w^2}{3},
\ee
respectively,
where $s_w$ is the sine of the weak mixing angle.
With $s_w^2 = 0.22339$~\cite{PDG2022},
Eq.~\eqref{tree} gives
$g_L^\mathrm{tree,SM} = -0.42554$ and $g_R^\mathrm{tree,SM} = 0.07446$.
After inclusion of radiative corrections,
the SM prediction for the couplings is~\cite{Field:1997gz}
\be
\label{SM}
g_L^\mathrm{SM} = -0.420875,
\qquad
g_R^\mathrm{SM} = 0.077362.
\ee
In the presence of New Physics (NP),
we define $\delta g_L$ and $\delta g_R$ through
\be
\label{delta}
\delta g_L = g_L - g_L^\mathrm{SM},
\quad
\delta g_R = g_R - g_R^\mathrm{SM}.
\ee

Experimentally,
$g_L$ and $g_R$ are obtained from the observable quantities $A_b$ and $R_b$;
their precise experimental definitions may be found
in Refs.~\cite{Field:1997gz,Haber:1999zh,freitas}
and in appendix~A of Ref.~\cite{Jurciukonis:2021wny}.
One has
\be
\label{Ab_exp}
A_b = \frac{2 r_b \sqrt{1 - 4 \mu_b}}{1
  - 4 \mu_b + \left( 1 + 2 \mu_b \right) r_b^2},
\ee
where $r_b = \left( g_L + g_R \right) \!
\left/ \left( g_L - g_R \right) \right.$
and $\mu_b = \left. m_b^2 \left( m_Z^2 \right) \right/ m_Z^2$.
We use the numerical values $m_b \left( m_Z^2 \right) = 3$\,GeV
and $m_Z = 91.1876$\,GeV~\cite{PDG2022}.
Furthermore,
\be
\label{Rb_exp}
R_b
= \frac{s_b\, c^\mathrm{QCD}\, c^\mathrm{QED}}{s_b\,
  c^\mathrm{QCD}\, c^\mathrm{QED} + s_c + s_u + s_s + s_d},
\ee
where $c^\mathrm{QCD} = 0.9953$ and $c^\mathrm{QED} = 0.99975$
are QCD and QED corrections,
respectively,

\be
\label{s_b}
s_b = \left( 1 - 6 \mu_b \right) \left( g_L - g_R \right)^2
+ \left( g_L + g_R \right)^2,
\ee
and $s_c + s_u + s_s + s_d = 1.3184$.
A recent overall fit of many electroweak observables gives~\cite{freitas}
\bs
\label{dez}
\ba
R_b^\mathrm{fit} &=& 0.21629 \pm 0.00066, \label{Rbfit}
\\
A_b^\mathrm{fit} &=& 0.923 \pm 0.020. \label{Abfit}
\ea
\es
On the other hand,
$A_b$ has been extracted by measuring the $Z$-pole
forward--backward asymmetry $A_{FB}^{0,b}$ at LEP1
and by measuring the left--right forward--backward asymmetry
$A_{LR}^{FB}$ at SLAC---for details see Ref.~\cite{freitas}
and appendix~A of Ref.~\cite{Jurciukonis:2021wny}.
The averaged result of those measurements is
\be
A_b^\mathrm{average} = 0.901 \pm 0.013. \label{Abtrue}
\ee
While the $A_b^\mathrm{fit}$ of Eq.~\eqref{Abfit}
deviates from the SM prediction $A_b^\mathrm{SM} = 0.9347$ by just 0.6$\sigma$,
the $A_b^\mathrm{average}$ of Eq.~\eqref{Abtrue}
displays a much larger disagreement 2.6$\sigma$. 
The $R_b^\mathrm{fit}$ of Eq.~\eqref{Rbfit}
is 0.7$\sigma$ above the SM value $R_b^\mathrm{SM} = 0.21582 \pm 0.00002$.

In this work we consider both the set of values~\eqref{dez},
which we denote through the superscript `fit',
and the set formed by values~\eqref{Rbfit} and~\eqref{Abtrue},
which we denote through the superscript `average'.
Plugging the central values of those two sets
into Eqs.~\eqref{Ab_exp} and~\eqref{Rb_exp},
we obtain four different solutions for $g_L$ and $g_R$---for details
see Ref.~\cite{Jurciukonis:2021wny}.
Two of those solutions may be discarded by both
theoretical and experimental arguments~\cite{Choudhury:2001hs},
while the other two solutions are reasonable;
they are given in table~\ref{table_solutions}. 
\begin{table}[ht]
  \begin{center}
    \begin{tabular}{|c|rr|rr|}
     \hline
     \multicolumn{1}{|c|}{solution} &
     \multicolumn{1}{c}{$g_L$} &
     \multicolumn{1}{c}{$g_R$} &
     \multicolumn{1}{|c}{$\delta g_L$} &
     \multicolumn{1}{c|}{$\delta g_R$} \\ \hline \hline
     1$^\mathrm{fit}$ & $-0.420206$ & $0.084172$ \hspace*{0.1mm} &
     \hspace*{0.1mm} $0.000669$ & $0.006810$ \\
     2$^\mathrm{fit}$ & $-0.419934$ & $-0.082806$ \hspace*{0.1mm} &
     \hspace*{0.1mm} $0.000941$ & $-0.160168$ \\
     \hline
     1$^\mathrm{average}$ & $-0.417814$ & $0.095496$ \hspace*{0.1mm} &
     \hspace*{0.1mm} $0.003061$ & $0.018134$ \\
     2$^\mathrm{average}$ & $-0.417504$ & $-0.094139$ \hspace*{0.1mm} &
     \hspace*{0.1mm} $0.003371$ & $-0.171501$ \\
     \hline
    \end{tabular}
  \end{center}
  \vspace*{-5mm}
  \caption{The results of Eqs.~\eqref{Ab_exp} and~\eqref{Rb_exp}
    for $g_L$ and $g_R$ and the corresponding values of $\delta g_L$
    and $\delta g_R$,
    extracted through Eqs.~\eqref{SM} and~\eqref{delta}.
    The superscript `fit' corresponds to the input values~\eqref{dez}
    while the superscript `average' corresponds
    to the input values~\eqref{Rbfit} and~\eqref{Abtrue}.
    \label{table_solutions}}
\end{table}
Notice that $\delta g_R$ seems to be much larger than $\delta g_L$.

\begin{figure}[ht]
  \begin{center}
    \includegraphics[width=1.0\textwidth]{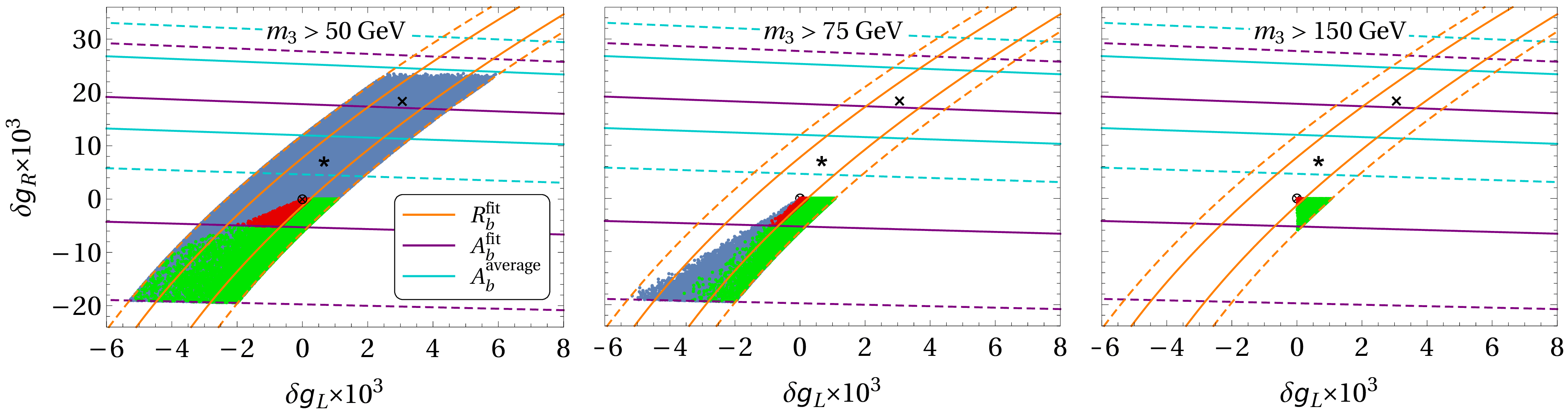}
  \end{center}
  \caption{Scatter plot of the values of $\delta g_L$ and $\delta g_R$
    in the aligned 2HDM.
    See Ref.~\cite{Jurciukonis:2021wny} for details.}
  \label{print39a}
\end{figure}

The two solutions have been studied in the context
of the two-Higgs-doublet model (2HDM)
and three-Higgs-doublet model (3HDM)~\cite{Fontes:2019fbz,Jurciukonis:2021wny};
it has been found that those extensions of the SM
do not improve significantly the fit of the $Zb\bar{b}$ vertex
relative to the SM---see Fig.~\ref{print39a}.

Two decades ago,
research has been carried out on the $Z \to b\bar{b}$ decay asymmetry
in simplified left--right models~\cite{Lee:1999kw,He:2002ha,He:2003qv}.
A model with an additional neutral gauge boson including two Higgs doublets,
a scalar singlet,
and one charged and one neutral vector-like singlets
has been studied in Ref.~\cite{Liu:2017xmc}.
The SM extended by an additional vector boson $Z'$
has been claimed to provide a good fit to $A_{FB}^{0,b}$ near the $Z$ pole
and to $R_b$ measured at energies
above that pole~\cite{Dermisek:2011xu,Dermisek:2012qx}.
In Refs.~\cite{Agashe:2006at,DaRold:2010as} it is shown that
some natural composite Higgs models
with a subgroup of the custodial symmetry $O(3)$ are able to solve
the $A_{FB}^{0,b}$ anomaly while reproducing the observed $R_b$.
There are also studies of contributions from models with extra dimensions
to the process
$Z \to b\bar{b}$~\cite{Papavassiliou:2000pq,Oliver:2002up,Jha:2014faa},
an analysis of $Z$-pole observables
in an effective theory~\cite{Choudhury:2013jta},
revised QCD effects on the $Z b\bar{b}$ forward--backward
asymmetry~\cite{dEnterria:2020cgt},
and a recent explanation~\cite{Crivellin:2020oup}
of the $Zb\bar{b}$ forward--backward asymmetry
by adding to the SM new heavy-quark multiplets---an $SU(2)_L$ doublet
with hypercharge $-5/6$ and an $SU(2)_L$ singlet
with hypercharge $-1/3$.

The discrepancies in $A_b$ may be evidence for NP,
but they may also be due to a statistical fluctuation
or to another experimental effect on one of asymmetries;
more precise experiments are needed.
Hadron colliders may cover the experimental regions
of the $Zb\bar{b}$ couplings of LEP1,
but with large uncertainties~\cite{Murphy:2015cha}.
Lepton colliders of the next generation offer better opportunities
for further studies of the $Zb\bar{b}$ vertex,
since they could collect a large amount of data
around the $Z$ pole~\cite{Gori:2015nqa}.
Some recent papers~\cite{Yan:2021veo,Dong:2022ayy,Yan:2021htf,Li:2021uww}
propose novel methods to probe the $Z b\bar b$ couplings
at both existing and future colliders.

In this paper we seek to reproduce solution~1
in table~\ref{table_solutions} by invoking a version of the left--right model
(LRM)~\cite{Pati:1974yy,Mohapatra:1974gc,
  Mohapatra:1974hk,Senjanovic:1975rk,Senjanovic:1978ev},
\ie\ a model with gauge group $SU(2)_L \times SU(2)_R \times U(1)$,
at the one-loop level. (We also comment on the possibility
of reproducing that solution at the tree level.)
We are inspired in this endeavour by the observation that,
since $\delta g_R$ appears to be much larger than $\delta g_L$,
then maybe a model with right-handed gauge interactions
provides a better fit to the $Z b \bar b$ vertex;
the same rationale was used before in Ref.~\cite{He:2003qv}.
The version of the LRM
that we use here is characterized by the following four features:
\begin{itemize}
\item The gauge coupling constants of $SU(2)_L$ and $SU(2)_R$
  are allowed to differ.\footnote{This is fully compatible
  with left--right symmetry at very high energies.
  Namely,
  that \emph{discrete} symmetry,
  which has been called $P$ in Ref.~\cite{Chang:1983fu}
  or $D$ in some other papers,
  may be broken at a very high energy
  while keeping the $SU(2)_L \times SU(2)_R \times U(1)$ gauge group
  intact~\cite{Chang:1983fu}.}$^,$\footnote{We do not take
  into account in this paper the constraints on the LRM
  derived in Refs.~\cite{Bernard:2020cyi,Karmakar:2022iip},
  since those papers assume left--right symmetry,
  at least in the scalar potential.}
  This is done in order to allow greater flexibility of the LRM
  in fitting the $Z b \bar b$ vertex.
\item The spontaneous symmetry breaking of $SU(2)_L$ and $SU(2)_R$
  is achieved exclusively by scalar doublets of those groups,
  avoiding the triplets that exist in other versions of the LRM.
\item We assume $CP$ conservation,
  both at the Lagrangian level and in the vacuum state.
\item The lightest neutral gauge boson couples to the left- and
right-handed fermions with exactly the same strength
as the $Z$ gauge boson of the SM.
\end{itemize}
While the first feature above complicates our LRM,
the other three features simplify it considerably.
The fourth feature is very helpful to our computation,
because it makes the infrared divergences
in the $Z_l b \bar b$ vertex of the LRM
exactly identical to the same divergences
in the $Z b \bar b$ vertex of the SM;
those divergences then disappear when comparing the vertices in the two models.
That feature should not constitute an unreasonable restriction,
because---as discussed below---previous studies of
(different versions of)
the LRM suggest that
the mixing of the two neutral massive gauge bosons of the LRM
should be extremely small anyway.

Another feature of our model is that
we only include in it the top and bottom quarks---we neglect
both all the other quarks and the leptons;
we do this because they are inessential for the $Z b \bar b$ vertex.

In this paper we carefully work out the renormalization of the vertex,
which is non-trivial because of the enlarged gauge group.
This forces us to be painstaking in the definition of the model and,
in particular,
of all its symmetries.
The paper is organized as follows.
In Sec.~\ref{sec:description} we describe
the gauge structure of the model.
Section~\ref{sec:tree-level} deals with
the fit of the $Z b \bar b$ vertex in the LRM at tree level.
Section~\ref{sec:scalars} proceeds with the description
of the scalar structure of the model.
In Sec.~\ref{sec:parameters} we collect all the parameters of the model
and outline our procedure for fitting the $Z b \bar b$ vertex.
Section~\ref{sec:renormalization} deals with the one-loop calculation
and the renormalization procedure.
In Sec.~\ref{sec:numerical} we give the practical results of our work.
Thereafter,
many appendices deal in detail with technical issues.

\section{Description of the model: gauge interactions and quarks}
\label{sec:description}

\subsection{Gauge coupling constants and covariant derivative}

\paragraph{Gauge coupling constants, $\theta_w$, and $\alpha$:}
We consider a \emph{$CP$-conserving} left--right model (LRM),
\textit{i.e.}\ a model with gauge group
$SU(2)_L \times SU(2)_R \times U(1)_X$.
The gauge coupling constant of $SU(2)_L$ is $g$;
the gauge coupling constant of $SU(2)_R$ is $l$;\footnote{The
left--right-symmetric model assumes $l = g$.
\emph{We allow\/ $l$ to be different from $g$} for the sake of generality.}
the gauge coupling constant of $U(1)_X$ is $h$.
We define
\bs
\label{10}
\ba
& G \equiv g^2, \qquad L \equiv l^2, \qquad H \equiv h^2, &
\\*[2mm]
\label{rhos1}
& \varrho_1 \equiv \sqrt{G L + G H + L H},
\qquad
\varrho_2 \equiv g \sqrt{L + H}, &
\\*[1mm]
\label{nfore0433}
& \displaystyle{c_w \equiv \frac{\varrho_2}{\varrho_1}, \quad
s_w \equiv \frac{- l h}{\varrho_1},
\qquad
c_\alpha \equiv \frac{g l}{\varrho_2}, \quad
s_\alpha \equiv \frac{- g h}{\varrho_2},} &
\ea
\es
where $c_w \equiv \cos{\theta_w}$,
$s_w \equiv \sin{\theta_w}$,
$c_\alpha \equiv \cos{\alpha}$,
and $s_\alpha \equiv \sin{\alpha}$.
The electromagnetic coupling constant $e$ is given by
\be
\label{eeeee}
e = \frac{- g l h}{\varrho_1}, \qquad E \equiv e^2.
\ee
Notice that
\be
L c_w^2 - G s_w^2 = \frac{G L^2}{\varrho_1^2} > 0.
\label{mvjfgop}
\ee
Using the measured value of $s_w$,
Eq.~\eqref{mvjfgop} produces the lower bound
\be
\label{nvkff0005}
\left| \frac{l}{g} \right| > \left| \frac{s_w}{c_w} \right| \approx 0.53.
\ee

\paragraph{Covariant derivative:}
The covariant derivative is
\ba
\label{covder}
D^\mu &=& \partial^\mu
- i g \left( T_L^+ W^{+ \mu} + T_L^- W^{- \mu} \right)
- i l \left( T_R^+ V^{+ \mu} + T_R^- V^{- \mu} \right)
\no & &
+ i e A^\mu Q - i\, \frac{g}{c_w}\, Z^\mu \left( T_{L3} - Q s_w^2 \right)
- i\ \frac{l}{c_\alpha}\, X^\mu \left( T_{R3} - Y s_\alpha^2  \right),
\ea
where
\begin{itemize}
\item $T_L^\pm$ and $T_R^\pm$ are the raising and lowering operators
  of $SU(2)_L$ and $SU(2)_R$,
  respectively;
\item $T_{L3}$ and $T_{R3}$ are the third generators
  of $SU(2)_L$ and $SU(2)_R$,
  respectively;
\item $Y = T_{R3} + X$ is the weak hypercharge
  and $Q = T_{L3} + Y$ is the electric charge;\footnote{$X$ is
  the quantum number that generates $U(1)_X$.
  It should not be confused with the gauge field $X^\mu$.}
\item $A^\mu$ is the photon field,
  which is the only massless gauge field.
\end{itemize}

\paragraph{Signs:}
It is clear in Eq.~\eqref{covder} that
\begin{itemize}
\item The sign of the field $W^{\pm \mu}$ may be chosen
  so that $g$ is positive.
\item The sign of the field $V^{\pm \mu}$ may be chosen
  so that $l$ is positive.
\item The sign of the field $A^\mu$ may be chosen
  so that $e$ is negative.
\item The sign of the field $Z^\mu$ may be chosen
  so that $g \left/ c_w \right.$ is positive.
\item The sign of the field $X^\mu$ may be chosen
  so that $l \left/ c_\alpha \right.$ is positive.
\end{itemize}
Accordingly,
from now on we shall assume that $g$,
$l$,
$- e$,
$c_w$,
and $c_\alpha$ are positive.
Equation~\eqref{nfore0433} then informs us that $\varrho_2$ and $\varrho_1$
are positive too.
Equation~\eqref{eeeee} tells us that $h$ is positive too.
Hence,
$\theta_w$ and $\alpha$ are both angles of the fourth quadrant
in our sign convention.

\subsection{Gauge-boson mixing}

The fields $Z_\mu$ and $X_\mu$ mix,
just as the fields $W^+_\mu$ and $V^+_\mu$.
We write
\be
\label{ufiee99843}
\left( \begin{array}{c} Z_\mu \\ X_\mu \end{array} \right)
=
\left( \begin{array}{cc} c_\psi & - s_\psi \\ s_\psi & c_\psi \end{array} \right)
\left( \begin{array}{c} Z_{l\mu} \\ Z_{h\mu} \end{array} \right),
\qquad
\left( \begin{array}{c} W^+_\mu \\*[1mm] V^+_\mu \end{array} \right)
=
\left( \begin{array}{cc} c_\xi & s_\xi \\*[1mm]
- s_\xi & c_\xi \end{array} \right)
\left( \begin{array}{c} W^+_{l\mu} \\*[1mm] W^+_{h\mu} \end{array} \right),
\ee
where $Z_{l\mu}$ and $Z_{h\mu}$ are neutral eigenstates
of mass with squared masses $M_l$ and $M_h$,
respectively,
and $W_{l\mu}^+$ and $W_{h\mu}^+$ are charged eigenstates
of mass with squared masses $\bar M_l$ and $\bar M_h$,
respectively.
By definition,
$M_l < M_h$ and $\bar M_l < \bar M_h$.
We identify $Z_l$ as the observed neutral gauge boson with mass
$m_Z = \sqrt{M_l} = 91.1876$\,GeV
and $W_l^\pm$ as the observed charged gauge bosons with mass
$m_W = \sqrt{\bar M_L} = 80.378$\,GeV.
In Eqs.~\eqref{ufiee99843} $c_\psi \equiv \cos{\psi}$,
$s_\psi \equiv \sin{\psi}$,
$c_\xi \equiv \cos{\xi}$,
and $s_\xi \equiv \sin{\xi}$,
where $\psi$ and $\xi$ are mixing angles.
Note that,
if our LRS model had not been assumed to be $CP$-invariant,
then the second Eq.~\eqref{ufiee99843}
would have contained a phase in the mixing matrix.
Without loss of generality,
one may choose the overall sign of $\left( Z_{l \mu},\ Z_{h \mu} \right)^T$
in such a way that $c_\psi$ is non-negative;\footnote{The \emph{relative}
sign of $Z_{l \mu}$ and $Z_{h \mu}$ is fixed by the requirement
that the mixing matrix
$\left( \begin{array}{cc} c_\psi & - s_\psi \\
  s_\psi & c_\psi \end{array} \right)$
has \emph{positive} determinant.}
similarly,
we also assume $c_\xi \ge 0$.
Equations~\eqref{ufiee99843} come about
because in the Lagrangian there are mass terms
\bs
\label{cieoggkl}
\ba
\mathcal{L} &=& \cdots +
\frac{1}{2}
\left( \begin{array}{cc} Z_\mu & X_\mu \end{array} \right)
M_n
\left( \begin{array}{c} Z^\mu \\ X^\mu \end{array} \right)
+
\left( \begin{array}{cc} W_\mu^- & V_\mu^- \end{array} \right)
M_c
\left( \begin{array}{c} W^{+\mu} \\ V^{+\mu} \end{array} \right)
\label{jvirr0w0e} \\ &=&
\frac{1}{2} \left( M_l\, Z_{l \mu} Z_l^\mu + M_h\, Z_{h \mu} Z_h^\mu \right)
+ \bar M_l\, W_{l \mu}^- W_l^{\mu +}
+ \bar M_h\, W_{h \mu}^- W_h^{\mu +}.
\ea
\es
In Eq.~\eqref{jvirr0w0e} the $2 \times 2$ matrices $M_n$ and $M_c$
are real and symmetric.

\subsection{Quarks}

\paragraph{Multiplets:}
In our simplified LRM we only consider the third-generation quarks,
\viz\ the left-handed $t_L$ and $b_L$
and the right-handed $t_R$ and $b_R$;\footnote{There should be no confusion
between the left-handed and right-handed bottom-quark fields---$b_L$ and $b_R$,
respectively---and the scalar field $b$.}
we disconsider both the lepton sector and the other two quark generations.
Under $SU(2)_L \times SU(2)_R \times U(1)_X$,
\be
\label{sym2}
\left( \begin{array}{c} t_L \\ b_L \end{array} \right)
\to \mathcal{U_L} \left( \begin{array}{c} t_L \\ b_L \end{array} \right)
e^{i \gamma / 6},
\qquad
\left( \begin{array}{c} t_R \\ b_R \end{array} \right)
\to \mathcal{U_R} \left( \begin{array}{c} t_R \\ b_R \end{array} \right)
e^{i \gamma / 6}.
\ee
The quantum numbers of the quark fields are in table~\ref{u9}.
\begin{table}[h]
\begin{center}
\begin{tabular}{|c||c|c|c|c|c|}
  \hline
  fields & $T_{L3}$ & $T_{R3}$ & $X$ & $Y=T_{R3}+X$ & $Q=T_{L3}+Y$ \\ \hline
  $t_L$ & $1/2$ & $0$ & $1/6$ & $1/6$ & $2/3$ \\
  $b_L$ & $-1/2$ & $0$ & $1/6$ & $1/6$ & $-1/3$ \\
  $t_R$ & $0$ & $1/2$ & $1/6$ & $2/3$ & $2/3$ \\
  $b_R$ & $0$ & $-1/2$ & $1/6$ & $-1/3$ & $-1/3$ \\
\hline
\end{tabular}
\end{center}
{\caption{The $U(1)$ quantum numbers of the quark fields.}\label{u9}}
\end{table}

\section{Fitting the $Z b \bar b$ coupling at tree level in the LRM}
\label{sec:tree-level}

\begin{figure}[ht]
  \begin{center}
    \includegraphics[width=1.0\textwidth]{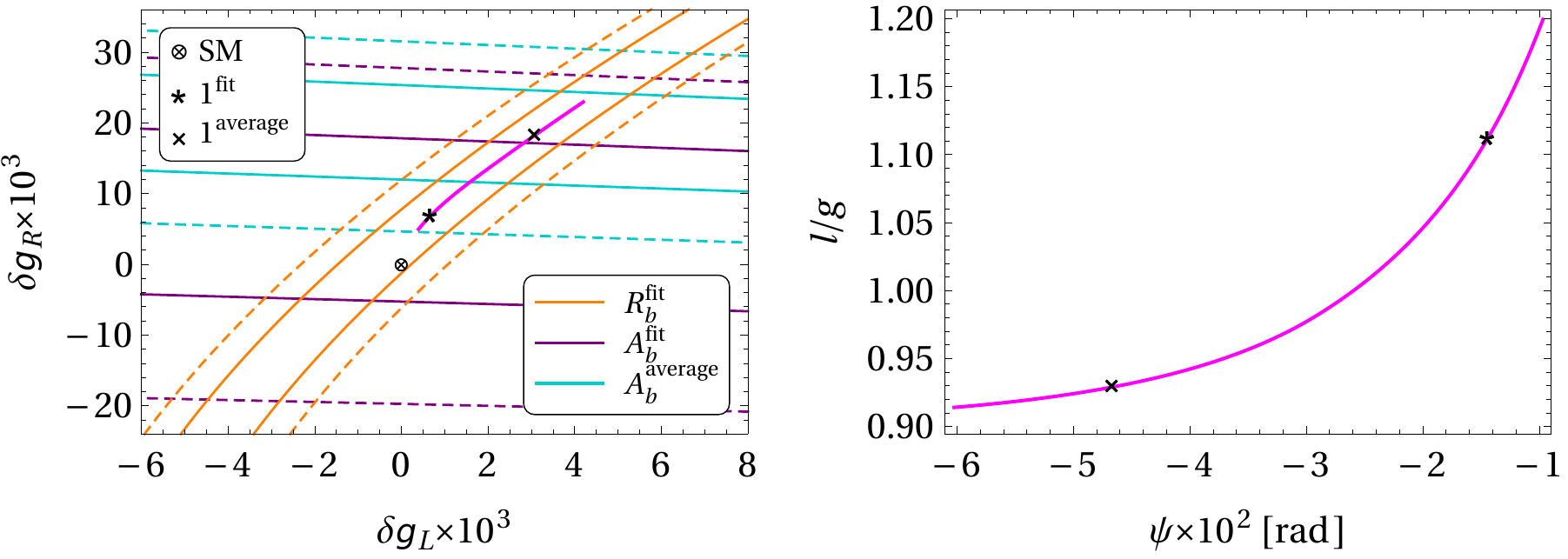}
  \end{center}
  \caption{Left panel:
    a smooth trajectory in the $\delta g_R$ \textit{vs}.\ $\delta g_L$ plane
    connecting the $1^\mathrm{fit}$ and $1^\mathrm{average}$ points
    of table~\ref{table_solutions}.
    Right panel:
    the same trajectory translated into the $l/g$ \textit{vs}.\ $\psi$ plane
    through solving Eqs. \eqref{ue93003}.
    We have used $c_w^2 = m_W^2 \left/ m_Z^2 \right.$.}
  \label{XXX}
\end{figure}

With the covariant derivative in Eq.~\eqref{covder}
and since the bottom quark has the quantum numbers in table~\ref{u9},
one has
\bs
\ba
\mathcal{L} &=& \cdots
+ \bar b_L \gamma_\mu \left[ \frac{g}{c_w}\, Z^\mu
  \left( \frac{s_w^2}{3} - \frac{1}{2} \right)
  - \frac{l s_\alpha^2}{6 c_\alpha}\, X^\mu \right] b_L
\no & &
+ \bar b_R \gamma_\mu \left[ \frac{g s_w^2}{3 c_w}\, Z^\mu
  + \frac{l}{c_\alpha}\, X^\mu
  \left( \frac{s_\alpha^2}{3} - \frac{1}{2} \right) \right] b_R.
\ea
\es
Using both Eq.~\eqref{nfore0433} and the first Eq.~\eqref{ufiee99843},
we find the coupling of the $b$ quark
to the light neutral gauge boson---which we identify
as the observed one---written in the form of Eq.~\eqref{mr003p} as
\be
\mathcal{L}_{Z_lb \bar b} = \frac{g}{c_w}\, Z_l^\mu\, \bar b\, \gamma_\mu
\left( g_L^\mathrm{tree,LRM} P_L + g_R^\mathrm{tree,LRM} P_R \right) b,
\ee
with
\bs
\label{jvfg044}
\ba
g_L^\mathrm{tree,LRM} &=&
c_\psi \left( \frac{s_w^2}{3} - \frac{1}{2} \right)
- s_\psi\, \frac{g s_w^2}{6 \sqrt{L c_w^2 - G s_w^2}},
\\
g_R^\mathrm{tree,LRM} &=&
c_\psi\, \frac{s_w^2}{3}
+ s_\psi\, \frac{g}{\sqrt{L c_w^2 - G s_w^2}} \left( \frac{s_w^2}{3} -
\frac{L c_w^2}{2 G} \right).
\ea
\es
One sees that,
besides the Weinberg angle $\theta_w$,
the two quantities $g_L^\mathrm{tree,LRM}$ and $g_R^\mathrm{tree,LRM}$
depend on two parameters of the LRM,
\viz\ $l/g$ and $\psi$.
It should be possible to adjust the latter in order to reproduce
the observed  $\delta g_L$ and $\delta g_R$
in either the `average' or `fit' solutions:
\bs
\label{ue93003}
\ba
\delta g_L &=&
\left( \frac{s_w^2}{3} - \frac{1}{2} \right)
\left( \cos{\psi} - 1 \right)
- \frac{s_w^2}{6} \left( \frac{l^2}{g^2}\, c_w^2 - s_w^2 \right)^{-1/2}
\sin{\psi},
\\
\delta g_R &=&
\frac{s_w^2}{3} \left( \cos{\psi} - 1 \right)
+
\left( \frac{s_w^2}{3} - \frac{l^2}{g^2}\, \frac{c_w^2}{2} \right)
\left( \frac{l^2}{g^2}\, c_w^2 - s_w^2 \right)^{-1/2}
\sin{\psi}.
\ea
\es
The solution to Eqs.~\eqref{ue93003}
is depicted in Fig.~\ref{XXX}.
In particular,
one sees that
\be
\frac{l}{g} = 1.112, \quad \psi = -0.0144 \quad
\mbox{for\ the\ `fit'\ solution}
\ee
and
\be
\frac{l}{g} = 0.929, \quad \psi = -0.0467 \quad
\mbox{for\ the\ `average'\ solution}.
\ee
Thus,
in order to fit the $Z b \bar b$ vertex in the LRM at tree level
one needs a mixing angle $\psi \lesssim -10^{-2}$;
moreover,
one should not be very far from the left--right-symmetric case $l = g$.

It remains to be seen whether values of $l/g$ and $\psi$
like those in the right panel of Fig.~\ref{XXX} are compatible
with phenomenology.
This matter goes beyond the scope of the present work.
We just refer to the rather old Ref.~\cite{Erler:2009jh}
and the more recent Refs.~\cite{Bobovnikov:2018fwt,Osland:2020onj,Osland:2022ryb};
their authors have investigated how large $\psi$ is allowed to be
in a left--right-symmetric model
(\viz\ a model with $l=g$).
They all conclude that $\left| \psi \right| \lesssim 10^{-3}$.
It is conceivable that allowing for $l/ g \neq 1$
may give $\left| \psi \right|$ some room for being larger;
but $\psi \sim -0.01$ seems far-fetched.
One thus concludes that using the LRM at tree level
for fitting the $Z b \bar b$ vertex does not work.
It remains to be seen whether the LRM at one-loop level can do a better job;
that is the aim of this work.
Unfortunately,
in order to do that job properly
one must carefully renormalize the LRM;
that forces us to define the model completely,
including its scalar sector.
That is what we do in the next section.

\section{Description of the model: scalars}
\label{sec:scalars}

\subsection{Multiplets}

The scalar multiplets of our LRM
consist of an $SU(2)_L$ doublet $H_L$,
an $SU(2)_R$ doublet $H_R$,\footnote{Other
left--right models use triplets of $SU(2)_L$ and $SU(2)_R$
instead of $H_L$ and $H_R$.}
and a `bi-doublet'---\textit{i.e.},
a doublet both of $SU(2)_L$ and of $SU(2)_R$---$\Phi$.
Thus,
\be
\label{mg9996}
H_L = \left( \begin{array}{c} m \\*[1mm] n \end{array} \right),
\quad
H_R = \left( \begin{array}{c} p \\*[1mm] q \end{array} \right),
\quad
\Phi = \left( \begin{array}{cc} b^\ast & c \\ - a^\ast & d \end{array} \right),
\quad
\tilde \Phi \equiv \tau_2 \Phi^\ast \tau_2 =
\left( \begin{array}{cc} d^\ast & a \\ - c^\ast & b \end{array} \right),
\ee
where $m$, $n$, $p$, $q$, $a$, $b$, $c$, and $d$
are complex Klein--Gordon fields.
The multiplets~\eqref{mg9996}
transform under $SU(2)_L \times SU(2)_R \times U(1)_X$ as
\be
\label{sym1}
H_L \to \mathcal{U_L}\, H_L\, e^{i \gamma/2},
\quad
H_R \to \mathcal{U_R}\, H_R\, e^{i \gamma/2},
\quad
\Phi \to \mathcal{U_L}\, \Phi\ \mathcal{U_R}^\dagger,
\quad
\tilde \Phi \to \mathcal{U_L}\, \tilde \Phi\ \mathcal{U_R}^\dagger,
\ee
where $\mathcal{U_L}$ and $\mathcal{U_R}$
are the $2 \times 2$ unitary matrices with determinant~1
representing the $SU(2)_L$ and $SU(2)_R$ transformations,
respectively,
in the doublet representation.
The phase $\gamma$ is the parameter of the $U(1)_X$ transformation:
both $H_L$ and $H_R$ have $X = 1/2$ while $\Phi$ has $X = 0$.
The $U(1)$ quantum numbers of the scalar fields are given in table~\ref{u1}.
\begin{table}[h]
\begin{center}
\begin{tabular}{|c||c|c|c|c|c|}
  \hline
  fields & $T_{L3}$ & $T_{R3}$ & $X$ & $Y=T_{R3}+X$ & $Q=T_{L3}+Y$ \\ \hline
  $m$ & $1/2$ & $0$ & $1/2$ & $1/2$ & $1$ \\
  $n$ & $-1/2$ & $0$ & $1/2$ & $1/2$ & $0$ \\
  $p$ & $0$ & $1/2$ & $1/2$ & $1$ & $1$ \\
  $q$ & $0$ & $-1/2$ & $1/2$ & $0$ & $0$ \\
  $a,\ c$ & $1/2$ & $1/2$ & $0$ & $1/2$ & $1$ \\
  $b,\ d$ & $-1/2$ & $1/2$ & $0$ & $1/2$ & $0$ \\
\hline
\end{tabular}
\end{center}
{\caption{The $U(1)$ quantum numbers of the scalar fields.}\label{u1}}
\end{table}

\subsection{VEVs}

The electromagnetism-conserving\footnote{We
\emph{assume} conservation of electromagnetism by the vacuum state.}
vacuum expectation values (VEVs) are
\bs
\label{VEVs}
\ba
&
\left\langle 0 \left| m \right| 0 \right\rangle
= \left\langle 0 \left| p \right| 0 \right\rangle
= \left\langle 0 \left| a \right| 0 \right\rangle
= \left\langle 0 \left| c \right| 0 \right\rangle
= 0,
&
\\*[1mm]
&
\left\langle 0 \left| n \right| 0 \right\rangle = u_L, \quad
\left\langle 0 \left| q \right| 0 \right\rangle = u_R, \quad
\left\langle 0 \left| b \right| 0 \right\rangle = v_1, \quad
\left\langle 0 \left| d \right| 0 \right\rangle = v_2.
&
\ea
\es
Since we assume our model to be $CP$-conserving,
$u_L$,
$u_R$,
$v_1$,
and $v_2$ are taken to be \emph{real}.
Without loss of generality,
one may choose the signs of the scalar multiplets to set $u_L$,
$u_R$,
and $v_2$ non-negative;
only the sign of $v_1$ remains free.
We define
\be
U_L \equiv u_L^2, \quad
U_R \equiv u_R^2, \quad
V_1 \equiv v_1^2, \quad
V_2 \equiv v_2^2.
\ee
One may interchange $\Phi$ and $\tilde \Phi$,
\textit{i.e.}\ one may make $a \leftrightarrow c$ and $b \leftrightarrow d$.
Thus,
from now on we shall assume $V_2$ to be larger than $V_1$.

\subsection{Mixing of the scalars}

\paragraph{Definition of the mixing matrices:}
We expand the neutral-scalar fields about their VEVs as
\be
\label{expansion}
n = u_L + \frac{\rho_L + i \eta_L}{\sqrt{2}}, \quad
q = u_R + \frac{\rho_R + i \eta_R}{\sqrt{2}}, \quad
b = v_1 + \frac{\rho_1 + i \eta_1}{\sqrt{2}}, \quad
d = v_2 + \frac{\rho_2 + i \eta_2}{\sqrt{2}},
\ee
where $\rho_{L,R,1,2}$ and $\eta_{L,R,1,2}$ are \emph{real} Klein--Gordon fields.
Because of the assumed $CP$ invariance,
the fields $\rho_{L,R,1,2}$
(\textit{viz.}\ the scalars)
mix among themselves,
but they do not mix with the fields $\eta_{L,R,1,2}$
(\textit{viz.}\ the pseudoscalars).
Thus,
\be
\label{odefinition}
\left( \begin{array}{c} \rho_1 \\ \rho_2 \\ \rho_L \\ \rho_R \end{array} \right)
= V_\rho
\left( \begin{array}{c} S^0_5 \\*[1mm] S^0_6 \\*[1mm] S^0_7 \\*[1mm] S^0_8
\end{array} \right),
\ee
where $S^0_{5,6,7,8}$ are real eigenstates of mass with masses $\mu_{5,6,7,8}$,
respectively.
The $4 \times 4$ matrix $V_\rho$ is real and orthogonal.
Analogously to Eq.~\eqref{odefinition},
\be
\label{vdefinition}
\left( \begin{array}{c} \eta_1 \\ \eta_2 \\ \eta_L \\ \eta_R \end{array} \right)
= V_\eta
\left( \begin{array}{c} G_l^0 \\*[1mm] G_h^0 \\*[1mm] S^0_3 \\*[1mm] S^0_4
\end{array} \right),
\ee
where $G_l^0$ and $G_h^0$ are the Goldstone bosons
that are `eaten' by $Z_{l \mu}$ and $Z_{h \mu}$,
respectively,
while $S^0_3$ and $S^0_4$ are physical pseudoscalars
with squared masses $M_{\eta 1}$ and $M_{\eta 2}$,
respectively.
The matrix $V_\eta$ is real and orthogonal.
Also,
\be
\label{tdefinition}
\left( \begin{array}{c} a \\ c \\ m \\ p \end{array} \right)
= V_\varphi
\left( \begin{array}{c} G_l^+ \\*[1mm] G_h^+ \\*[1mm] H_3^+ \\*[1mm] H_4^+
\end{array} \right),
\ee
where $G_l^+$ and $G_h^+$ are Goldstone bosons
that are `eaten' by $W_{l \mu}^+$ and $W_{h \mu}^+$,
respectively,
while $H_3^+$ and $H_4^+$ are physical charged scalars
with squared masses $M_{\varphi 1}$ and $M_{\varphi 2}$,
respectively.
The matrix $V_\varphi$ is real and orthogonal
because of the assumed $CP$ conservation.

\paragraph{Parameterization of $V_\rho$:}
The parameterization that we use for the matrix $V_\rho$
in Eq.~\eqref{odefinition} is the following:
\ba
V_\rho &=&
\left( \begin{array}{cccc}
  1 & 0 & 0 & 0 \\
  0 & 1 & 0 & 0 \\
  0 & 0 & c_3 & - s_3 \\
  0 & 0 & s_3 & c_3
\end{array} \right)
\times
\left( \begin{array}{cccc}
  1 & 0 & 0 & 0 \\
  0 & c_2 & -s_2 & 0 \\
  0 & s_2 & c_2 & 0 \\
  0 & 0 & 0 & 1
\end{array} \right)
\times
\left( \begin{array}{cccc}
  1 & 0 & 0 & 0 \\
  0 & 1 & 0 & 0 \\
  0 & 0 & c_6 & s_6 \\
  0 & 0 & s_6 & -c_6
\end{array} \right)
\no & &
\times
\left( \begin{array}{cccc}
  c_1 & s_1 & 0 & 0 \\
  s_1 & -c_1 & 0 & 0 \\
  0 & 0 & 1 & 0 \\
  0 & 0 & 0 & 1
\end{array} \right)
\times
\left( \begin{array}{cccc}
  1 & 0 & 0 & 0 \\
  0 & s_4 & 0 & c_4 \\
  0 & 0 & 1 & 0 \\
  0 & c_4 & 0 & -s_4
\end{array} \right)
\times
\left( \begin{array}{cccc}
  1 & 0 & 0 & 0 \\
  0 & 0 & c_5 & s_5 \\
  0 & 0 & s_5 & -c_5 \\
  0 & 1 & 0 & 0
\end{array} \right), \hspace*{7mm}
\label{uidfr30}
\ea
where $c_i \equiv \cos{\theta_i}$ and $s_i \equiv  \sin{\theta_i}$
for $i = 1, \ldots, 6$.
In this way one obtains a matrix $V_\rho$ with a simple first row
and a simple first column:
\be
\label{c93kf}
V_\rho = \left( \begin{array}{cccc}
  c_1 & s_1 c_4 & s_1 s_4 c_5 & s_1 s_4 s_5 \\
  s_1 c_2 & \left( V_\rho \right)_{22} &
  \left( V_\rho \right)_{23} &
  \left( V_\rho \right)_{24} \\
  s_1 s_2 c_3 & \left( V_\rho \right)_{32} &
  \left( V_\rho \right)_{33} &
  \left( V_\rho \right)_{34} \\
  s_1 s_2 s_3 & \left( V_\rho \right)_{42} &
  \left( V_\rho \right)_{43} &
  \left( V_\rho \right)_{44}
\end{array} \right).
\ee
The sign of $S^0_5$ is chosen in such a way
that $\det{V_\rho} = +1$ is positive.
We choose the signs of $S^0_6$,
$S^0_7$,
and $S^0_8$ in such a way that $c_4$,
$c_5$,
and $s_5$ are non-negative.

\paragraph{Goldstone bosons:}
Since the scalar doublets of $SU(2)_L$ are
\be
\label{su2l}
\left( \begin{array}{c} m \\ n \end{array} \right),
\quad
\left( \begin{array}{c} a \\ b \end{array} \right),
\quad \mbox{and} \quad
\left( \begin{array}{c} c \\ d \end{array} \right),
\ee
and the scalar doublets of $SU(2)_R$ are
\be
\label{su2r}
\left( \begin{array}{c} p \\ q \end{array} \right),
\quad
\left( \begin{array}{c} - a \\ d^\ast \end{array} \right),
\quad \mbox{and} \quad
\left( \begin{array}{c} - c \\ b^\ast \end{array} \right),
\ee
the neutral fields
\be
u_L \eta_L + v_1 \eta_1 + v_2 \eta_2
\quad \mbox{and} \quad
u_R \eta_R - v_1 \eta_1 - v_2 \eta_2
\label{gb0}
\ee
are Goldstone bosons.
The charged fields
\be
u_L m + v_1 a + v_2 c \quad \mbox{and} \quad u_R p - v_2 a - v_1 c
\label{gb+}
\ee
are Goldstone bosons too.
Thus,
there are two neutral Goldstone bosons and two charged Goldstone bosons,
and it is a non-trivial problem to find out how they mix
to form the states that are `swallowed' by the neutral gauge bosons
$Z_{l \mu}$ and $Z_{h \mu}$
and by the charged gauge bosons $W_{l \mu}^+$ and $W_{h \mu}^+$,
respectively.
This problem is addressed in Appendix~\ref{app:goldstonebosons}.
The results of that appendix may be summarized as follows:
\be
\label{mf049lg}
\begin{array}{ll}
  \displaystyle{\left( V_\eta \right)_{11} =
  \frac{v_1\left( c_\psi \varrho_1 - s_\psi L \right)}{\sqrt{2 M_l}\,
    \sqrt{L + H}},}
  & \displaystyle{\left( V_\eta \right)_{12} =
  \frac{v_1\left( - s_\psi \varrho_1 - c_\psi L \right)}{\sqrt{2 M_h}\,
    \sqrt{L + H}},}
  \\*[3mm]
  \displaystyle{\left( V_\eta \right)_{21} =
  \frac{v_2 \left( c_\psi \varrho_1 - s_\psi L \right)}{\sqrt{2 M_l}\,
    \sqrt{L + H}},}
  & \displaystyle{\left( V_\eta \right)_{22} =
  \frac{v_2 \left( - s_\psi \varrho_1 - c_\psi L \right)}{\sqrt{2 M_h}\,
    \sqrt{L + H}},}
  \\*[3mm]
  \displaystyle{\left( V_\eta \right)_{31} =
  \frac{u_L \left( c_\psi \varrho_1 + s_\psi H \right)}{\sqrt{2 M_l}\,
    \sqrt{L + H}},}
  &
  \displaystyle{\left( V_\eta \right)_{32} =
  \frac{u_L \left( - s_\psi \varrho_1 + c_\psi H \right)}{\sqrt{2 M_h}\,
    \sqrt{L + H}},}
  \\*[3mm]
  \displaystyle{\left( V_\eta \right)_{41} =
    s_\psi\, \frac{u_R\, \sqrt{L + H}}{\sqrt{2 M_l}},}
  & \displaystyle{\left( V_\eta \right)_{42} = c_\psi\, \frac{u_R\,
    \sqrt{L + H}}{\sqrt{2 M_h}},}
\end{array}
\ee
and
\be
\label{tt1}
\begin{array}{ll}
  \displaystyle{\left( V_\varphi \right)_{11}
  = \frac{c_\xi g v_1 + s_\xi l v_2}{\sqrt{2 \bar M_l}},}
  &
  \displaystyle{\left( V_\varphi \right)_{12}
  = \frac{s_\xi g v_1 - c_\xi l v_2}{\sqrt{2 \bar M_h}},}
  \\*[3mm]
  \displaystyle{\left( V_\varphi \right)_{21}
  = \frac{c_\xi g v_2 + s_\xi l v_1}{\sqrt{2 \bar M_l}},}
  &
  \displaystyle{\left( V_\varphi \right)_{22}
  = \frac{s_\xi g v_2 - c_\xi l v_1}{\sqrt{2 \bar M_h}},}
  \\*[3mm]
  \displaystyle{\left( V_\varphi \right)_{31} =
    \frac{c_\xi g u_L}{\sqrt{2 \bar M_l}},}
  &
  \displaystyle{\left( V_\varphi \right)_{32} =
    \frac{s_\xi g u_L}{\sqrt{2 \bar M_h}},}
  \\*[3mm]
  \displaystyle{\left( V_\varphi \right)_{41} =
    \frac{- s_\xi l u_R}{\sqrt{2 \bar M_l}},}
  &
  \displaystyle{\left( V_\varphi \right)_{42} =
    \frac{c_\xi l u_R}{\sqrt{2 \bar M_h}}.}
\end{array}
\ee

\paragraph{The physical pseudoscalars:}

Two orthonormalized linear combinations of the $\eta$ fields
that are orthogonal to the Goldstone bosons~\eqref{gb0} are
\be
\eta_a \equiv \frac{v_2 \eta_1 - v_1 \eta_2}{\sqrt{T_1}},
\qquad
\eta_b \equiv
\frac{T_1 \left( u_L \eta_R + u_R \eta_L \right)
+ u_L u_R \left( v_1 \eta_1 + v_2 \eta_2 \right)}{\sqrt{T_1} \sqrt{T_2}},
\label{etaaetab}
\ee
where
\be
\label{T1T2}
T_1 \equiv V_1 + V_2,
\qquad
T_2 \equiv U_L U_R + T_1 \left( U_L + U_R \right).
\ee
Note that $\sqrt{T_1}$ and $\sqrt{T_2}$ are positive---this
corresponds to the definition of the signs of $\eta_a$ and $\eta_b$.
The Lagrangian contains mass terms for the pseudoscalar fields as
\be
\mathcal{L} = \cdots - \frac{1}{2}
\left( \begin{array}{cc} \eta_a, & \eta_b \end{array} \right)
M_\eta
\left( \begin{array}{c} \eta_a \\ \eta_b \end{array} \right),
\label{Meta}
\ee
where $M_\eta$ is a real $2 \times 2$ symmetric matrix.
This matrix is diagonalized as
\be
\label{diageta}
\left( \begin{array}{cc} c_\eta & - s_\eta \\ s_\eta & c_\eta \end{array} \right)
M_\eta
\left( \begin{array}{cc} c_\eta & s_\eta \\ - s_\eta & c_\eta \end{array} \right)
=
\left( \begin{array}{cc} M_{\eta 1} & 0 \\ 0 & M_{\eta 2} \end{array} \right),
\ee
where $c_\eta \equiv \cos{\eta}$,
$s_\eta \equiv \sin{\eta}$,
and $M_{\eta 1} < M_{\eta 2}$.
From Eq.~\eqref{diageta},
\bs
\label{theetas}
\ba
\left( M_\eta \right)_{11} &=& M_{\eta 1} c_\eta^2 + M_{\eta 2} s_\eta^2,
\\
\left( M_\eta \right)_{22} &=& M_{\eta 1} s_\eta^2 + M_{\eta 2} c_\eta^2,
\\
\left( M_\eta \right)_{12} &=& \left( M_{\eta 2} - M_{\eta 1} \right) c_\eta s_\eta.
\ea
\es
Since
\be
\left( \begin{array}{c} \eta_a \\ \eta_b \end{array} \right)
=
\left( \begin{array}{cc} c_\eta & s_\eta \\ - s_\eta & c_\eta \end{array} \right)
\left( \begin{array}{c} S^0_3 \\ S^0_4 \end{array} \right),
\label{mf00ree}
\ee
one has
\be
\begin{array}{ll}
  \displaystyle{\left( V_\eta \right)_{13} =
  c_\eta\, \frac{v_2}{\sqrt{T_1}}
  - s_\eta\, \frac{v_1 u_L u_R}{\sqrt{T_1 T_2}},}
  &
  \displaystyle{\left( V_\eta \right)_{14} =
  s_\eta\, \frac{v_2}{\sqrt{T_1}}
  + c_\eta\, \frac{v_1 u_L u_R}{\sqrt{T_1 T_2}},}
  \\*[3mm]
  \displaystyle{\left( V_\eta \right)_{23} =
  c_\eta\, \frac{- v_1}{\sqrt{T_1}}
  - s_\eta\, \frac{v_2 u_L u_R}{\sqrt{T_1 T_2}},}
  &
  \displaystyle{\left( V_\eta \right)_{24} =
  s_\eta\, \frac{- v_1}{\sqrt{T_1}}
  + c_\eta\, \frac{v_2 u_L u_R}{\sqrt{T_1 T_2}},}
  \\*[3mm]
  \displaystyle{\left( V_\eta \right)_{33} =
  s_\eta u_R \sqrt{\frac{T_1}{T_2}},}
  &
  \displaystyle{\left( V_\eta \right)_{34} =
  - c_\eta u_R \sqrt{\frac{T_1}{T_2}},}
  \\*[3mm]
  \displaystyle{\left( V_\eta \right)_{43} = - s_\eta u_L \sqrt{\frac{T_1}{T_2}},}
  &
  \displaystyle{\left( V_\eta \right)_{44} =
  c_\eta u_L \sqrt{\frac{T_1}{T_2}}.}
\end{array}
\ee
In Eq.~\eqref{mf00ree},
we choose the sign of $\left( S^0_3,\ S^0_4 \right)^T$
in such a way that $c_\eta$ is non-negative.\footnote{The relative sign
of $S^0_3$ and $S^0_4$ is fixed by the condition
that the determinant of the mixing matrix
$\left( \begin{array}{cc} c_\eta & s_\eta \\
  - s_\eta & c_\eta \end{array} \right)$
is positive.}

\paragraph{The physical charged scalars:}
Two orthonormalized fields
that are orthogonal to the charged Goldstone bosons \eqref{gb+} are
\bs
\label{phiab}
\ba
\label{phia}
\varphi_a^+ &\equiv& \frac{\left( V_1 - V_2 \right) m
  - v_1 u_L a + v_2 u_L c}{\sqrt{K_1}},
\\
\label{phib}
\varphi_b^+ &\equiv& \frac{2 v_1 v_2 u_L u_R m - K_1 p
  + v_2 u_R \left( V_1 - V_2 - U_L \right) a
  + v_1 u_R \left( V_2 - V_1 - U_L \right) c}{\sqrt{K_1} \sqrt{K_2}},
\hspace*{7mm}
\ea
\es
where
\be
\label{K1K2}
K_1 \equiv \left( V_1 - V_2 \right)^2
+ \left( V_1 + V_2 \right) U_L,
\qquad
K_2 \equiv U_L U_R + K_1 + \left( V_1 + V_2 \right) U_R.
\ee
In Eqs.~\eqref{phiab},
the normalization factors $\sqrt{K_1}$ and $\sqrt{K_2}$
are positive---this corresponds
to a convention for the signs of $\varphi_a^+$ and $\varphi_b^+$.
The Lagrangian contains mass terms for the charged-scalar fields as
\be
\label{mvarphi}
\mathcal{L} = \cdots - 
\left( \begin{array}{cc} \varphi_a^-, & \varphi_b^- \end{array} \right)
M_\varphi
\left( \begin{array}{c} \varphi_a^+ \\ \varphi_b^+ \end{array} \right),
\ee
where $M_\varphi$ is a real
(because our model is $CP$-conserving)
$2 \times 2$ symmetric matrix.
That matrix is diagonalized as
\be
\label{diagvarphi}
\left( \begin{array}{cc} c_\varphi & - s_\varphi \\
  s_\varphi & c_\varphi \end{array} \right)
M_\varphi
\left( \begin{array}{cc} c_\varphi & s_\varphi \\
  - s_\varphi & c_\varphi \end{array} \right)
=
\left( \begin{array}{cc} M_{\varphi 1} & 0 \\
  0 & M_{\varphi 2} \end{array} \right),
\ee
where $c_\varphi \equiv \cos{\varphi}$,
$s_\varphi \equiv \sin{\varphi}$,
and $M_{\varphi 1} < M_{\varphi 2}$.
It follows from Eqs.~\eqref{diagvarphi} that
\bs
\label{cjigur8}
\ba
\left( M_\varphi \right)_{11} &=& M_{\varphi 1} c_\varphi^2
+ M_{\varphi 2} s_\varphi^2,
\\
\left( M_\varphi \right)_{22} &=& M_{\varphi 1} s_\varphi^2
+ M_{\varphi 2} c_\varphi^2,
\\
\left( M_\varphi \right)_{12} &=& \left( M_{\varphi 2} - M_{\varphi 1} \right)
c_\varphi s_\varphi.
\ea
\es
Since
\be
\left( \begin{array}{c} \varphi_a^+ \\ \varphi_b^+ \end{array} \right)
=
\left( \begin{array}{cc} c_\varphi & s_\varphi \\
  - s_\varphi & c_\varphi \end{array} \right)
\left( \begin{array}{c} H^+_3 \\ H^+_4 \end{array} \right),
\label{if9344t}
\ee
one has
\bs
\ba
\left( V_\varphi \right)_{13} &=& \frac{- c_\varphi v_1 u_L}{\sqrt{K_1}}
+ \frac{s_\varphi v_2 u_R \left( V_2 - V_1 + U_L \right)}{\sqrt{K_1 K_2}},
\\
\left( V_\varphi \right)_{23} &=& \frac{c_\varphi v_2 u_L}{\sqrt{K_1}}
+ \frac{s_\varphi v_1 u_R \left( V_1 - V_2 + U_L \right)}{\sqrt{K_1 K_2}},
\\
\left( V_\varphi \right)_{33} &=&
\frac{c_\varphi \left( V_1 - V_2 \right)}{\sqrt{K_1}}
- \frac{2 s_\varphi v_1 v_2 u_L u_R}{\sqrt{K_1 K_2}},
\\
\left( V_\varphi \right)_{43} &=& s_\varphi\, \sqrt{\frac{K_1}{K_2}},
\ea
\es
\bs
\ba
\left( V_\varphi \right)_{14} &=& \frac{- s_\varphi v_1 u_L}{\sqrt{K_1}}
+ \frac{c_\varphi v_2 u_R \left( V_1 - V_2 - U_L \right)}{\sqrt{K_1 K_2}},
\\
\left( V_\varphi \right)_{24} &=& \frac{s_\varphi v_2 u_L}{\sqrt{K_1}}
+ \frac{c_\varphi v_1 u_R \left( V_2 - V_1 - U_L \right)}{\sqrt{K_1 K_2}},
\\
\left( V_\varphi \right)_{34} &=&
\frac{s_\varphi \left( V_1 - V_2 \right)}{\sqrt{K_1}}
+ \frac{2 c_\varphi v_1 v_2 u_L u_R}{\sqrt{K_1 K_2}},
\\
\left( V_\varphi \right)_{44} &=& - c_\varphi\, \sqrt{\frac{K_1}{K_2}}.
\ea
\es
In Eq.~\eqref{if9344t},
we choose the sign of $\left( H_3^+,\ H_4^+ \right)$
such that $c_\varphi \ge 0$.\footnote{The relative sign of $H_3^+$ and $H_4^+$
is not free,
since altering it would change the sign
of the determinant of the mixing matrix
$\left( \begin{array}{cc} c_\varphi & s_\varphi \\
  - s_\varphi & c_\varphi \end{array} \right)$.}

\subsection{Gauge-fixing terms}

The terms in the Lagrangian that are bi-linear
in either the gauge-boson fields or the scalar fields are
\ba
\label{93834kmf}
\mathcal{L} &=& \cdots
+ \frac{1}{2} \left(
\partial_\mu G_l^0\ \partial^\mu G_l^0
+ \partial_\mu G_h^0\ \partial^\mu G_h^0
+ \partial_\mu S_3^0\ \partial^\mu S_3^0
+ \partial_\mu S_4^0\ \partial^\mu S_4^0
\right)
\no & &
- \frac{1}{4}
\left( \partial_\mu Z_{l \nu} - \partial_\nu Z_{l \mu} \right) 
\left( \partial^\mu Z_l^\nu - \partial^\nu Z_l^\mu \right) 
\no & &
- \frac{1}{4}
\left( \partial_\mu Z_{h \nu} - \partial_\nu Z_{h \mu} \right) 
\left( \partial^\mu Z_h^\nu - \partial^\nu Z_h^\mu \right) 
\no & &
- \frac{1}{4}
\left( \partial_\mu A_\nu - \partial_\nu A_\mu \right) 
\left( \partial^\mu A^\nu - \partial^\nu A^\mu \right)
\no & &
- \frac{M_{\eta_1}}{2} \left( S_3^0 \right)^2
- \frac{M_{\eta_2}}{2} \left( S_4^0 \right)^2
+ \frac{M_l}{2}\, Z_{l \mu}\, Z_l^\mu
+ \frac{M_h}{2}\, Z_{h \mu}\, Z_h^\mu
\no & &
+ \sqrt{M_l}\, Z_{l \mu}\, \partial^\mu G_l^0
+ \sqrt{M_h}\, Z_{h \mu}\, \partial^\mu G_h^0,
\ea
and
\ba
\label{93834kmf2}
\mathcal{L} &=& \cdots
+ \partial_\mu G_l^-\ \partial^\mu G_l^+
+ \partial_\mu G_h^-\ \partial^\mu G_h^+
+ \partial_\mu H_3^-\ \partial^\mu H_3^+
+ \partial_\mu H_4^-\ \partial^\mu H_4^+
\no & &
- \left( \partial^\mu W_l^{+ \nu} \right) \left( \partial_\mu W_{l \nu}^- \right)
+ \left( \partial^\mu W_l^{+ \nu} \right) \left( \partial_\nu W_{l \mu}^- \right)
\no & &
- \left( \partial^\mu W_h^{+ \nu} \right) \left( \partial_\mu W_{h \nu}^- \right)
+ \left( \partial^\mu W_h^{+ \nu} \right) \left( \partial_\nu W_{h \mu}^- \right),
\no & &
- M_{\varphi_1} H_3^- H_3^+
- M_{\varphi_2} H_4^- H_4^+
+ \bar M_l\, W_{l \mu}^- W_l^{\mu +}
+ \bar M_h\, W_{h \mu}^- W_h^{\mu +}
\no & &
+ i \left(
\sqrt{\bar M_l}\ W_{l \mu}^-\, \partial^\mu G_l^+
+ \sqrt{\bar M_h}\ W_{h \mu}^-\, \partial^\mu G_h^+
- \mathrm{H.c.}
\right)
\ea
\textit{cf.}\ Eq.~\eqref{jvfi000}.
We add to them the gauge-fixing terms\footnote{We restrict ourselves
to $R_\xi$ gauges.}
\bs
\ba
\mathcal{L}_\mathrm{gf,0} &=&
- \frac{\left( \partial_\mu Z_l^\mu - \xi_l\, \sqrt{M_l}\, G_l^0
  \right)^2}{2 \xi_l}
- \frac{\left( \partial_\mu Z_h^\mu - \xi_h\, \sqrt{M_h}\, G_h^0
  \right)^2}{2 \xi_h}
- \frac{\left( \partial_\mu A^\mu \right)^2}{2 \xi_A},
\\
\mathcal{L}_\mathrm{gf,\pm} &=&
- \frac{\left( \partial_\mu W_l^{+ \mu}
  - i \bar \xi_l\, \sqrt{\bar M_l}\, G_l^+ \right)
  \left( \partial_\nu W_l^{- \nu}
  + i \bar \xi_l\, \sqrt{\bar M_l}\, G_l^- \right)}{\bar \xi_l}
\no & &
- \frac{\left( \partial_\mu W_h^{+ \mu}
  - i \bar \xi_h\, \sqrt{\bar M_h}\, G_h^+ \right)
  \left( \partial_\nu W_h^{+ \nu}
  + i \bar \xi_h\, \sqrt{\bar M_h}\, G_h^- \right)}{\bar \xi_h}.
\ea
\es

\subsection{Scalar potential}

The scalar potential appears in the Lagrangian
as $\mathcal{L} = \cdots - V$ and is in our model of the form
$V = V_H + V_\Phi + V_{H\Phi}$,
where
\bs
\label{pott}
\ba
V_H &=&
\mu_L\, H_L^\dagger H_L + \mu_R\, H_R^\dagger H_R
\no & &
+ \lambda_L\, H_L^\dagger H_L\, H_L^\dagger H_L
+ \lambda_R\, H_R^\dagger H_R\, H_R^\dagger H_R
+ \lambda_{LR}\, H_L^\dagger H_L\, H_R^\dagger H_R,
\label{vh} \\
V_\Phi &=&
\mu_1\, \mathrm{tr} \left( \Phi^\dagger \Phi \right)
+ \mu_2\, \mathrm{tr}
\left( \tilde \Phi^\dagger \Phi + \Phi^\dagger \tilde \Phi \right)
\no & &
+ \lambda_1 \left[ \mathrm{tr} \left( \Phi^\dagger \Phi \right) \right]^2
+ \lambda_2 \left\{
\left[ \mathrm{tr} \left( \Phi^\dagger \tilde \Phi \right) \right]^2
+ \mathrm{H.c.} \right\}
+ \lambda_3 \left| \mathrm{tr} \left( \Phi^\dagger \tilde \Phi \right) \right|^2
\no & &
+ \lambda_4\, \mathrm{tr} \left( \Phi^\dagger \Phi \right) \mathrm{tr}
\left( \tilde \Phi^\dagger \Phi + \Phi^\dagger \tilde \Phi \right),
\label{vphi} \\
V_{H\Phi} &=&
m_1 \left( H_L^\dagger \Phi H_R + H_R^\dagger \Phi^\dagger H_L \right)
+ m_2
\left( H_L^\dagger \tilde \Phi H_R + H_R^\dagger \tilde \Phi^\dagger H_L \right)
\no & &
+ \lambda_{3L}\, H_L^\dagger \Phi \Phi^\dagger H_L
+ \lambda_{3R}\, H_R^\dagger \Phi^\dagger \Phi H_R
+ \lambda_{4L}\, H_L^\dagger \tilde \Phi \tilde \Phi^\dagger H_L
+ \lambda_{4R}\, H_R^\dagger \tilde \Phi^\dagger \tilde \Phi H_R
\no & &
+ \lambda_{5L}\, H_L^\dagger
\left( \Phi \tilde \Phi^\dagger + \tilde \Phi \Phi^\dagger \right) H_L
+ \lambda_{5R}\, H_R^\dagger
\left( \Phi^\dagger \tilde \Phi + \tilde \Phi^\dagger \Phi \right) H_R.
\label{vhphi}
\ea
\es
The parameters
$\lambda_L$,
$\lambda_R$,
$\lambda_{LR}$,
$\lambda_1$,
$\lambda_2$,
$\lambda_3$,
$\lambda_4$,
$\lambda_{3L}$,
$\lambda_{3R}$,
$\lambda_{4L}$,
$\lambda_{4R}$,
$\lambda_{5L}$,
and $\lambda_{5R}$ are dimensionless;
the parameters $m_1$ and $m_2$ have mass dimension;
the parameters $\mu_L$,
$\mu_R$,
$\mu_1$,
and $\mu_2$ have mass-squared dimension.
All these parameters are real because of the assumed $CP$ conservation.
Notice that we have \emph{not} imposed the parity symmetry
$H_L \leftrightarrow H_R,\ \Phi \to \Phi^\dagger,\
\tilde \Phi \to \tilde \Phi^\dagger$ on $V$.
In component fields,
$V = V_{(2)} + V_{(3)} + V_{(4)}$,
where
\bs
\label{pott2}
\ba
\label{v2}
V_{(2)} &=&
\mu_L \left( \left| m \right|^2 + \left| n \right|^2 \right)
+ \mu_R \left( \left| p \right|^2 + \left| q \right|^2 \right)
\no & &
+ \mu_1 \left( \left| a \right|^2 + \left| b \right|^2
+ \left| c \right|^2 + \left| d \right|^2\right)
+ 2 \mu_2 \left( a^\ast c + a c^\ast + b^\ast d + b d^\ast \right),
\\
\label{v3}
V_{(3)} &=& m_1 \left( - a^\ast n^\ast p - a n p^\ast
+ b^\ast m^\ast p + b m p^\ast
+ c m^\ast q + c^\ast m q^\ast
+ d n^\ast q + d^\ast n q^\ast \right)
\no & &
+ m_2 \left( - c^\ast n^\ast p - c n p^\ast
+ d^\ast m^\ast p + d m p^\ast
+ a m^\ast q + a^\ast m q^\ast
+ b n^\ast q + b^\ast n q^\ast \right),
\hspace*{10mm}
\\
\label{v4}
V_{(4)} &=&
\lambda_L \left( \left| m \right|^2 + \left| n \right|^2 \right)^2
+ \lambda_R \left( \left| p \right|^2 + \left| q \right|^2 \right)^2
+ \lambda_{LR} \left( \left| m \right|^2 + \left| n \right|^2 \right)
\left( \left| p \right|^2 + \left| q \right|^2 \right)
\no & &
+ \lambda_1
\left( \left| a \right|^2 + \left| b \right|^2
+ \left| c \right|^2 + \left| d \right|^2\right)^2
\no & &
+ 4 \lambda_2 \left( a^2 {c^\ast}^2 + b^2 {d^\ast}^2
+ 2 a b c^\ast d^\ast + \mathrm{H.c.} \right)
\no & &
+ 4 \lambda_3 \left( \left| a c \right|^2 +  \left| b d \right|^2
  + a d b^\ast c^\ast + a^\ast d^\ast b c \right)
\no & &
+  2 \lambda_4 \left( \left| a \right|^2 + \left| b \right|^2
+ \left| c \right|^2 + \left| d \right|^2 \right)
\left( a^\ast c + b^\ast d + a c^\ast + b d^\ast \right)
\no & &
+ \lambda_{3L} \left\{
  \left( \left| b \right|^2 + \left| c \right|^2 \right) \left| m \right|^2
  + \left( \left| a \right|^2 + \left| d \right|^2 \right) \left| n \right|^2
  + 2\, \mathrm{Re} \left[ \left( c d^\ast - a b^\ast \right) m^\ast n \right]
  \right\}
\no & &
+ \lambda_{4L} \left\{ \left( \left| b \right|^2
+ \left| c \right|^2 \right) \left| n \right|^2 + \left( \left| a \right|^2
+ \left| d \right|^2 \right) \left| m \right|^2
- 2\, \mathrm{Re} \left[ \left( c d^\ast
  - a b^\ast \right) m^\ast n \right]
\right\}
\no & &
+ \lambda_{5L} \left( \left| m \right|^2 + \left| n \right|^2 \right)
\left( a c^\ast + a^\ast c + b d^\ast + b^\ast d \right)
\no & &
+ \lambda_{3R} \left\{
  \left( \left| a \right|^2 + \left| b \right|^2 \right) \left| p \right|^2
  + \left( \left| c \right|^2 + \left| d \right|^2 \right) \left| q \right|^2
  + 2\, \mathrm{Re} \left[ \left( b c - a d \right) p^\ast q \right] \right\}
\no & &
+ \lambda_{4R} \left\{
  \left( \left| a \right|^2 + \left| b \right|^2 \right) \left| q \right|^2
  + \left( \left| c \right|^2 + \left| d \right|^2 \right) \left| p \right|^2
  - 2\, \mathrm{Re} \left[ \left( b c - a d \right) p^\ast q \right] \right\}
\no & &
+ \lambda_{5R}  \left( \left| p \right|^2 + \left| q \right|^2 \right)
\left( a c^\ast + a^\ast c + b d^\ast + b^\ast d \right).
\ea
\es
The parameters in $V_{(4)}$ are constrained
by the unitarity and bounded-from-below conditions;
these are worked out in Appendices~\ref{app:unitarity} and~\ref{app:BFB},
respectively.
Additional constraints derive from the condition that
the assumed minimum of the potential is a global,
not just local,
minimum;
they are partially given in Appendix~\ref{app:extra}.
Still other constraints have to do with the observed couplings
of the scalar of mass 125\,GeV,
which we assume to be $S_5^0$,\footnote{We do not in general assume,
though,
$S_5^0$ to be \emph{the lightest} scalar.}
to pairs of gauge bosons or to quark pairs;
they are worked out in Appendix~\ref{app:kappa}.

\subsection{Yukawa couplings and quark masses}

\paragraph{Yukawa couplings:} The Yukawa couplings are given by
\bs
\label{vvfkgfor0}
\ba
\label{lghfpgk}
\mathcal{L}_\mathrm{Yukawa} &=& - \left( \begin{array}{cc}
  \bar t_L, & \bar b_L \end{array} \right)
\left( y_1 \Phi + y_2 \tilde \Phi \right)
\left( \begin{array}{c} t_R \\ b_R \end{array} \right)
+ \mathrm{H.c.}
\\ &=& - \left( y_1 b^\ast + y_2 d^\ast \right) \bar t_L t_R
- \left( y_1 c + y_2 a \right) \bar t_L b_R
\no & &
+ \left( y_1 a^\ast + y_2 c^\ast \right) \bar b_L t_R
- \left( y_1 d + y_2 b \right) \bar b_L b_R + \mathrm{H.c.},
\label{uf83344}
\ea
\es
where the Yukawa coupling constants $y_1$ and $y_2$ are real
because of the assumed $CP$ invariance of the model.

\paragraph{Quark masses:}
When $b$ and $d$ acquire real VEVs $v_1$ and $v_2$,
respectively,
Eq.~\eqref{uf83344} gives rise to quark masses
\be
\label{mbmt}
m_t = y_1 v_1 + y_2 v_2,
\qquad
m_b = y_1 v_2 + y_2 v_1.
\ee
From Eqs.~\eqref{mbmt},
\be
\label{yuks}
y_1 = \frac{- v_1 m_t + v_2 m_b}{V_2 - V_1},
\qquad
y_2 = \frac{v_2 m_t - v_1 m_b}{V_2 - V_1}.
\ee
Without loss of generality,
we fix the relative sign of $\left( t_L,\ b_L \right)^T$
and $\left( t_R,\ b_R \right)^T$
in such a way that $y_2 \ge 0$.
Since we have already fixed $V_2 \ge V_1$,
this means that we always use $m_t \ge 0$.
The sign of $m_b$---just as the sign of $v_1$---remains free.

\section{Parameter counting and procedure}
\label{sec:parameters}

\paragraph{Counting of parameters:}
Our left--right model has in its Lagrangian the following parameters:
\begin{itemize}
\item The gauge coupling constants $g$,
  $l$,
  and $h$.
\item The parameters of the potential $\mu_L$,
  $\mu_R$,
  $\mu_1$,
  $\mu_2$,
  $m_1$,
  $m_2$,
  $\lambda_L$,
  $\lambda_R$,
  $\lambda_{LR}$,
  $\lambda_1$,
  $\lambda_2$,
  $\lambda_3$,
  $\lambda_4$,
  $\lambda_{3L}$,
  $\lambda_{3R}$,
  $\lambda_{4L}$,
  $\lambda_{4R}$,
  $\lambda_{5L}$,
  and $\lambda_{5R}$.
\item The Yukawa couplings $y_1$ and $y_2$.
\end{itemize}
This makes 24 real parameters.
(There are other parameters in the model,
but they are dependent on these 24.
For instance,
$\theta_w$ and $\alpha$ depend on the gauge coupling constants;
the VEVs $u_{L,R}$ and $v_{1,2}$ depend on the parameters of the potential.)

\paragraph{Counting of observables:}
We use observable quantities as input of the renormalization procedure.
We choose these observables to be exclusively masses,
mixing angles,
and the electromagnetic coupling contant.
The quantities at our disposal are:
\begin{itemize}
\item The squared electromagnetic coupling constant,
  \viz\ $E$.
\item The squared masses of the neutral gauge bosons,
  \viz\ $M_l$ and $M_h$.
\item The squared masses of the charged gauge bosons,
  \viz\ $\bar M_l$ and $\bar M_h$.
\item The mixing angle $\psi$ between the two neutral gauge bosons.
\item The mixing angle $\xi$ between the two charged gauge bosons.
\item The masses $\mu_5$,
  $\mu_6$,
  $\mu_7$,
  and $\mu_8$ of the four scalars.
\item The six mixing angles $\theta_i$
  ($i = 1, \ldots, 6$)
  that parameterize the mixing matrix $V_\rho$.
\item The squared masses of the two pseudoscalars,
  \viz\ $M_{\eta 1}$ and $M_{\eta 2}$.
\item The mixing angle $\eta$ between the two pseudoscalars.
\item The squared masses of the two charged scalars,
  \viz\ $M_{\varphi 1}$ and $M_{\varphi 2}$.
\item The mixing angle $\varphi$ between the two charged scalars.
\item The masses $m_t$ and $m_b$ of the top and bottom quarks.
\end{itemize}
This makes 25 real observables.
There is,
thus,
one more observable than there are parameters in the model.
This means that there must be one constraint among the 25 observables;
that constraint is derived in Appendix~\ref{app:scalarmasses},
\viz\ it is in Eq.~\eqref{meta2}.

\paragraph{Procedure:} Our practical procedure is the following.
\begin{enumerate}
\item We input the following 24 quantities:
  \begin{itemize}
  \item The squared electromagnetic coupling constant,
    \textit{viz.}\ $E = 4 \pi \left/ 137.035\,999\,084 \right.$.
  \item The squared masses of the neutral gauge bosons,
    \viz\ $M_l$ and $M_h$.
    We choose $\sqrt{M_l} = 91.1876\,\mathrm{GeV}$
    and $\sqrt{M_h} \in \left[ 0.75,\, 4 \right]$\, TeV.\footnote{Actually,
    a more realistic lower bound
    on the masses of the new gauge bosons of the LRM
    would be 2\,TeV
    or 3\,TeV~\cite{TWIST:2011jfx,Civitarese:2016hbg,Babu:1993hx,
	Dekens:2014ina,Bertolini:2014sua,ValeSilva:2016dcg,Blanke:2011ry,
	Babu:2020bgz,CMS:2019gwf,ThomasArun:2021rwf,CMS:2021mux,Osland:2022ryb}.
	(The precise bound depends on the ratio $g/l$
	between the gauge coupling constants of $SU(2)_L$ and $SU(2)_R$.)
	We opt for the lax lower bound 750\,GeV
	in order to explore all the possibilities to fit
	the $Z b \bar b$ vertex.}
  \item The squared masses of the charged gauge bosons,
    \viz\ $\bar M_l$ and $\bar M_h$.
    We choose $\sqrt{\bar M_l} = 80.377\,\mathrm{GeV}$
    and $\sqrt{\bar M_h} \in \left[ 0.75\, \mathrm{TeV},\,
      \sqrt{M_h} \right]$.\footnote{In our LRM $\bar M_h$
    must always smaller than $M_h$,
    see Appendix~\ref{app:GLH}.}
  \item The mixing angle between the two neutral gauge bosons is
    chosen $\psi = 0$.
    In this way the neutral gauge boson $Z_l$ of the LRM
    has the same interactions,
    at the tree level,
    as the observed boson of mass $\sqrt{M_l} = 91.1876$\,GeV.
    As a consequence,
    the infrared divergences---due to the zero masses of the photon
    and of the gluons---in the one-loop diagrams
    for the vertex $Z_l b \bar b$ in the LRM
    cancel out when one subtracts from those diagrams
    the analogous diagrams for the vertex $Z b \bar b$ in the SM,
    \viz\ when one compares the LRM to the SM
    in order to compute $\delta g_L$ and $\delta g_R$.
  \item The mixing angle between the two charged gauge bosons,
    \viz\ $\xi \in \left[ - 0.01,\, + 0.01 \right]$.
    In practice,
    this angle cannot be larger than 0.005,
    because of the lower bound $0.75\, \mathrm{TeV}$
    that we impose on $\sqrt{\bar M_h}$---see Appendix~\ref{app:results}.
  \item The masses $m_t = 172.69$\,GeV
    and $m_b = \pm 4.18$\,GeV of the top and bottom quarks,
    respectively.
    We choose $m_t$ positive;
    the sign of $m_b$ may be either positive or negative.
  \item The masses of the four physical scalars,
    \textit{viz.}\ $\mu_5 = 125.25\,\mathrm{GeV}$
    and $\mu_6 < \mu_7 < \mu_8 < 1$\,TeV.\footnote{We allow
    for very light scalars of mass as low as 10\,GeV,
    although these are in practice most likely excluded by experiment.
    We do this in order to explore whether this radical possibility
    might allow us to fit $g_L$ and $g_R$ adequately.}
  \item The mass of the lightest pseudoscalar,
    \viz\ $\sqrt{M_{\eta 1}} < 1$\,TeV.
  \item The masses of the two physical charged scalars,
    \viz\ $\sqrt{M_{\varphi 1}} < \sqrt{M_{\varphi 2}} < 1\,$TeV.
  \item The mixing angle between the two pseudoscalars,
    \viz\ $\eta \in \left[ - \pi / 2,\, + \pi / 2 \right]$.
  \item The mixing angle between the two charged scalars,
    \viz\ $\varphi \in \left[ - \pi / 2,\, + \pi / 2 \right]$.
  \item The six mixing angles that parameterize the mixing matrix
    of the scalars $V_\rho$,
    \viz\ the $\theta_i$ ($i = 1, 2, \ldots, 6$).
    We choose $\theta_4 \in \left[ - \pi / 2,\, + \pi / 2 \right]$
    and $\theta_5 \in \left[ 0,\, + \pi / 2 \right]$;
    the other four mixing angles are in principle free,
    but in practice $\theta_{1,2,3}$ are strongly constrained
    by the experimental constraints of Appendix~\ref{app:kappa}.
  \end{itemize}
\item Following the procedure outlined in Appendix~\ref{app:GLH}
  for the case $\psi = 0$,
  we determine $G$,
  $L$,
  $H$,
  $V_1$,
  $V_2$,
  $U_L$,
  and $U_R$.
  In order for the procedure to run smoothly,
  the inequalities~\eqref{mfoder0ewe9} must hold,
  and $\left( a_s \right)^2 - 4 x y D^2$ in Eqs.~\eqref{uf8e9r9ee}
  must be positive;
  furthermore,
  all seven final quantities must turn out positive.
  If any of these does not happen,
  then the input values are inadequate and must be discarded.
  This requirement alone forces $\xi$ to be very small.
\item We fix $v_2 = \sqrt{V_2}$,
  $u_L = \sqrt{U_L}$,
  and $u_R = \sqrt{U_R}$.
  We also fix $v_1 = \pm \sqrt{V_1}$;
  its sign is opposite to the one of $\xi$,
  \textit{cf.}~Eq.~\eqref{nvkdfps0}.
  The gauge coupling constants $g = \sqrt{G}$,
  $l = \sqrt{L}$,
  and $h = \sqrt{H}$ are positive.
\item We compute the Yukawa coupling constants by means of Eqs.~\eqref{yuks}.
  We check that they are not much too large,
  \textit{viz.}\ that $\left| y_1 \right| \lesssim M$
  and $\left| y_2 \right| \lesssim M$ with $M \approx 4 \pi$.
\item We compute $T_1$,
  $T_2$,
  $K_1$,
  and $K_2$ by means of Eqs.~\eqref{T1T2} and~\eqref{K1K2}.
\item We compute the matrix elements of $M_\varphi$
  by using Eqs.~\eqref{cjigur8}.
\item We compute $M_{\eta 2}$ by using Eq.~\eqref{meta2}.
  If we obtain $M_{\eta 2} < M_{\eta 1}$,
  then we make $M_{\eta 2} \leftrightarrow M_{\eta 1}$
  together with $\eta \to \eta - \pi / 2$.
\item We compute the matrix elements of $M_\eta$
  by using Eqs.~\eqref{theetas}.
\item We compute the matrix elements of $M_\rho$
  by using Eq.~\eqref{computemrho}.
\item We compute the parameters of $V_{(3)}$ and $V_{(4)}$
  in Eqs.~\eqref{pott2} by following the steps in the last paragraph
  of Appendix~\ref{app:scalarmasses}.
\item We check the unitarity conditions on the parameters of $V_{(4)}$.
\item We check the bounded-from-below conditions
  on the parameters of $V_{(4)}$.
\item We compute $\mu_L$,
  $\mu_R$,
  $\mu_1$,
  $\mu_2$,
  and $V_0$ by using Eqs.~\eqref{mu2} and~\eqref{vzero}.
\item We check the extra conditions in Appendix~\ref{app:extra}.
\item We compute the experimental parameters $\kappa$
  of Appendix~\ref{app:kappa}.
  We enforce the conditions~\cite{ATLAS:2020qdt,CMS:2018uag}
  $\kappa_W \in \left[ 0.59,\, 1.46 \right]$,
  $\kappa_Z \in \left[ 0.63,\, 1.32 \right]$,
  $\kappa_t \in \left[ 0.81,\, 1.47 \right]$,
  and $\left| \kappa_b \right| \in \left[ 0.11,\, 1.79 \right]$.
  Note that:
  \begin{description}
  \item We allow $\kappa_b$ to be either positive or negative,
    since experiment is as yet unable to fix its sign.
  \item We require the parameters $\kappa$ to be in their $3 \sigma$ ranges,
    because we do not want to miss out any possibility that the LRM
    might offer to fit $g_L$ and $g_R$.
  \item In our model $\kappa_W$ and $\kappa_Z$
    always turn out to be smaller than~1 and almost equal to each other,
    for reasons explained in Appendix~\ref{app:kappa},
    \textit{cf.}~Appendix~\ref{app:results}.
  \end{description}
\end{enumerate}

\section{Calculation and renormalization}
\label{sec:renormalization}

We now describe the calculation
of the renormalized one-loop process $Z_l \to b \bar b$.
We do it by using \FM~\cite{Fontes:2019wqh,Fontes:2021iue},
which resorts to \ts{FeynRules}~\cite{Christensen:2008py,Alloul:2013bka},
\ts{QGRAF}~\cite{Nogueira:1991ex},
and \ts{FeynCalc}~\cite{Mertig:1990an,Shtabovenko:2016sxi,Shtabovenko:2020gxv}.
More specifically,
we use \FMS to generate the Feynman rules of the model
(both for the renormalized interactions and for the counterterms),
to generate the Feynman diagrams,
and to calculate the one-loop amplitudes and the counterterms.

The ultraviolet (UV) renormalized one-loop process
(denoted $i\, \hat{\Gamma}_{\mu}^{Z_l b \bar b}$)
is the sum of the non-renormalized one-loop process
(denoted $i\, {\Gamma}_{\mu}^{Z_l b \bar b}$)
and the counterterms of the process
(denoted $i \left. {\Gamma}_{\mu}^{Z_l b \bar b} \right|_{\text{CT}}$);
that is,
\be
i\, \hat{\Gamma}_{\mu}^{Z_l b \bar b}
=
i\, {\Gamma}_{\mu}^{Z_l b \bar b} + 
i \left. {\Gamma}_{\mu}^{Z_l b \bar b} \right|_{\text{CT}}.
\label{eq:base}
\ee
The terms in the right-hand side of Eq.~\eqref{eq:base}
are determined when considering the theory up to the one-loop level;
more specifically,
one must take the original Lagrangian of the theory,
identify each parameter and each field as a bare quantity,
and then split it into a renormalized quantity and a
counterterm;\fn{For details see e.g.\ Ref.~\cite{Fontes:2021kue}.}
thus,
for a generic bare parameter $p_{(0)}$ and a generic bare field $\phi_{(0)}$,
we write
\be
p_{(0)} = p + \delta p,
\qquad
\phi_{(0)} = \phi + \dfrac{1}{2}\, \delta Z_{\phi}\, \phi,
\label{eq:generic-expansion}
\ee
where $p$ is the renormalized parameter,
$\delta p$ is the corresponding counterterm,
$\phi$ is the renormalized field,
and $\delta Z_{\phi}$ is its counterterm.

The Feynman rules for the renormalized parameters are obtained
by expanding the Lagrangian and keeping only the terms with power zero
in the counterterms.
They are used to calculate the interactions of the theory;
in particular,
it is this set of rules that is used to calculate the
(non-renormalized, one-loop)
diagrams contributing to $i\, {\Gamma}_{\mu}^{Z_l b \bar b}$.

The Feynman rules for the counterterms
are obtained by expanding the Lagrangian
and keeping only the terms with power one in the counterterms.
The set of all those terms that contribute to $Z_l \to b \bar b$
constitutes $i \left. {\Gamma}_{\mu}^{Z_l b \bar b} \right|_{\text{CT}}$.
By using \FM\ it is straightforward to conclude that
\be
i \left. {\Gamma}_{\mu}^{Z b \bar b} \right|_{\text{CT}}
= i \gamma_\mu \left( F_L P_L + F_R P_R \right),
\ee
where,
with $\psi = 0$,
\bs
\ba
F_L &=&
\dfrac{s_w\, h\, \delta c_\alpha
  + c_\alpha\, h\, \delta s_w + c_\alpha\, s_w\, \delta h}{6}
- \dfrac{g\, \delta c_w + c_w\, \delta g}{2}
\no & &
+ \dfrac{s_\alpha\, h}{6}\, \delta s_\psi
+ \left(
\dfrac{{\delta Z_{Z_lZ_l}}}{2}
+ \dfrac{{{{\delta Z_{33}^{dq,\mathrm{L}}}}^*}}{2}
+ \dfrac{{\delta Z_{33}^{dq,\mathrm{L}}}}{2}
\right) i \left. {\Gamma}_{\mu,\mathrm{L}}^{Z_l b \bar b} \right|_\mathrm{tree}
\no & &
+ \dfrac{{\delta Z_{A Z_l}}}{2}\, i \left. {\Gamma}_{\mu,\mathrm{L}}^{A b \bar b}
\right|_\mathrm{tree}
+ \dfrac{{\delta Z_{Z_h Z_l}}}{2}\, i \left. {\Gamma}_{\mu,\mathrm{L}}^{Z_h b \bar b}
\right|_\mathrm{tree},
\\[5mm]
F_R &=&
\dfrac{s_w\, h\, \delta c_\alpha
  + c_\alpha\, h\, \delta s_w + c_\alpha\, s_w\, \delta h}{6}
+ \dfrac{s_w\, l\, \delta s_\alpha
  + s_\alpha\, l\, \delta s_w
  + s_\alpha\, s_w\, \delta l}{2}
\no & &
+ \left( \dfrac{s_\alpha\, h}{6} - \dfrac{c_\alpha\, l}{2} \right) \delta s_\psi
+ \left(
\dfrac{{\delta Z_{Z_lZ_l}}}{2}
+ \dfrac{{{{\delta Z_{33}^{dq,\mathrm{R}}}}^*}}{2}
+ \dfrac{{\delta Z_{33}^{dq,\mathrm{R}}}}{2}
\right) i \left. {\Gamma}_{\mu,\mathrm{R}}^{Z_l b \bar b} \right|_\mathrm{tree}
\no & &
+ \dfrac{{\delta Z_{AZ_l}}}{2}\, i \left.
{\Gamma}_{\mu,\mathrm{R}}^{A b \bar b} \right|_\mathrm{tree}
+ \dfrac{{\delta Z_{Z_hZ_l}}}{2}\, i \left.
{\Gamma}_{\mu,\mathrm{R}}^{Z_h b \bar b} \right|_\mathrm{tree}.
\ea
\label{eq:FLFR}
\es
Notice that,
when $\psi = 0$,
\bs
\ba
i \left. {\Gamma}_{\mu,\mathrm{L}}^{Z_l b \bar b} \right|_\mathrm{tree}
&=& \dfrac{c_\alpha\, s_w\, h}{6} - \dfrac{c_w\, g}{2},
\\
i \left. {\Gamma}_{\mu,\mathrm{L}}^{A b \bar b} \right|_\mathrm{tree}
&=& \dfrac{c_\alpha\, c_w\, h}{6} + \dfrac{s_w\, g}{2},
\\
i \left. {\Gamma}_{\mu,\mathrm{L}}^{Z_h b \bar b} \right|_\mathrm{tree}
&=& \dfrac{s_\alpha\, h}{6},
\\
i \left. {\Gamma}_{\mu,\mathrm{R}}^{Z_l b \bar b} \right|_\mathrm{tree}
&=& \dfrac{c_\alpha\, s_w\, h}{6} + \dfrac{s_\alpha\, s_w\, l}{2},
\\
i \left. {\Gamma}_{\mu,\mathrm{R}}^{A b \bar b} \right|_\mathrm{tree}
&=& \dfrac{c_\alpha\, c_w\, h}{6} + \dfrac{s_\alpha\, c_w\, l}{2},
\\
i \left. {\Gamma}_{\mu,\mathrm{R}}^{Z_h b \bar b} \right|_\mathrm{tree}
&=& \dfrac{s_\alpha\, h}{6} - \dfrac{c_\alpha\, l}{2}.
\ea
\es
It is important to understand the role
of the neutral-gauge-boson mixing angle $\psi$ in these equations.
As mentioned in the previous section,
since we identify the LRM gauge boson $Z_l$
with the observed neutral boson of mass 91.1876\,GeV,
we set the renormalized $\psi$ to zero.
However,
there is no symmetry of the theory that implies $\psi=0$;
the choice $\psi=0$ is just a particular solution
of a model where $\psi$ is in general nonzero.
Therefore,
when renormalizing that model a nonzero bare parameter $\psi_{(0)}$
must be allowed and renormalized;
the circumstance that we consider a particular solution of the model
where the renormalized $\psi$ vanishes
does not change the fact that the bare $\psi_{(0)}$ is in general nonzero
and has a nonzero counterterm $\delta \psi$.\fn{Actually,
  $\delta \psi$ is crucial to absorb the divergences
  of the process $Z_l \to b \bar b$.
  The non-inclusion of that counterterm
  would lead to the same kind of inconsistency
  as the one pointed out in Ref.~\cite{Fontes:2021znm}.}

We need to compute the counterterms that appear in Eqs.~\eqref{eq:FLFR}.
We perform that computation by using on-shell subtraction (OSS),
except for the independent mixing angles;
the latter are fixed through
a symmetry relation~\cite{Fontes:2021iue}.\fn{For details
  see Ref.~\cite{Fontes:2021kue}.
  Note that in this paper we use $\eta_g = \eta_f = 1$,
  while Refs.~\cite{Denner:1991kt,Denner:2019vbn}
  have used $\eta_g = - \eta_f = - 1$ instead.
  (The purely conventional parameters $\eta_g$ and $\eta_f$
  are defined in Eqs.~(F.23) of Ref.~\cite{Fontes:2021kue}).}

Let us start with the field counterterms.
Through a trivial generalization of the field counterterms
calculated through OSS in the SM,
we find:
\bs
\ba
\delta Z_{V_i V_i} &=&
\left. \widetilde{\operatorname{Re}}\,
\dfrac{\partial\, \Sigma_{\mathrm{T}}^{V_i V_i} \left( k^2 \right)}{\partial k^2}
\right|_{k^2 = m_{V_i}^2},
\\
\delta Z_{V_i V_j}
&=&
2\,
\widetilde{\operatorname{Re}}\,
\dfrac{\Sigma_\mathrm{T}^{V_iV_j} \left( m_{V_j}^2 \right)}{m_{V_j}^2
  - m_{V_i}^2} \quad \Leftarrow\ i \neq j,
\\
\delta Z_{33}^{dq, \mathrm{L}}
&=&
- \widetilde{\operatorname{Re}}\,
\Sigma_{\mathrm{L}}^{b \bar b} \left( m_b^2 \right)
\no & &
- m_b \left. \frac{\partial}{\partial p^2}\,
\widetilde{\operatorname{Re}}
\left\{ m_b \left[ \Sigma_{\mathrm{L}}^{b \bar b} \left( p^2 \right)
  + \Sigma_{\mathrm{R}}^{b \bar b} \left( p^2 \right) \right]
+ \Sigma_{\mathrm{l}}^{b \bar b} \left( p^2 \right)
+ \Sigma_{\mathrm{r}}^{b \bar b} \left( p^2 \right) \right\}
  \right|_{p^2 = m_b^2},
\label{Chap-Reno:eq:reno-ferm-diag-L}
\\
\delta Z_{33}^{dq, \mathrm{R}}
&=&
- \widetilde{\operatorname{Re}}\,
\Sigma_{\mathrm{R}}^{b \bar b} \left( m_b^2 \right)
\no & &
- m_b \left.
\frac{\partial}{\partial p^2}\,
\widetilde{\operatorname{Re}} \left\{
m_b \left[ \Sigma_{\mathrm{L}}^{b \bar b} \left( p^2 \right)
  + \Sigma_{\mathrm{R}}^{b \bar b} \left( p^2 \right) \right]
+ \Sigma_{\mathrm{l}}^{b \bar b} \left( p^2 \right)
+ \Sigma_{\mathrm{r}}^{b \bar b} \left( p^2 \right) \right\}
\right|_{p^2 = m_b^2}.
\hspace*{5mm}
\label{eq:field-CTs-1}
\ea
\es
Here,
$\Sigma_{k}^{xy} \left( p^2 \right)$ represents the $k^\mathrm{th}$ component
of the non-renormalized one-loop two-point function
for the fields $x,y$ with 4-momentum $p$.\fn{For details
  see section 5.4.1 of Ref.~\cite{Fontes:2021kue}.
  We are assuming the tadpole scheme dubbed PRTS
  in Ref.~\cite{Fontes:2021kue}.
  This implies that the Green's functions considered here
  do not include the one-loop tadpoles.}
The operator $\widetilde{\operatorname{Re}}$
discards the absorptive parts of the loop integrals
but keeps the imaginary parts of complex parameters.\fn{This detail
  is actually irrelevant in this paper,
  since in our LRM there are no complex parameters.
  Hence,
  in this case we might have used $\operatorname{Re}$
  instead of $\widetilde{\operatorname{Re}}$.}

We now turn to the parameter counterterms.
The first thing to notice is that
not all the parameters which intervene in Eqs.~\eqref{eq:FLFR}
are independent,
\ie\ not all of them are counterterms of the independent parameters.
Indeed,
we choose as independent parameters the squared masses of $Z_l$,
$Z_h$,
$W_l$,
and $W_h$ together with $e$,
$\psi$,
and $\xi$.
One may use the relations of the model to rewrite Eqs.~\eqref{eq:FLFR}
in terms of the counterterms of the independent parameters
(we omit this here as the rewritten expressions are extremely long).
Those counterterms are defined,
as usual,
by the splitting of the bare version of the independent parameters:
\bs
\ba
M_{l(0)} &=& M_l + \delta M_l,
\\
M_{h(0)} &=& M_h + \delta M_h,
\\
\bar M_{l(0)} &=& \bar M_l + \delta \bar M_l,
\\
\bar M_{h(0)} &=& \bar M_h + \delta \bar M_h,
\\
e_{(0)} &=& \left( 1 + \delta Z_e \right) e,
\\
\psi_{(0)} &=& \delta \psi,
\\
\xi_{(0)} &=& \xi + \delta \xi.
\ea
\es
The mass counterterms in the OSS scheme are
\be
\delta m_{V_i}^2 = - \widetilde{\operatorname{Re}}\,
\Sigma_{\mathrm{T}}^{V_iV_i} \left(m_{V_i}^2 \right).
\ee

The counterterm $\delta Z_e$,
which \textit{a priori}\/ has a very complicated expression,
may be significantly simplified by using a Ward identity.
That identity was derived in detail,
for the case of the SM,
in appendix~F of Ref.~\cite{Fontes:2021kue}.
By generalizing it to the LRM,
we obtain:
\be
\delta Z_e =
- \frac{1}{2}\, \delta Z_{AA}
- \frac{1}{2 c_w}\, \delta Z_{Z_lA}
\left( s_w\, c_\psi + \frac{s_\alpha}{c_\alpha}\, s_\psi \right) 
- \frac{1}{2 c_w}\, \delta Z_{Z_hA}
\left( \frac{s_\alpha}{c_\alpha}\, c_\psi - s_w\, s_\psi \right),
\label{74834sd}
\ee
which we use after setting $\psi = 0$.
Note that Eq.~\eqref{74834sd} reduces to its SM version,
\viz\ to equation~(F.60) of Ref.~\cite{Fontes:2021kue},
when $\alpha = \psi = 0$ and $Z_l = Z$.

In order to fix the counterterms $\delta \psi$ and $\delta \xi$
we apply a method similar to the one described in Ref.~\cite{Fontes:2021iue}
to obtain:\fn{Whereas the counterterms
  of the mixing parameters in Ref.~\cite{Fontes:2021iue}
  were fixed in Feynman gauge,
  the counterterms $\delta \psi$ and $\delta \xi$ in this paper
  are not fixed in any specific gauge.
  This has to do with the tadpole scheme.
  Namely,
  Ref.~\cite{Fontes:2021iue} used the tadpole scheme dubbed FJTS,
  wherein the parameter counterterms
  are required to behave as gauge-independent
  (\ie\ not to change when the gauge is changed)
  in order for the observables to be gauge-independent;
  and while that naturally happens for the parameter counterterms
  fixed through on-shell or minimal subtraction,
  one needs to enforce it in the case of parameter counterterms
  fixed by symmetry relations.
  In the present paper,
  by contrast,
  the tadpole scheme at stake is PRTS,
  wherein the parameter counterterms are in general gauge dependent;
  that gauge dependence ends up cancelling in the renormalized functions,
  as long as the whole set of parameter counterterms is renormalized
  by using a momentum-subtraction scheme.
  Since the latter condition is satisfied in this paper,
  the renormalized functions of this paper are gauge-independent,
  as we have explicitly checked (numerically).
  For details see Ref.~\cite{Fontes:2021kue}.}
\bs
\label{eq:psi-reno}
\ba
\delta \psi &=& \dfrac{1}{4} \left[
  \delta Z_{Z_l Z_h} - \delta Z_{Z_h Z_l}
  + \dfrac{s_w}{c_w} \left( \delta Z_{Z_h A} - \delta Z_{A Z_h} \right)
  \right],
\\
\delta \xi &=& \dfrac{1}{4} \left( \delta Z_{W_h W_l} - \delta Z_{W_l W_h} \right).
\ea
\es
Using Eqs.~\eqref{eq:FLFR} to~\eqref{eq:psi-reno}
we calculate $i \left. {\Gamma}_{\mu}^{Z_l b \bar b} \right|_{\text{CT}}$
and hence Eq.~\eqref{eq:base}.

Using \FM\ we have numerically checked
both the UV-finiteness and the gauge-independence
of $i\, \hat{\Gamma}_{\mu}^{Z_l b \bar b}$.

\section{Discussion}
\label{sec:numerical}

\begin{figure}[ht]
  \begin{center}
    \includegraphics[width=1.0\textwidth]{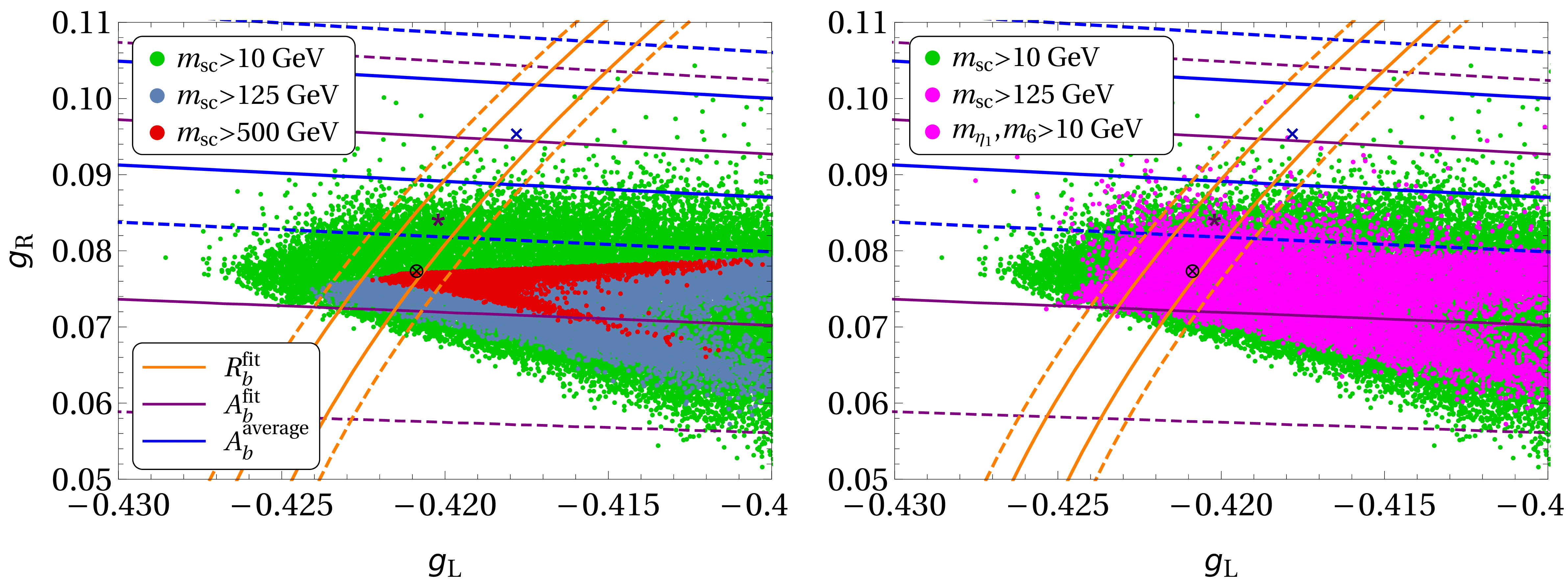}
  \end{center}
  \caption{Scatter plot of the values of $g_L$ and $g_R$ in our model.
    The crossed circle marks the SM prediction~\eqref{SM};
    the star marks the best-fit point of solution~1$^\mathrm{fit}$
    and the cross the best-fit point of solution~1$^\mathrm{average}$,
    \textit{cf.}\ table~\ref{table_solutions}.
    The orange lines mark the $1\sigma$ (full lines)
    and $2\sigma$ (dashed lines) boundaries
    of the region determined by the experimental value~\eqref{Rbfit};
    similarly,
    the violet lines correspond to the value~\eqref{Abfit}
    and the blue lines to the value~\eqref{Abtrue}.
    Left panel:
    the three cases where the lower bound on the masses of the scalars
    are 10\,GeV (green points),
    125\,GeV (blue points),
    and  500\,GeV (red points).
    Right panel:
    the green points are the same as in the left panel;
    the magenta points have only the masses
    $m_{\eta1}$ of the lightest pseudoscalar
    and $m_6$ of the lightest neutral scalar above 10\,GeV,
    while all the other scalars have masses above 125\,GeV. 
    }
  \label{fig_gLgR}
\end{figure}

In this section we discuss the results obtained by using
the procedure in section~\ref{sec:parameters}.
In our numerical analysis
the scalar masses were assumed to be lower than 1\,TeV;
this bound is mostly irrelevant,
because the contributions of the new scalars to $\delta g_L$ and $\delta g_R$
tend to zero when the scalars become very heavy.
On the other hand,
the lower bound on the masses of the scalars is important,
because scalars with very low masses allow one to fit the $Z b \bar b$ vertex,
as has already been found out
in the context of the 2HDM and 3HDM~\cite{Jurciukonis:2021wny}.
For this reason,
in this section we display four different fits:
\begin{description}
\item In the first fit
  (for which we have used green points in our scatter plots)
  we have assumed a lower bound 10\,GeV
  on the masses of all the scalars.\footnote{This is avowedly
  a lax lower bound.
  In particular,
  there is a recent experimental lower bound of 150\,GeV
  on the masses of charged scalars~\cite{Khachatryan:2015qxa,Arbey:2017gmh};
  but it is also true that low-mass charged scalars
  are \emph{not} very useful to adequately fit
  the $Z b \bar b$ vertex.
  We have used this lower bound to illustrate
  that neutral scalars with \emph{very} low masses
  are really \emph{needed} to fit the $Z b \bar b$ vertex
  in our LRM.}
\item In the second fit
  (displayed through blue points in the scatter plots),
  the lower bound on the scalar masses is 125\,GeV.
\item In the third fit
  (red points)
  the lower bound is 500\,GeV.
\item In the fourth fit
  (displayed exclusively in the right panel of Fig.~\ref{fig_gLgR}),
  the lower bound on all the scalar masses is 125\,GeV,
  \emph{except} only the lightest scalar and the lightest pseudoscalar,
  which are allowed to have mass as low as 10\,GeV.
\end{description}
We emphasize that we have attempted to explore
whether very small masses allow one to fit $g_L$ and $g_R$ adequately,
irrespective of whether such low masses are realistic or not;
we do \emph{not} claim that our very-low-mass scalars
may ever be compatible with the experimental data.

We depict in Fig.~\ref{fig_gLgR}
the confrontation between experiment and the values of $g_L$ and $g_R$
attainable in our LR model.
In the left panel one sees that,
when one forces the scalar masses to be larger than 125\,GeV,
the LRM is unable to achieve a better agreement
with experiment than the SM---it even does not attain
the $2\sigma$ interval on the value of $A_b$~\eqref{Abtrue}.
Only when one allows very low scalar masses $\gtrsim 10$\,GeV
are the central values of both solutions 1$^\mathrm{fit}$
and 1$^\mathrm{average}$ attainable.
In the right panel of Fig.~\ref{fig_gLgR} one sees that,
if both the masses of the lightest pseudoscalar,
\viz~$m_{\eta1}$,
and lightest neutral scalar,
\viz~$m_6$,
are $\gtrsim 10$\,GeV,
while all the other charged and neutral scalars are above 125\,GeV,
then the central values of both solutions 1$^\mathrm{fit}$
and 1$^\mathrm{average}$ are attainable too.
This shows that only one light scalar and one light pseudoscalar
are needed in order to correct $g_R$.
However,
the other scalars are important too;
when the lower bound on their masses is increased to 200\,GeV
we are already unable to reach solution 1$^\mathrm{fit}$
(points for that fit are not shown).

The unexpected conclusion of this paper is thus that,
in spite of its many parameters---most of them in the scalar sector---and
in spite of the existence of right-handed currents,
our LRM with $\psi = 0$ is basically unable to fit the experimental values
of $g_L$ and---mainly---$g_R$,
unless one resorts to scalars with extremely low masses.
In particular,
our LRM does not do a better job in fitting the $Z b \bar b$ vertex
than the much simpler 2HDM or 3HDM.

We should emphasize once again that we
have assumed the renormalized mixing angle $\psi$ to be zero.
This assumption has a crucial technical advantage:
instead of having to deal with the effect of soft photons and gluons
on the $Z b \bar b$ vertex,
and on the way they operate to cancel the infrared
divergences in the loops---both in the SM
and in the LRM---we could simply
subtract $g_L$ and $g_R$ in both models to eliminate
those divergences.
On the one hand,
this assumption is certainly a weakness of our computation.
On the other hand,
taking into account the soft photons and gluons would certainly entail
various technical difficulties; moreover,
since recent analyses~\cite{Bobovnikov:2018fwt,Osland:2020onj,Osland:2022ryb}
suggest that $\left| \psi \right|$ cannot be much larger than $10^{-3}$ anyway,
while at the tree level one needs $\psi \sim 10^{-2}$
to fit the $Z b \bar b$ vertex---see
Section~\ref{sec:tree-level}---we believe that
allowing $\psi \neq 0$ would not change our final conclusions substantially.

We may also comment on the ratio $l/g$ between the gauge coupling constants
of $SU(2)_L$ and $SU(2)_R$.
The right panel of Fig.~\ref{XXX} suggests that $l/g$ should not be
very far from 1 in a fit of the $Z b \bar b$ vertex at tree level.
At loop level we see,
in Fig.~\ref{fig_gauge},
that $l/g$ is much freer;
we have found points
(not depicted in that figure)
with $l/g$ as high as~6.
We emphasize,
however,
that in making Fig.~\ref{fig_gauge} we just require the points
to adequately fit $R_b$,
while in Fig.~\ref{XXX} what is at stake is fitting both $R_b$
\emph{and} $A_b$.

In Appendix~\ref{app:results}
we give more information on the ranges of the parameters of the LRM.

\vspace*{5mm}

\paragraph{Acknowledgements:}
We thank Kristjan Kannike and Per Osland for explanations
about Refs.~\cite{Kannike:2021fth}
and~\cite{Bobovnikov:2018fwt,Osland:2020onj,Osland:2022ryb},
respectively.
The work of D.F.\ is supported by the United States Department of Energy
under Grant Contract DE-SC0012704.
The work of D.J.\ is supported
by the Lithuanian Academy of Sciences through project DaFi2021
and by a COST STSM grant through action CA1620.
The work of L.L.\ is supported by the Portuguese
Foundation for Science and Technology through projects UIDB/00777/2020,
UIDP/00777/2020,
CERN/FIS-PAR/0008/2019,
CERN/FIS-PAR/0002/2021,
CERN/FIS-PAR/0004/2019,
and CERN/FIS-PAR/0019/2021.

\newpage

\begin{appendix}

\setcounter{equation}{0}
\renewcommand{\theequation}{A\arabic{equation}}

\section{The Goldstone bosons}
\label{app:goldstonebosons}

\paragraph{Gauge-boson mass matrices:}
From Eq.~\eqref{covder} and Table~\ref{u1},
the covariant derivatives of the scalar fields are
\bs
\label{pripri}
\ba
D^\mu m &=& \partial^\mu m
- \frac{i g}{\sqrt{2}}\, W^{+ \mu} n
+ i e A^\mu m
+ \frac{i g \left( s_w^2 - c_w^2 \right)}{2 c_w}\, Z^\mu m
+ \frac{i l s_\alpha^2}{2 c_\alpha}\, X^\mu m, \label{dmum2}
\\
D^\mu n &=& \partial^\mu n
- \frac{i g}{\sqrt{2}}\, W^{- \mu} m
+ \frac{i g}{2 c_w}\, Z^\mu n
+ \frac{i l s_\alpha^2}{2 c_\alpha}\, X^\mu n, \label{dmun2}
\\
D^\mu p &=& \partial^\mu p
- \frac{i l}{\sqrt{2}}\, V^{+ \mu} q
+ i e A^\mu p
+ \frac{i g s_w^2}{c_w}\, Z^\mu p
+ \frac{i l \left( s_\alpha^2 - c_\alpha^2 \right)}{2 c_\alpha}\, X^\mu p,
\label{dmup2}
\\
D^\mu q &=& \partial^\mu q
- \frac{i l}{\sqrt{2}}\, V^{- \mu} p
+ \frac{i l}{2 c_\alpha}\, X^\mu q, \label{dmuq2}
\\
D^\mu a &=& \partial^\mu a
- \frac{i g}{\sqrt{2}}\, W^{+ \mu} b
+ \frac{i l}{\sqrt{2}}\, V^{+ \mu} {d}^\ast
+ i e A^\mu a
+ \frac{i g \left( s_w^2 - c_w^2 \right)}{2 c_w}\, Z^\mu a
- \frac{i l c_\alpha}{2}\, X^\mu a, \label{dmua2}
\hspace*{7mm} \\
D^\mu b &=& \partial^\mu b
- \frac{i g}{\sqrt{2}}\, W^{- \mu} a
- \frac{i l}{\sqrt{2}}\, V^{+ \mu} c^\ast
+ \frac{i g}{2 c_w}\, Z^\mu b
- \frac{i l c_\alpha}{2}\, X^\mu b, \label{dmub2}
\\
D^\mu c &=& D^\mu a \left(
a \to c,\,
b \to d,\,
{d}^\ast \to {b}^\ast
\right), \label{dmuc2}
\\
D^\mu d &=& D^\mu b \left(
b \to d,\,
a \to c,\,
c^\ast \to a^\ast
\right). \label{dmud2}
\ea
\es
When the scalar fields in Eqs.~\eqref{pripri}
get substituted by their VEVs \eqref{VEVs},
one obtains
\bs
\label{ue1}
\ba
D^\mu m &\to& - \frac{i g u_L}{\sqrt{2}}\, W^{+ \mu}, \label{dmum}
\\
D^\mu n &\to& \frac{i u_L}{2} \left( \frac{g}{c_w}\, Z^\mu
+ \frac{l s_\alpha^2}{c_\alpha}\, X^\mu \right), \label{dmun}
\\
D^\mu p &\to& - \frac{i l u_R}{\sqrt{2}}\, V^{+ \mu}, \label{dmup}
\\
D^\mu q &\to& \frac{i l u_R}{2 c_\alpha}\, X^\mu, \label{dmuq}
\\
D^\mu a &\to& \frac{i}{\sqrt{2}} \left( - g W^{+ \mu} v_1
+ l V^{+ \mu} v_2 \right), \label{dmua}
\\
D^\mu b &\to& \frac{i v_1}{2} \left( \frac{g}{c_w}\, Z^\mu
- l c_\alpha X^\mu \right), \label{dmub}
\\
D^\mu c &\to& \frac{i}{\sqrt{2}} \left( - g W^{+ \mu} v_2
+ l V^{+ \mu} v_1 \right), \label{dmuc}
\\
D^\mu d &\to& \frac{i v_2}{2} \left( \frac{g}{c_w}\, Z^\mu
- l c_\alpha X^\mu \right). \label{dmud}
\ea
\es
From Eqs.~\eqref{dmum},
\eqref{dmup},
\eqref{dmua},
and~\eqref{dmuc} one gets the charged-gauge-boson mass matrix $M_c$,
given by
\be
\label{nfde9re00}
A \equiv \left( M_c \right)_{11} =
\frac{G V_L^2}{2},
\qquad
B \equiv \left( M_c \right)_{22} =
\frac{L V_R^2}{2},
\qquad
D \equiv \left( M_c \right)_{12} = - g l v_1 v_2,
\ee
where
\be
V_L \equiv \sqrt{V_1 + V_2 + U_L}, \qquad V_R \equiv \sqrt{V_1 + V_2 + U_R}
\ee
are positive by definition.
From Eqs.~\eqref{dmun},
\eqref{dmuq},
\eqref{dmub},
and~\eqref{dmud} one has,
for the neutral-gauge-boson mass matrix $M_n$,
the formulas
\bs
\label{7fuf8}
\ba
R \equiv \left( M_n \right)_{11} &=&
\frac{G L + G H + L H}{2 \left( L + H \right)}\, V_L^2,
\label{mn11} \\
S \equiv \left( M_n \right)_{22} &=&
\frac{L^2 \left( V_1 + V_2 \right) + H^2 U_L + \left( L + H \right)^2
  U_R}{2 \left( L + H \right)},
\label{mn22} \\
T \equiv \left( M_n \right)_{12} &=&
\frac{\varrho_1}{2 \left( L + H \right)}
\left[ H U_L - L \left( V_1 + V_2 \right) \right].
\label{mn12}
\ea
\es
Note that $\left( M_n \right)_{12}$ cannot be made zero in any natural
(\textit{i.e.}\ enforceable through a symmetry)
way.
This means that
the mixing angle $\psi$ cannot be made zero through any symmetry.

\paragraph{$G_W^\pm$ and $G_V^\pm$:}
We define the normalized states
\be
\label{nidf0}
G_W^- \equiv \frac{u_L m^\ast+ v_1 a^\ast + v_2 c^\ast}{V_L},
\qquad
G_V^- \equiv \frac{u_R p^\ast- v_2 a^\ast - v_1 c^\ast}{V_R}.
\ee
Then,
from Eqs.~\eqref{dmum2},
\eqref{dmup2},
\eqref{dmua2},
and~\eqref{dmuc2},
\be
\label{mmmmmd}
\mathcal{L} = \cdots - i \left(
\sqrt{A}\ W_\mu^+\, \partial^\mu G_W^-
+
\sqrt{B}\ V_\mu^+\, \partial^\mu G_V^-
\right) + \mathrm{H.c.},
\ee
where
\be
\label{jforwo}
\sqrt{A} = \frac{g V_L}{\sqrt{2}},
\qquad
\sqrt{B} = \frac{l V_R}{\sqrt{2}}
\ee
are positive by definition.

\paragraph{Diagonalization of $M_c$:}
From Eqs.~\eqref{ufiee99843} and~\eqref{cieoggkl},
\be
\label{djfire00}
\left( \begin{array}{cc} c_\xi & - s_\xi \\*[1mm]
s_\xi & c_\xi \end{array} \right)
\left( \begin{array}{cc} A & D \\ D & B \end{array} \right)
\left( \begin{array}{cc} c_\xi & s_\xi \\*[1mm]
  - s_\xi & c_\xi \end{array} \right)
= \left( \begin{array}{cc} \bar M_l & 0 \\ 0 & \bar M_h \end{array} \right).
\ee
Therefore,
\bs
\label{xixi}
\ba
2 c_\xi s_\xi &=& \frac{2 D}{\bar M_h - \bar M_l}, \label{xi1} \\*[1mm]
c_\xi^2 - s_\xi^2 &=& \frac{B - A}{\bar M_h - \bar M_l}, \\*[1mm]
\bar M_h - \bar M_l &=& \sqrt{\left( A - B \right)^2 + 4 D^2}, \\
\bar M_l &=& \frac{A + B - \left( \bar M_h - \bar M_l \right)}{2}, \\*[1mm]
\bar M_h &=& \frac{A + B + \left( \bar M_h - \bar M_l \right)}{2}.
\ea
\es
From Eqs.~\eqref{nfde9re00} and~\eqref{xi1},
\be
c_\xi s_\xi \left( \bar M_h - \bar M_l \right) = - g l v_1 v_2,
\label{nvkdfps0}
\ee
Since $c_\xi$,
$\bar M_h - \bar M_l$,
$g$,
$l$,
and $v_2$ are positive (without lack of generality),
Eq.~\eqref{nvkdfps0} means that the sign of $s_\xi$ must be chosen
opposite to the sign of $v_1$.

\paragraph{$G_l^\pm$ and $G_h^\pm$:}
We now define
\bs
\label{99}
\ba
G_l^- &\equiv& \frac{c_\xi\, \sqrt{A}\, G_W^-
  - s_\xi\, \sqrt{B}\, G_V^-}{\sqrt{\bar M_l}},
\\
G_h^- &\equiv& \frac{s_\xi\, \sqrt{A}\, G_W^-
  + c_\xi\, \sqrt{B}\, G_V^-}{\sqrt{\bar M_h}},
\ea
\es
where $\sqrt{\bar M_l}$ and $\sqrt{\bar M_h}$ are positive by definition.
Then Eq.~\eqref{mmmmmd} is rewritten as
\be
\label{jvfi000}
\mathcal{L} = \cdots - i \left(
\sqrt{\bar M_l}\ W_{l \mu}^+\ \partial^\mu G_l^-
+ \sqrt{\bar M_h}\ W_{h \mu}^+\ \partial^\mu G_h^-
\right) + \mathrm{H.c.},
\ee
which goes into the last line of Eq.~\eqref{93834kmf2}.
Equations~\eqref{99} are the definition of $G_l^\pm$ and $G_h^\pm$.
Note that these states are orthogonal:
\bs
\ba
\sqrt{\bar M_l \bar M_h}\ G_l^- \cdot G_h^-
&=& c_\xi s_\xi \left( A - B \right)
+ \left( c_\xi^2 - s_\xi^2 \right) \sqrt{A B}\ G_W^- \cdot G_V^-
\\ &=& \frac{D}{\bar M_h - \bar M_l} \left( A - B \right)
+ \frac{B - A}{\bar M_h - \bar M_l}\ \sqrt{A B}\ \frac{- 2 v_1 v_2}{V_L V_R}
\\ &=& \frac{D}{\bar M_h - \bar M_l} \left( A - B \right)
+ \frac{B - A}{\bar M_h - \bar M_l}\ \sqrt{A B}\ \frac{2 D}{g l V_L V_R}
\\ &=& \frac{D}{\bar M_h - \bar M_l} \left( A - B \right)
+ \frac{B - A}{\bar M_h - \bar M_l}\, D
\\ &=& 0.
\ea
\es
They are also correctly normalized:
\bs
\ba
\bar M_l\, G_l^- \cdot G_l^-
&=& c_\xi^2 A + s_\xi^2 B - 2 c_\xi s_\xi \sqrt{A B}\, G_W^- \cdot G_V^-
\\ &=& \frac{1 + c_\xi^2 - s_\xi^2}{2}\, A
+ \frac{1 - c_\xi^2 + s_\xi^2}{2}\, B
- 2 c_\xi s_\xi \sqrt{A B}\, \frac{- 2 v_1 v_2}{V_L V_R}
\\ &=& \frac{A + B}{2} + \frac{B - A}{2 \left( \bar M_h - \bar M_l \right)}
\left( A - B \right)
- \frac{2 D}{\bar M_h - \bar M_l}\, \sqrt{A B}\, \frac{D}{\sqrt{A B}}
\hspace*{7mm} \\ &=& \frac{A + B}{2}
- \frac{\left( A - B \right)^2}{2 \left( \bar M_h - \bar M_l \right)}
- \frac{2 D^2}{\bar M_h - \bar M_l}
\\ &=& \bar M_l,
\ea
\es
\bs
\ba
\bar M_h\, G_h^- \cdot G_h^-
&=& s_\xi^2 A + c_\xi^2 B + 2 c_\xi s_\xi \sqrt{A B}\, G_W^- \cdot G_V^-
\\ &=& \frac{1 - c_\xi^2 + s_\xi^2}{2}\, A
+ \frac{1 + c_\xi^2 - s_\xi^2}{2}\, B
+ 2 c_\xi s_\xi \sqrt{A B}\, \frac{- 2 v_1 v_2}{V_L V_R}
\\ &=& \frac{A + B}{2} - \frac{B - A}{2 \left( \bar M_h - \bar M_l \right)}
\left( A - B \right)
+ \frac{2 D}{\bar M_h - \bar M_l}\, \sqrt{A B}\, \frac{D}{\sqrt{A B}}
\hspace*{10mm}
\\ &=& \frac{A + B}{2} + \frac{\left( A - B \right)^2}{2
\left( \bar M_h - \bar M_l \right)}
+ \frac{2 D^2}{\bar M_h - \bar M_l}
\\ &=& \bar M_h.
\ea
\es

\paragraph{$G_Z$ and $G_X$:}
We define
\be
G_Z \equiv \frac{u_L \eta_L + v_1 \eta_1 + v_2 \eta_2}{V_L},
\qquad
G_X^\prime \equiv \frac{u_R \eta_R - v_1 \eta_1 - v_2 \eta_2}{V_R}.
\ee
These states are normalized,
but they are not orthogonal to each other:
\be
G_Z \cdot G_Z = G^\prime_X \cdot G^\prime_X = 1,
\qquad
G_Z \cdot G_X^\prime = \frac{- V_1 - V_2}{V_L V_R}.
\ee
From Eqs.~\eqref{dmun2},
\eqref{dmuq2},
\eqref{dmub2},
and~\eqref{dmud2},
\be
\label{mdksel4}
\mathcal{L} = \cdots +
\left( \frac{g}{\sqrt{2} c_w}\, Z^\mu
+ \frac{l s_\alpha^2}{\sqrt{2} c_\alpha}\, X^\mu \right) V_L\, \partial_\mu G_Z
+ \frac{l}{\sqrt{2}  c_\alpha}\, X^\mu\, V_R\, \partial_\mu G_X^\prime.
\ee
Using the quantities $R$ and $S$ in Eqs.~\eqref{7fuf8},
one may rewrite Eq.~\eqref{mdksel4} as
\be
\label{7u86596}
\mathcal{L} = \cdots + \sqrt{R}\, Z^\mu\, \partial_\mu G_Z
+ \sqrt{S}\, X^\mu\, \partial_\mu G_X,
\ee
where
\be
\sqrt{R} = \frac{g V_L}{\sqrt{2} c_w}
\ee
and
\be
G_X \equiv \frac{1}{\sqrt{S}}\, \frac{l}{\sqrt{2} c_\alpha}
\left( s_\alpha^2 V_L G_Z + V_R G^\prime_X \right).
\ee
Notice that both $\sqrt{R}$ and $\sqrt{S}$ are,
by definition,
positive.
One easily ascertains that
\be
G_X \cdot G_X = 1, \qquad G_Z \cdot G_X = \frac{T}{\sqrt{R S}},
\ee
where $T$ is in Eq.~\eqref{mn12}.

\paragraph{$G_l^0$ and $G_h^0$:}
We define
\be
\label{3o00jj0}
G_l^0 \equiv \frac{c_\psi \sqrt{R}\, G_Z + s_\psi \sqrt{S}\, G_X}{\sqrt{M_l}},
\qquad
G_h^0 \equiv \frac{- s_\psi \sqrt{R}\, G_Z + c_\psi \sqrt{S}\, G_X}{\sqrt{M_h}}.
\ee
Applying the first Eq.~\eqref{ufiee99843} to Eq.~\eqref{7u86596},
we obtain
\be
\mathcal{L} = \cdots + \sqrt{M_l}\, Z_l^\mu\, \partial_\mu G_l^0
+ \sqrt{M_h}\, Z_h^\mu\, \partial_\mu G_h^0,
\ee
as in the last line of Eq.~\eqref{93834kmf}.
The diagonalization of the neutral-gauge-boson mass matrix proceeds as
\be
\label{78df09rv}
\left( \begin{array}{cc} R & T \\ T & S \end{array} \right)
=
\left( \begin{array}{cc} c_\psi & - s_\psi \\ s_\psi & c_\psi \end{array} \right)
\left( \begin{array}{cc} M_l & 0 \\ 0 & M_h \end{array} \right)
\left( \begin{array}{cc} c_\psi & s_\psi \\ - s_\psi & c_\psi \end{array} \right),
\ee
\textit{i.e.}
\bs
\label{uf9995}
\ba
R &=& M_l\, c_\psi^2 + M_h\, s_\psi^2, \\
S &=& M_l\, s_\psi^2 + M_h\, c_\psi^2, \\
T &=& \left( M_l - M_h \right) c_\psi s_\psi. \label{tttttt}
\ea
\es
Hence,
\be
G_l^0 \cdot G_l^0 = G_h^0 \cdot G_h^0 = 1, \qquad G_l^0 \cdot G_h^0 = 0,
\ee
\textit{viz.}\ the states $G_l^0$ and $G_h^0$ are orthonormal as they should.
Equations~\eqref{3o00jj0} lead to Eqs.~\eqref{mf049lg};
for instance,
defining $v_\eta \equiv v_1 \eta_1 + v_2 \eta_2$,
\bs
\ba
G_l^0
&=& \frac{1}{\sqrt{M_l}} \left\{
c_\psi \sqrt{R}\ \frac{v_\eta + u_L \eta_L}{V_L}
+ s_\psi\, \frac{l}{\sqrt{2} c_\alpha}
\left[ s_\alpha^2 \left( v_\eta + u_L \eta_L \right)
  + u_R \eta_R - v_\eta \right] \right\}
\\ &=& \frac{1}{\sqrt{2 M_l}} \left[
c_\psi\, \frac{g}{c_w} \left( v_\eta + u_L \eta_L \right)
+ s_\psi \left( \frac{l}{c_\alpha}\, u_R \eta_R - l c_\alpha v_\eta \right)
\right]
\\ &=& \frac{1}{\sqrt{2 M_l}} \left[
c_\psi\, \frac{\varrho_1}{\sqrt{L + H}} \left( v_\eta + u_L \eta_L \right)
+ s_\psi \left( \sqrt{L + H}\, u_R \eta_R
- \frac{L}{\sqrt{L + H}}\, v_\eta \right)
\right].
\hspace*{7mm}
\ea
\es

\setcounter{equation}{0}
\renewcommand{\theequation}{B\arabic{equation}}

\section{Unitarity conditions}
\label{app:unitarity}

The unitarity conditions constrain
the quartic part of the scalar potential in Eq.~\eqref{v4}.

\paragraph{The states with $T_{L3} = T_{R3} = X = 0$:}
Let us consider the scattering among themselves
of the twelve two-scalar states with quantum numbers $T_{L3} = T_{R3} = X = 0$.
Those states are
\be
\label{states}
a a^\ast,\ c c^\ast,\ b b^\ast,\ d d^\ast,\ a c^\ast,\ c a^\ast,\
b d^\ast,\ d b^\ast,\ m m^\ast,\ n n^\ast,\ p p^\ast,
\ \mbox{and} \ \, q q^\ast
\ee
in this fixed,
conventional order.
The coefficients for their scattering among themselves
are given by the $12 \times 12$ symmetric matrix
\be
M_{12 \times 12} = \left( \begin{array}{cccccc}
  A_{2 \times 2} & B_{2 \times 2} & C_{2 \times 2} &
  D_{2 \times 2} & E_{2 \times 2} & F_{2 \times 2} \\
  B_{2 \times 2} & A_{2 \times 2} & D_{2 \times 2} &
  C_{2 \times 2} & G_{2 \times 2} & F_{2 \times 2} \\
  C_{2 \times 2} & D_{2 \times 2} & H_{2 \times 2} &
  I_{2 \times 2} & J_{2 \times 2} & K_{2 \times 2} \\
  D_{2 \times 2} & C_{2 \times 2} & I_{2 \times 2} &
  H_{2 \times 2} & J_{2 \times 2} & K_{2 \times 2} \\
  E_{2 \times 2} & G_{2 \times 2} & J_{2 \times 2} &
  J_{2 \times 2} & L_{2 \times 2} & M_{2 \times 2} \\
  F_{2 \times 2} & F_{2 \times 2} & K_{2 \times 2} &
  K_{2 \times 2} & M_{2 \times 2} & N_{2 \times 2}
\end{array} \right),
\ee
where
\bs
\ba
A_{2 \times 2} =
\left( \begin{array}{cc} 4 \lambda_1 & 2 \lambda_1 + 4 \lambda_3 \\
  2 \lambda_1 + 4 \lambda_3 & 4 \lambda_1 \end{array} \right),
& &
B_{2 \times 2} =
\left( \begin{array}{cc} 2 \lambda_1 & 2 \lambda_1 \\
  2 \lambda_1 & 2 \lambda_1 \end{array} \right),
\\
C_{2 \times 2} =
\left( \begin{array}{cc} 4 \lambda_4 & 4 \lambda_4 \\
  4 \lambda_4 & 4 \lambda_4 \end{array} \right),
& &
D_{2 \times 2} =
\left( \begin{array}{cc} 2 \lambda_4 & 2 \lambda_4 \\
  2 \lambda_4 & 2 \lambda_4 \end{array} \right),
\\
E_{2 \times 2} =
\left( \begin{array}{cc} \lambda_{4L} & \lambda_{3L} \\
  \lambda_{3L} & \lambda_{4L} \end{array} \right),
& &
F_{2 \times 2} =
\left( \begin{array}{cc} \lambda_{3R} & \lambda_{4R} \\
  \lambda_{4R} & \lambda_{3R} \end{array} \right),
\\
G_{2 \times 2} =
\left( \begin{array}{cc} \lambda_{3L} & \lambda_{4L} \\
  \lambda_{4L} & \lambda_{3L} \end{array} \right),
& &
H_{2 \times 2} =
\left( \begin{array}{cc} 2 \lambda_1 + 4 \lambda_3 & 16 \lambda_2 \\
  16 \lambda_2 & 2 \lambda_1 + 4 \lambda_3 \end{array} \right),
\hspace*{7mm} \\
I_{2 \times 2} =
\left( \begin{array}{cc} 4 \lambda_3 & 8 \lambda_2 \\
  8 \lambda_2 & 4 \lambda_3 \end{array} \right),
& &
J_{2 \times 2} =
\left( \begin{array}{cc} \lambda_{5L} & \lambda_{5L} \\
  \lambda_{5L} & \lambda_{5L} \end{array} \right),
\\
K_{2 \times 2} =
\left( \begin{array}{cc} \lambda_{5R} & \lambda_{5R} \\
  \lambda_{5R} & \lambda_{5R} \end{array} \right),
& &
L_{2 \times 2} =
\left( \begin{array}{cc} 4 \lambda_L & 2 \lambda_L \\
  2 \lambda_L & 4 \lambda_L \end{array} \right),
\\
M_{2 \times 2} =
\left( \begin{array}{cc} \lambda_{LR} & \lambda_{LR} \\
  \lambda_{LR} & \lambda_{LR} \end{array} \right),
& &
N_{2 \times 2} =
\left( \begin{array}{cc} 4 \lambda_R & 2 \lambda_R \\
  2 \lambda_R & 4 \lambda_R \end{array} \right).
\ea
\es
It is readily found that $M_{12 \times 12}$ is equivalent to the direct sum
of the six matrices
\bs
\label{12}
\ba
\hspace*{-8mm} & &
\left( \begin{array}{c} 2 \lambda_1 - 8 \lambda_2 \end{array} \right),
\qquad
\left( \begin{array}{c} 2 \lambda_1 - 24 \lambda_2 + 8 \lambda_3
\end{array} \right),
\qquad
\left( \begin{array}{cc} 2 \lambda_1 + 4 \lambda_3 &
  4 \lambda_4 \\
  4 \lambda_4 &
  2 \lambda_1 + 8 \lambda_2 \end{array} \right),
\\*[1mm]
\hspace*{-8mm} & &
\left( \begin{array}{cc} 2 \lambda_1 - 4 \lambda_3 &
  \sqrt{2} \left( \lambda_{3L} - \lambda_{4L} \right) \\
  \sqrt{2} \left( \lambda_{3L} - \lambda_{4L} \right) &
  2 \lambda_L \end{array} \right),
\qquad
\left( \begin{array}{cc} 2 \lambda_1 - 4 \lambda_3 &
  \sqrt{2} \left( \lambda_{3R} - \lambda_{4R} \right) \\
  \sqrt{2} \left( \lambda_{3R} - \lambda_{4R} \right) &
  2 \lambda_R \end{array} \right),
\hspace*{12mm}
\\*[1mm]
\hspace*{-8mm} & &
\left( \begin{array}{cccc}
  2 \lambda_1 + 24 \lambda_2 + 8 \lambda_3 & 12 \lambda_4 &
  2 \sqrt{2} \lambda_{5L} & 2 \sqrt{2} \lambda_{5R} \\
  12 \lambda_4 & 10 \lambda_1 + 4 \lambda_3 &
  \sqrt{2} \left( \lambda_{3L} + \lambda_{4L} \right) &
  \sqrt{2} \left( \lambda_{3R} + \lambda_{4R} \right) \\
  2 \sqrt{2} \lambda_{5L} &
  \sqrt{2} \left( \lambda_{3L} + \lambda_{4L} \right) &
  6 \lambda_L & 2 \lambda_{LR} \\
  2 \sqrt{2} \lambda_{5R} &
  \sqrt{2} \left( \lambda_{3R} + \lambda_{4R} \right) &
  2 \lambda_{LR} & 6 \lambda_R
\end{array} \right).
\label{151c}
\ea
\es

\paragraph{Other states:}
Besides the 12 states~\eqref{states},
there are other sets
(with quantum numbers other than $T_{L3} = T_{R3} = X = 0$)
of two-scalar states that scatter among themselves.
Most of them yield matrices of scattering coefficients
that are equivalent to one or more of the matrices~\eqref{12}.
But there are a few extra scattering matrices,
\textit{viz.}
\begin{itemize}
\item the four states with $T_{L3} = 0$ and $T_{R3} = X = 1/2$,
  \textit{i.e.}\ $a n$,
  $c n$,
  $b m$,
  and $d m$ produce a $4 \times 4$ matrix that is equivalent to the direct sum
  of the two $2 \times 2$ matrices
  \be
  \label{2}
  \left( \begin{array}{cc}
    \lambda_{3L} & \lambda_{5L} \\ \lambda_{5L} & \lambda_{4L}
  \end{array} \right),
  \qquad
  \left( \begin{array}{cc}
    2 \lambda_{3L} - \lambda_{4L} & \lambda_{5L} \\
    \lambda_{5L} & 2 \lambda_{4L} - \lambda_{3L}
  \end{array} \right);
  \ee
\item the four states with $T_{R3} = 0$ and $T_{L3} = X = 1/2$
  produce the same scattering matrices as in Eq.~\eqref{2}
  but with all the sub-indices $L \to R$;
\item the scattering of the single state $m n$ produces the matrix
  $\left( \begin{array}{c} 2 \lambda_L \end{array} \right)$;
\item the scattering of the single state $p q$ produces the matrix
  $\left( \begin{array}{c} 2 \lambda_R \end{array} \right)$;
\item the scattering of the single state $m p$ produces the matrix
  $\left( \begin{array}{c} \lambda_{LR} \end{array} \right)$.
\end{itemize}
All the other scatterings just reproduce one or more of the matrices above.

\paragraph{Summary of the unitarity conditions:}
Unitarity means that no scattering has too large an amplitude.
This implies that the eigenvalues of all the scattering matrices
in the previous two paragraphs are no larger,
in modulus,
than a certain number $M$;
we use $M = 8 \pi$.
Thus,
the unitarity constraints on $V_4$ are\footnote{Clearly,
some of the conditions~\eqref{uytoewe} are redundant---for instance,
inequality~\eqref{r2} is stronger than
inequality~\eqref{r1}---but we do not have to care much about that.}
\bs
\label{uytoewe}
\ba
2 \left| \lambda_L \right| &<& M; \\
2 \left| \lambda_R \right| &<& M; \\
\left| \lambda_{LR} \right| &<& M; \\
\left| \lambda_{3L} + \lambda_{4L} \right|
+ \sqrt{\left( \lambda_{3L} - \lambda_{4L} \right)^2
  + 4 \left( \lambda_{5L} \right)^2} &<& 2 M; \label{r1} \\
\left| \lambda_{3L} + \lambda_{4L} \right|
+ \sqrt{9 \left( \lambda_{3L} - \lambda_{4L} \right)^2
  + 4 \left( \lambda_{5L} \right)^2} &<& 2 M; \label{r2} \\
\left| \lambda_{3R} + \lambda_{4R} \right|
+ \sqrt{\left( \lambda_{3R} - \lambda_{4R} \right)^2
  + 4 \left( \lambda_{5R} \right)^2} &<& 2 M; \\
\left| \lambda_{3R} + \lambda_{4R} \right|
+ \sqrt{9 \left( \lambda_{3R} - \lambda_{4R} \right)^2
  + 4 \left( \lambda_{5R} \right)^2} &<& 2 M; \\
2 \left| \lambda_1 - 4 \lambda_2 \right| &<& M; \\
2 \left| \lambda_1 - 12 \lambda_2 + 4 \lambda_3 \right| &<& M; \\
\left| 2 \lambda_1 + 4 \lambda_2 + 2 \lambda_3 \right|
+ \sqrt{\left( 4 \lambda_2 - 2 \lambda_3 \right)^2
  + 16 \left( \lambda_4 \right)^2} &<& M; \\
\left| 2 \lambda_1 - 4 \lambda_3 + 2 \lambda_L \right|
+ \sqrt{\left( 2 \lambda_1 - 4 \lambda_3 - 2 \lambda_L \right)^2
  + 8 \left( \lambda_{3L} - \lambda_{4L} \right)^2} &<& 2 M; \\
\left| 2 \lambda_1 - 4 \lambda_3 + 2 \lambda_R \right|
+ \sqrt{\left( 2 \lambda_1 - 4 \lambda_3 - 2 \lambda_R \right)^2
  + 8 \left( \lambda_{3R} - \lambda_{4R} \right)^2} &<& 2 M.
\ea
\es
Moreover,
the moduli of the four (potentially complex) eigenvalues of the
$4 \times 4$ matrix~\eqref{151c} must be smaller than $M$.

\setcounter{equation}{0}
\renewcommand{\theequation}{C\arabic{equation}}

\section{Bounded-from-below conditions}
\label{app:BFB}

The bounded-from-below (BFB) conditions state that
the quartic part of the scalar potential,
\textit{viz.}\ Eq.~\eqref{v4},
must be positive for any values of the fields.
In order to derive the BFB conditions
on $V_{(4)}$ we follow Ref.~\cite{Kannike:2021fth}.

\paragraph{Auxiliary quantities:}
We define the quantities:
\bs
\ba
r_L &\equiv& \left| m \right|^2 + \left| n \right|^2,
\\
r_R &\equiv& \left| p \right|^2 + \left| q \right|^2,
\\
r_0 &\equiv& \left| a \right|^2 + \left| b \right|^2
+ \left| c \right|^2 + \left| d \right|^2,
\\
r_1 &\equiv& a^\ast c + a c^\ast + b^\ast d + b d^\ast,
\\
r_2 &\equiv& - i \left( a^\ast c - a c^\ast + b^\ast d - b d^\ast \right),
\\
k &\equiv& \frac{r_1}{r_0},
\\
\rho_L &\equiv& \frac{
  \left( \left| b \right|^2 + \left| c \right|^2 \right)
  \left| m \right|^2
  + \left( \left| a \right|^2 + \left| d \right|^2 \right)
  \left| n \right|^2
  + 2\, \mathrm{Re}
  \left[ \left( c d^\ast - a b^\ast \right) m^\ast n \right]}{r_0 r_L},
\\
\rho_R &\equiv& \frac{
  \left( \left| a \right|^2 + \left| b \right|^2 \right) \left| p \right|^2
  + \left( \left| c \right|^2 + \left| d \right|^2 \right) \left| q \right|^2
  + 2\, \mathrm{Re}
  \left[ \left( b c - a d \right) p^\ast q \right]}{r_0 r_R}.
\ea
\es
Equation~\eqref{v4} is then rewritten
\ba
V_{(4)} &=&
\lambda_L \left( r_L \right)^2
+ \lambda_R \left( r_R \right)^2
+ \lambda_{LR} r_L r_R
\no & &
+ \lambda_1 \left( r_0 \right)^2
+ \left( \lambda_3 + 2 \lambda_2 \right) \left( r_1 \right)^2
+ \left( \lambda_3 - 2 \lambda_2 \right) \left( r_2 \right)^2
+ 2 \lambda_4\, r_0\, r_1
\no & &
+ \lambda_{3L} \rho_L r_0 r_L
+ \lambda_{4L} \left( 1 - \rho_L \right) r_0 r_L
+ \lambda_{5L} r_1 r_L
\no & &
+ \lambda_{3R} \rho_R r_0 r_R
+ \lambda_{4R} \left( 1 - \rho_R \right) r_0 r_R
+ \lambda_{5R} r_1 r_R.
\ea

\paragraph{Elimination of $r_2$:}
We note that
\ba
\left( r_0 \right)^2 - \left( r_1 \right)^2 - \left( r_2 \right)^2
&=&
\left( \left| a \right|^2 + \left| b \right|^2
+ \left| c \right|^2 + \left| d \right|^2 \right)^2
- 4 \left| a^\ast c + b^\ast d \right|^2
\no &=&
\left( \left| a \right|^2 - \left| c \right|^2 \right)^2
+ \left( \left| b \right|^2 - \left| d \right|^2 \right)^2
+ 2 \left( \left| a b^\ast - c d^\ast \right|^2
+ \left| a d - b c \right|^2 \right)
\no &\ge& 0.
\ea
Therefore,
$\left( r_2 \right)^2 \le \left( r_0 \right)^2 - \left( r_1 \right)^2$.
This implies that
\be
\mathrm{min} \left[ \left( \lambda_3 - 2 \lambda_2 \right)
  \left( r_2 \right)^2 \right]
=
\left( \lambda_3 - 2 \lambda_2 \right)
\Theta \left( 2 \lambda_2 - \lambda_3 \right)
\left[ \left( r_0 \right)^2 - \left( r_1 \right)^2 \right],
\ee
where $\Theta$ is Heaviside's theta function.
Therefore,
\ba
\label{930323}
\mathrm{min} \left[ V_{(4)} \right] &=&
\lambda_L \left( r_L \right)^2
+ \lambda_R \left( r_R \right)^2
+ \lambda_{LR} r_L r_R
\no & &
+ \left[
  \lambda_1
  + \left( \lambda_3 + 2 \lambda_2 \right) k^2
  + \left( \lambda_3 - 2 \lambda_2 \right)
  \Theta \left( 2 \lambda_2 - \lambda_3 \right)
  \left( 1 - k^2 \right)
  + 2 \lambda_4 k
  \right] \left( r_0 \right)^2
\no & &
+ \left[ \lambda_{3L} \rho_L + \lambda_{4L} \left( 1 - \rho_L \right)
  + \lambda_{5L} k \right] r_0 r_L
\no & &
+ \left[ \lambda_{3R} \rho_R + \lambda_{4R} \left( 1 - \rho_R \right)
  + \lambda_{5R} k \right] r_0 r_R.
\ea
This must be positive for any realistic values of $\rho_L$,
$\rho_R$,
$k$,
$r_L$,
$r_R$,
and $r_0$.
We define the matrix
\be
\Lambda = \left( \begin{array}{ccc}
  2\, \lambda_{00} &
  \lambda_{L0} &
  \lambda_{R0} \\
  \lambda_{L0} &
  2\, \lambda_L &
  \lambda_{LR} \\
  \lambda_{R0} &
  \lambda_{LR} &
  2\, \lambda_R
\end{array} \right),
\ee
where
\bs
\ba
\lambda_{00} &\equiv& \lambda_1
+ \left( \lambda_3 + 2 \lambda_2 \right) k^2
+ \left( \lambda_3 - 2 \lambda_2 \right)
\Theta \left( 2 \lambda_2 - \lambda_3 \right)
\left( 1 - k^2 \right)
+ 2 \lambda_4 k,
\\
\lambda_{L0} &\equiv& \lambda_{3L} \rho_L
+ \lambda_{4L} \left( 1 - \rho_L \right)
+ \lambda_{5L} k,
\label{lambdal0} \\
\lambda_{R0} &\equiv& \lambda_{3R} \rho_R
+ \lambda_{4R} \left( 1 - \rho_R \right)
+ \lambda_{5R} k
\ea
\es
are functions of $\rho_L$,
$\rho_R$,
and $k$.
Equation~\eqref{930323} then reads
\be
\label{8gir0swreir}
\mathrm{min} \left[ V_{(4)} \right] =
\frac{1}{2} \left( \begin{array}{ccc} r_0, & r_L, & r_R \end{array} \right)
\Lambda
\left( \begin{array}{c} r_0 \\ r_L \\ r_R \end{array} \right).
\ee
Since $r_0$,
$r_L$,
and $r_R$ are non-negative,
the BFB condition is equivalent to the requirement
that the matrix $\Lambda$ be co-positive~\cite{Kannike:2012pe},
\textit{viz.}\ that
\bs
\label{fullconditions}
\ba
\lambda_{00} &>& 0, \label{183a}
\\
\lambda_L &>& 0, \label{183b}
\\
\lambda_R &>& 0, \label{183c}
\\
\bar{\lambda}_{L0} &>& 0, \label{183d}
\\
\bar{\lambda}_{R0} &>& 0, \label{183e}
\\
\bar{\lambda}_{LR} &>& 0, \label{183f}
\\
\sqrt{\lambda_{00} \lambda_L \lambda_R}
+ \frac{\lambda_{L0}}{2}\, \sqrt{\lambda_R}
+ \frac{\lambda_{R0}}{2}\, \sqrt{\lambda_L}
+ \frac{\lambda_{LR}}{2}\, \sqrt{\lambda_{00}}
+ \sqrt{2 \bar \lambda_{L0} \bar \lambda_{R0} \bar \lambda_{LR}} &>& 0,
\label{183g}
\ea
\es
where
\bs
\ba
\bar{\lambda}_{L0}
&\equiv& \frac{\lambda_{L0}}{2}
+ \sqrt{\lambda_L \lambda_{00}},
\\
\bar{\lambda}_{R0}
&\equiv& \frac{\lambda_{R0}}{2}
+ \sqrt{\lambda_R \lambda_{00}},
\\
\bar{\lambda}_{LR} &\equiv& \frac{\lambda_{LR}}{2} + \sqrt{\lambda_L \lambda_R}.
\ea
\es
The inequalities~\eqref{fullconditions} must hold
for all realistic values of $\rho_L$,
$\rho_R$,
and $k$.
The inequalities~\eqref{183b},
\eqref{183c},
and~\eqref{183f} do not depend on those parameters;
the same does not apply to the other four inequalities.

\paragraph{The inequality~\eqref{183a}:}
The quantity $\lambda_{00}$ does not depend on $\rho_L$ and $\rho_R$,
it only depends on $k$.
Clearly,
\be
\left( r_0 \right)^2 - \left( r_1 \right)^2 - \left( r_2 \right)^2
\ge 0\ \Rightarrow\ \left| r_1 \right| \le \left| r_0 \right|\
\Leftrightarrow\ k \in \left[ -1,\, +1 \right].
\ee
Thus,
the inequality~\eqref{183a} means that $\lambda_{00}$ must be positive
for any value of $k \in \left[ -1,\, +1 \right]$.
Since $\lambda_{00}$ is a quadratic polynomial in $k$,
it is easy to derive the conditions for this to happen.
Kannike~\cite{Kannike:2021fth} proposed
\bs
\label{bidoublet_BFB}
\ba
\lambda_{1} &>& 0,
\\
\lambda_{1} - 2 \left| \lambda_{2} \right| + \lambda_{3} &>& 0,
\label{ebidoublet_BFB:1}
\\
\lambda_{1} + 2 \lambda_{2} + \lambda_{3} - 2 \left| \lambda_{4}\right| &>& 0,
\label{bidoublet_BFB:2}
\\
\lambda_{1} - 2 \lambda_{2} - \lambda_{3} + 
\sqrt{\left( \lambda_{1} + 2 \lambda_{2} + \lambda_{3} \right)^2
  - 4 \left( \lambda_{4} \right)^2} &>& 0,
\label{bidoublet_BFB:3}
\\
\lambda_{1} - 6 \lambda_{2} + \lambda_{3} +
\sqrt{\left( \lambda_{1} + 2 \lambda_{2} + \lambda_{3} \right)^2
  - 4 \left( \lambda_{4} \right)^2} &>& 0.
\label{bidoublet_BFB:4}
\ea
\es
Alternatively,
Chauhan~\cite{Chauhan:2019fji} gave
\bs
\label{ch2}
\ba
\lambda_1 &>& 0,
\\
\lambda_1 + 2 \lambda_2 + \lambda_3 - 2 \left| \lambda_4 \right| &>& 0,
\\
\lambda_1 - \frac{\left( \lambda_4 \right)^2}{2 \lambda_2 + \lambda_3} &>& 0
\quad \Leftarrow \quad 2 \lambda_2 + \lambda_3 > \left| \lambda_4 \right|,
\\
\lambda_1 - 2 \lambda_2 + \lambda_3
- \frac{\left( \lambda_4 \right)^2}{4 \lambda_2} &>& 0
\quad \Leftarrow \quad 4 \lambda_2 > \left| \lambda_4 \right|.
\ea
\es
We instead suggest
\bs
\label{lav2}
\ba
\lambda_1 + 2 \lambda_2 + \lambda_3 - 2 \left| \lambda_4 \right| &>& 0,
\\
\lambda_1 - 2 \lambda_2 + \lambda_3
- \frac{\left( \lambda_4 \right)^2}{4 \lambda_2} &>& 0
\quad \Leftarrow \quad \left( 4 \lambda_2 > \left| \lambda_4 \right|\
\mbox{and}\ \, 2 \lambda_2 > \lambda_3 \right),
\\
\lambda_1 - \frac{\left( \lambda_4 \right)^2}{2 \lambda_2 + \lambda_3} &>& 0
\quad \Leftarrow \quad \left(
2 \lambda_2 + \lambda_3 > \left| \lambda_4 \right|\
\mbox{and}\ \, 2 \lambda_2 < \lambda_3 \right).
\ea
\es
The sets of inequalities~\eqref{bidoublet_BFB},
\eqref{ch2},
and~\eqref{lav2} are equivalent to each other and they are equivalent to
$\lambda_{00} > 0,\ \forall k \in \left[ -1,\, +1 \right]$.

\paragraph{The inequalities~\eqref{183d}, \eqref{183e}, and~\eqref{183g}:}
For any values of the scalar fields,
one may use an $SU(2)_L$ transformation to make $m = 0$
and then perform an $SU(2)_R$ transformation to render $a = 0$.
In the gauge $a = m = 0$ one has
\bs
\ba
r_L &=& \left| n \right|^2,
\\
r_0 &=& \left| b \right|^2 + \left| c \right|^2 + \left| d \right|^2,
\\
r_1 &=& b^\ast d + b d^\ast,
\\
\rho_L &=& \frac{\left| d \right|^2}{\left| b \right|^2
  + \left| c \right|^2 + \left| d \right|^2}.
\ea
\es
Then,
\ba
\left( r_0 \right)^2 \left[ k^2 + \left( 2 \rho_L - 1 \right)^2 - 1 \right]
&=&
\left( r_1 \right)^2 + 4 \left( \rho_L r_0 \right)
\left[ \left( \rho_L - 1 \right) r_0 \right]
\no &=&
4 \left[ \mathrm{Re} \left( b^\ast d \right) \right]^2
- 4 \left| d \right|^2 \left( \left| b \right|^2 + \left| c \right|^2 \right)
\no &=&
- 4 \left| c d \right|^2 - 4 \left\{ \left| b d \right|^2
- \left[ \mathrm{Re} \left( b^\ast d \right) \right]^2 \right\}
\no &=&
- 4 \left| c d \right|^2
- 4 \left[ \mathrm{Im} \left( b^\ast d \right) \right]^2
\no &\le& 0.
\ea
Therefore,
\be
k^2 + \left( 2 \rho_L - 1 \right)^2 \le 1.
\label{wl}
\ee
Inequality~\eqref{wl} tells us that,
for any $k \in \left[ -1,\ +1 \right]$,
\be
\frac{ 1 - \sqrt{1 - k^2}}{2} \le \rho_L \le \frac{1 + \sqrt{1 - k^2}}{2}.
\ee
The quantity $\bar \lambda_{L0}$ is,
for any fixed value of $k$,
a linear function of $\rho_L$.
Therefore,
the inequality~\eqref{183d} holds for every possible $\rho_L$
provided it holds for
\be
\rho_L = \frac{ 1 + \left( - 1 \right)^{n_L} \sqrt{1 - k^2}}{2},
\label{rol}
\ee
where $n_L$ may be either $0$ or $1$.
Similarly,
the inequality~\eqref{183e} holds for every $\rho_R$
provided it holds for
\be
\rho_R = \frac{1 + \left( - 1 \right)^{n_R} \sqrt{1 - k^2}}{2},
\label{ror}
\ee
where $n_R$ may be either $0$ or $1$.
At last,
the inequality~\eqref{183g} holds for every $\rho_L$ and $\rho_R$
provided it holds for the $\rho_L$ and $\rho_R$
given by eqs.~\eqref{rol} and~\eqref{ror},
respectively.

\paragraph{Recipe for BFB:}
Our recipe for ascertaining the boundedness-from-below of $V_{(4)}$
consists of the following three steps:
\begin{enumerate}
\item Firstly we check that inequalities~\eqref{183b},
  \eqref{183c},
  and~\eqref{183f} hold.
\item Secondly we check that inequalities~\eqref{ch2} hold.
\item Thirdly we make a scan over $k$ from $k=-1$ to $k=+1$
  in steps of $0.001$,
  \textit{i.e.}\ we make $k = -1 + 0.001 n_s$
  for $n_s = 0, 1, \ldots, 2000$.
  For every such value of $k$,
  we consider the two values of $\rho_L$ and the two values of $\rho_R$
  given by eqs.~\eqref{rol} and~\eqref{ror},
  respectively,
  depending on whether $n_L = 0$ or $n_L = 1$
  and on whether $n_R = 0$ or $n_R = 1$.
  We check that inequalities~\eqref{183d},
  \eqref{183e},
  and~\eqref{183g} hold for all those four values
  of the pair $\left( \rho_L,\ \rho_R \right)$.
\end{enumerate}
Of course,
step~3 is just an approximation for considering every
$k \in \left[ -1,\ +1 \right]$,
but we have numerically checked that it is an almost perfect approximation
because the step $0.001$ is sufficiently small.
Provided the approximation is perfect,
it is clear from our derivation that the three requirements above
are \emph{necessary and sufficient conditions}
for the boundedness-from-below of $V_4$.

\setcounter{equation}{0}
\renewcommand{\theequation}{D\arabic{equation}}

\section{Other conditions on the scalar potential}
\label{app:extra}

\paragraph{Vacuum equations:}
In the potential~\eqref{pott2}
we substitute the fields by their VEVs~\eqref{VEVs} to obtain
the VEV of the potential
\ba
\label{v0}
\left\langle 0 \left| V \right| 0 \right\rangle \equiv V_0 &=&
\mu_1 \left( V_1 + V_2 \right)
+ 4 \mu_2 v_1 v_2
+ \mu_L U_L 
+ \mu_R U_R 
\no & &
+ 2 \left( m_1 v_2 + m_2 v_1 \right) u_L u_R
\no & &
+ \lambda_L U_L^2
+ \lambda_R U_R^2
+ \lambda_{LR} U_L U_R
\no & &
+ \lambda_1 \left( V_1 + V_2 \right)^2
+ \left( 8 \lambda_2 + 4 \lambda_3 \right) V_1 V_2
+ 4 \lambda_4 \left( V_1 + V_2 \right) v_1 v_2
\no & &
+ \lambda_{3L} V_2 U_L
+ \lambda_{4L} V_1 U_L
+ 2 \lambda_{5L} U_L v_1 v_2
\no & &
+ \lambda_{3R} V_2 U_R
+ \lambda_{4R} V_1 U_R
+ 2 \lambda_{5R} U_R v_1 v_2.
\ea
The equations of vacuum stability are
\bs
\label{vs}
\ba
0 = \frac{1}{2}\, \frac{\partial V_0}{\partial v_1} &=&
\mu_1 v_1
+ 2 \mu_2 v_2
+ m_2 u_L u_R
\no & &
+ 2 \lambda_1 \left( V_1 + V_2 \right) v_1
+ \left( 8 \lambda_2 + 4 \lambda_3 \right) V_2 v_1
+ 2 \lambda_4 \left( 3 V_1 + V_2 \right) v_2
\no & &
+ \lambda_{4L} U_L v_1
+ \lambda_{5L} U_L v_2
+ \lambda_{4R} U_R v_1
+ \lambda_{5R} U_R v_2,
\label{mu1a} \\
0 = \frac{1}{2}\, \frac{\partial V_0}{\partial v_2} &=&
\mu_1 v_2
+ 2 \mu_2 v_1
+ m_1 u_L u_R
\no & &
+ 2 \lambda_1 \left( V_1 + V_2 \right) v_2
+ \left( 8 \lambda_2 + 4 \lambda_3 \right) V_1 v_2
+ 2 \lambda_4 \left( V_1 + 3 V_2 \right) v_1
\no & &
+ \lambda_{3L} U_L v_2
+ \lambda_{5L} U_L v_1
+ \lambda_{3R} U_R v_2
+ \lambda_{5R} U_R v_1,
\label{mu2a} \\
0 = \frac{1}{2}\, \frac{\partial V_0}{\partial u_L} &=&
\mu_L u_L
+ \left( m_1 v_2 + m_2 v_1 \right) u_R
\no & &
+ 2 \lambda_L U_L u_L
+ \lambda_{LR} U_R u_L
+ \lambda_{3L} V_2 u_L
+ \lambda_{4L} V_1 u_L
+ 2 \lambda_{5L} v_1 v_2 u_L,
\label{mul} \\
0 = \frac{1}{2}\, \frac{\partial V_0}{\partial u_R} &=&
\mu_R u_R
+ \left( m_1 v_2 + m_2 v_1 \right) u_L
\no & &
+ 2 \lambda_R U_R u_R
+ \lambda_{LR} U_L u_R
+ \lambda_{3R} V_2 u_R
+ \lambda_{4R} V_1 u_R
+ 2 \lambda_{5R} v_1 v_2 u_R.
\label{mur}
\ea
\es
Solving Eqs.~\eqref{vs} for the $\mu$ parameters,
one obtains
\bs
\label{mu2}
\ba
\mu_1 &=&
\frac{m_1 v_2 - m_2 v_1}{V_1 - V_2}\, u_L u_R
- 2 \lambda_1 \left( V_1 + V_2 \right)
- 4 \lambda_4 v_1 v_2
\no & &
+ \frac{\lambda_{3L} V_2 - \lambda_{4L} V_1}{V_1 - V_2}\, U_L
+ \frac{\lambda_{3R} V_2 - \lambda_{4R} V_1}{V_1 - V_2}\, U_R,
\label{mu12} \\
\mu_2 &=& \frac{1}{2}
\left[
  \frac{m_1 v_1 - m_2 v_2}{V_2 - V_1}\, u_L u_R
  - \left( 8 \lambda_2 + 4 \lambda_3 \right) v_1 v_2
  - 2 \lambda_4 \left( V_1 + V_2 \right)
  \right. \no & & \left.
  - \frac{\lambda_{3L} - \lambda_{4L}}{V_1 - V_2}\, U_L v_1 v_2
  - \frac{\lambda_{3R} - \lambda_{4R}}{V_1 - V_2}\, U_R v_1 v_2
  - \lambda_{5L} U_L - \lambda_{5R} U_R
  \right],
\label{mu22} \\
\mu_L &=&
- \left( m_1 v_2 + m_2 v_1 \right) \frac{u_R}{u_L}
\no & &
- 2 \lambda_L U_L
- \lambda_{LR} U_R
- \lambda_{3L} V_2
- \lambda_{4L} V_1
- 2 \lambda_{5L} v_1 v_2,
\label{mul2} \\
\mu_R &=&
- \left( m_1 v_2 + m_2 v_1 \right) \frac{u_L}{u_R}
\no & &
- 2 \lambda_R U_R
- \lambda_{LR} U_L
- \lambda_{3R} V_2
- \lambda_{4R} V_1
- 2 \lambda_{5R} v_1 v_2.
\label{mur2}
\ea
\es
Plugging Eqs.~\eqref{mu2} back into Eq.~\eqref{v0} one obtains
\ba
\label{vzero}
V_0 &=& - \left( m_1 v_2 + m_2 v_1 \right) u_L u_R
\no & &
- \lambda_L U_L^2 - \lambda_R U_R^2 - \lambda_{LR} U_L U_R
\no & &
- \lambda_1 \left( V_1 + V_2 \right)^2
- \left( 8 \lambda_2 + 4 \lambda_3 \right) V_1 V_2
- 4 \lambda_4 \left( V_1 + V_2 \right) v_1 v_2
\no & &
- \left( \lambda_{3L} V_2 + \lambda_{4L} V_1 + 2 \lambda_{5L} v_1 v_2 \right) U_L 
\no & &
- \left( \lambda_{3R} V_2 + \lambda_{4R} V_1 + 2 \lambda_{5R} v_1 v_2 \right) U_R 
\ea

\paragraph{Alternative vacua:}

There are terms in the potential~\eqref{pott2} that may induce VEVs
of fields that \textit{a priori}\/ did not have a VEV.
They are the terms that are linear in any field,
\textit{viz.}
\bs
\ba
& & \left[ 2 \mu_2
  + 2 \lambda_4 \left( \left| a \right|^2 + \left| b \right|^2
  + \left| c \right|^2 + \left| d \right|^2 \right)
  \right. \no & & \left. \hspace*{10mm}
  + \lambda_{5L} \left( \left| m \right|^2 + \left| n \right|^2 \right)
  + \lambda_{5R} \left( \left| p \right|^2 + \left| q \right|^2 \right)
\right] \left( a^\ast c + b^\ast d + \mathrm{H.c.} \right);
\\
& & 8 \lambda_2 \left( a b c^\ast d^\ast + \mathrm{H.c.} \right);
\\
& & 4 \lambda_3 \left( a b^\ast c^\ast d + \mathrm{H.c.} \right);
\\
& &
m_1 \left( - a n p^\ast + b m p^\ast + c q m^\ast + d q n^\ast + \mathrm{H.c.}
\right);
\\
& &
m_2 \left( - c n p^\ast + d m p^\ast + a q m^\ast + b q n^\ast + \mathrm{H.c.}
\right);
\\
& &
\left( \lambda_{3L} - \lambda_{4L} \right)
\left[ m^\ast n \left( c d^\ast - a b^\ast \right) + \mathrm{H.c.} \right];
\\
& &
\left( \lambda_{3R} - \lambda_{4R} \right)
\left[ p^\ast q \left( b c - a d \right) + \mathrm{H.c.} \right].
\ea
\es
The presence in the potencial of these terms implies that
the only possible vacua wherein some of the eight scalar fields
are identically equal to zero are the following:
\begin{enumerate}
\item All the fields are zero.
\item Only $p$ is nonzero;
  or, equivalently, only $q$ is nonzero;
  or, equivalently, only $p$ and $q$ are nonzer.
  (The three situations are equivalent to each other
  through gauge transformations.)
\item Only $m$ is nonzero;
  or, equivalently, only $n$ is nonzero
  or, equivalently, only $m$ and $n$ are nonzero.
\item Only $a$ and $c$ are nonzero;
  or, equivalently, only $b$ and $d$ are nonzero.
\item Only $a$, $b$, $c$, and $d$ are nonzero.
\item Only $a$, $c$, $n$, and $p$ are nonzero;
  or, equivalently, only $b$, $d$, $m$, and $p$ are nonzero;
  or, equivalently, only $a$, $c$, $m$, and $q$ are nonzero;
  or, equivalently, only $b$, $d$, $n$, and $q$ are nonzero.
\end{enumerate}
Cases~5 and~6 are impossible to treat analytically.
There is yet another case,
wherein all eight fields are nonzero in the vacuum,
and this is also impossible to treat analytically.
So,
in the following we just consider cases 1 through 4.

\paragraph{Case~1:}
In case~1 the minimum of the potential has
\be
\left\langle 0 \left| V \right| 0 \right\rangle = 0 \equiv V^{(1)}.
\ee

\paragraph{Case~2:}
In case~2 the minimum of the potential has
\be
\left\langle 0 \left| V \right| 0 \right\rangle
= - \frac{\left( \mu_R \right)^2}{4 \lambda_R} \equiv V^{(2)}.
\ee

\paragraph{Case~3:}
In case~3 the minimum of the potential has
\be
\left\langle 0 \left| V \right| 0 \right\rangle
= - \frac{\left( \mu_L \right)^2}{4 \lambda_L} \equiv V^{(3)}.
\ee

\paragraph{Case~4:}
Case~4 gives
(with $b$ and $d$ nonzero)
\bs
\ba
V &=&
\mu_1 \left( B + D \right) + \lambda_1 \left( B + D \right)^2
+ 4 \lambda_3 B D + 8 \lambda_2 B D \cos{\left( 2 \theta \right)}
\no & &
+ 4 \left[ \mu_2 + \lambda_4 \left( B + D \right) \right]
\sqrt{B D} \cos{\theta}
\\ &=&
\mu_1 \left( B + D \right) + \lambda_1 \left( B + D \right)^2
+ 4 \left( \lambda_3 - 2 \lambda_2 \right) B D
\no & &
+ 16 \lambda_2 B D \cos^2{\theta}
+ 4 \hat \mu \sqrt{B D} \cos{\theta},
\label{fjvidfr}
\ea
\es
where $B \equiv \left| b \right|^2$,
$D \equiv \left| d \right|^2$,
$\theta \equiv \arg{\left( b^\ast d \right)}$,
and $\hat \mu \equiv \mu_2 + \lambda_4 \left( B + D \right)$.
One must find the value of Eq.~\eqref{fjvidfr} when
\bs
\label{174}
\ba
\frac{\partial V}{\partial \theta} =
\left( 4 \hat \mu \sqrt{B D}
+ 32 \lambda_2 B D \cos{\theta} \right)
\left( - \sin{\theta} \right)&=& 0,
\label{174a} \\
\frac{\partial V}{\partial B} =
\mu_1 + 2 \lambda_1 \left( B + D \right)
+ 4 \left( \lambda_3 - 2 \lambda_2 \right) D
+ 16 \lambda_2 D \cos^2{\theta}
& & \no
+ 2 \hat \mu \sqrt{\frac{D}{B}} \cos{\theta}
+ 4 \lambda_4 \sqrt{BD} \cos{\theta} &=& 0,
\label{174b} \\
\frac{\partial V}{\partial D} =
\mu_1 + 2 \lambda_1 \left( B + D \right)
+ 4 \left( \lambda_3 - 2 \lambda_2 \right) B
+ 16 \lambda_2 B \cos^2{\theta}
& & \no
+ 2 \hat \mu \sqrt{\frac{B}{D}} \cos{\theta}
+ 4 \lambda_4 \sqrt{BD} \cos{\theta} &=& 0.
\label{174c}
\ea
\es
Equation~\eqref{174a} has three possible solutions:
\bs
\label{175}
\ba
\cos{\theta} &=& - \frac{\hat \mu}{8 \lambda_2 \sqrt{B D}}; \label{175a}
\\
\theta &=& 0; \label{175b}
\\
\theta &=& \pi. \label{175c}
\ea
\es
Equations~\eqref{174b} and~\eqref{174c} may be subtracted from one another,
producing
\be
\left( D - B \right)
\left[ 4 \left( \lambda_3 - 2 \lambda_2 \right)
+ 16 \lambda_2 \cos^2{\theta}
+ \frac{2 \hat \mu \cos{\theta}}{\sqrt{B D}} \right]
= 0.
\ee
Therefore,
eqs.~\eqref{174b} and~\eqref{174c} have two possible solutions.
One of them is
\bs
\label{177}
\ba
B - D &=& 0,
\\
\mu_1 + 2 \hat \mu \cos{\theta} &=&
- 4 \left( \lambda_1 + \lambda_3 - 2 \lambda_2 \right) B
- 16 \lambda_2 B \cos^2{\theta}
- 4 \lambda_4 B \cos{\theta};
\ea
\es
and the other one is
\bs
\label{178}
\ba
\mu_1 &=& - 2 \lambda_1 \left( B + D \right)
- 4 \lambda_4 \sqrt{BD} \cos{\theta},
\\
\hat \mu \cos{\theta} &=& - 8 \lambda_2 \sqrt{B D} \cos^2{\theta}
- 2 \left( \lambda_3 - 2 \lambda_2 \right) \sqrt{B D}.
\ea
\es
Combining Eqs.~\eqref{175} with either Eqs.~\eqref{177} or Eqs.~\eqref{178},
there are altogether six cases:
\begin{enumerate}
\item Equations~\eqref{175b} and~\eqref{177} give
  \bs
  \ba
  \theta &=& 0,
  \\
  D = B &=& - \frac{\mu_1 + 2 \mu_2}{4 \left( \lambda_1 + 2 \lambda_2
    + \lambda_3 + 2 \lambda_4 \right)},
  \\
  V &=& - \frac{\left( \mu_1 + 2 \mu_2 \right)^2}{4 \left( \lambda_1
    + 2 \lambda_2 + \lambda_3 + 2 \lambda_4 \right)} \equiv V^{(4)}.
  \ea
  \es
\item Equations~\eqref{175c} and~\eqref{177} give
  \bs
  \ba
  \theta &=& \pi,
  \\
  D = B &=& - \frac{\mu_1 - 2 \mu_2}{4 \left( \lambda_1 + 2 \lambda_2
    + \lambda_3 - 2 \lambda_4 \right)},
  \\
  V &=& - \frac{\left( \mu_1 - 2 \mu_2 \right)^2}{4 \left( \lambda_1
    + 2 \lambda_2 + \lambda_3 - 2 \lambda_4 \right)} \equiv V^{(5)}.
  \ea
  \es
\item Equations~\eqref{175b} and~\eqref{178} give
  \bs
  \ba
  \theta &=& 0,
  \\
  B + D &=& \frac{- \hat \lambda \mu_1
    + 4 \lambda_4 \mu_2}{2 \lambda_1 \hat \lambda
    - 4 \left( \lambda_4 \right)^2},
  \\
  \sqrt{BD} &=& \frac{\lambda_4 \mu_1
    - 2 \lambda_1 \mu_2}{2 \lambda_1 \hat \lambda
    - 4 \left( \lambda_4 \right)^2},
  \\
  V &=&  \frac{- \hat \lambda \left( \mu_1 \right)^2
    - 8 \lambda_1 \left( \mu_2 \right)^2
    + 8 \lambda_4 \mu_1 \mu_2}{4 \left[ \lambda_1 \hat \lambda
      - 2 \left( \lambda_4 \right)^2 \right]} \equiv V^{(6)},
  \ea
  \es
  with $\hat \lambda \equiv 4 \lambda_2 + 2 \lambda_3$.
  This solution only exists if
  \be
  \label{bjgidr995}
  2 \left| \lambda_4 \mu_1 - 2 \lambda_1 \mu_2 \right|
  \le - \hat \lambda \mu_1 + 4 \lambda_4 \mu_2.
  \ee
\item Equations~\eqref{175c} and~\eqref{178} give the same result as
  number~3,
  except that the sign of $\sqrt{BD}$ gets inverted.
\item Equation~\eqref{175a} and~\eqref{177} give
  \bs
  \ba
  \cos{\theta} &=& - \frac{1}{4}\, \frac{2 \lambda_4 \mu_1
    - \bar \lambda \mu_2}{2 \lambda_2 \mu_1 - \lambda_4 \mu_2},
  \\
  D = B &=& \frac{1}{2}\, \frac{2 \lambda_2 \mu_1
    - \lambda_4 \mu_2}{\left( \lambda_4 \right)^2 - \lambda_2 \bar \lambda},
  \\
  V &=& \frac{4 \lambda_2 \left( \mu_1 \right)^2
    + \bar \lambda \left( \mu_2 \right)^2
    - 4 \lambda_4 \mu_1 \mu_2}{4 \left[ \left( \lambda_4 \right)^2
    - \lambda_2 \bar \lambda \right]} \equiv V^{(7)},
  \ea
  \es
  where $\bar \lambda \equiv 4
  \left( \lambda_1 - 2 \lambda_2 + \lambda_3 \right)$.
\item Equations~\eqref{175a} and~\eqref{178} give
  $\left( \lambda_3 - 2 \lambda_2 \right) \sqrt{B D} = 0$,
  which is a contradiction because we are assuming
  that both $B$ and $D$ are nonzero.
\end{enumerate}

\paragraph{The conditions:}
One must require
\be
\label{a21}
V_0 < V^{(1)}, \qquad
V_0 < V^{(2)}, \qquad
V_0 < V^{(3)},
\ee
and also
\bs
\label{a22}
\ba
V_0 < V^{(4)} &\Leftarrow&
\frac{\mu_1 + 2 \mu_2}{\lambda_1 + 2 \lambda_2
  + \lambda_3 + 2 \lambda_4} < 0,
\\
V_0 < V^{(5)} &\Leftarrow&
\frac{\mu_1 - 2 \mu_2}{\lambda_1 + 2 \lambda_2
  + \lambda_3 - 2 \lambda_4} < 0,
\\
V_0 < V^{(6)} &\Leftarrow&
2 \left| \lambda_4 \mu_1 - 2 \lambda_1 \mu_2 \right|
\le \left| - \hat \lambda \mu_1 + 4 \lambda_4 \mu_2 \right|
\quad \mbox{and} \quad \frac{- \hat \lambda \mu_1
  + 4 \lambda_4 \mu_2}{\lambda_1 \hat \lambda
  - 2 \left( \lambda_4 \right)^2} \ge 0, \hspace*{7mm}
\\
V_0 < V^{(7)} &\Leftarrow&
\left| 2 \lambda_4 \mu_1 - \bar \lambda \mu_2 \right|
\le 4 \left| 2 \lambda_2 \mu_1 - \lambda_4 \mu_2 \right|
\quad \mbox{and} \quad \frac{2 \lambda_2 \mu_1
  - \lambda_4 \mu_2}{\left( \lambda_4 \right)^2
  - \lambda_2 \bar \lambda} \ge 0.
\ea
\es
In the conditions~\eqref{a21} and~\eqref{a22},
the $\mu$ parameters must be substituted by their expressions
in terms of the $m$ and $\lambda$ parameters by using Eqs.~\eqref{mu2}.

\setcounter{equation}{0}
\renewcommand{\theequation}{E\arabic{equation}}

\section{Experimental constraints}
\label{app:kappa}

\paragraph{$\kappa_Z$ and $\kappa_W$:}
From Eq.~\eqref{pripri},
\ba
\mathcal{L} &=& \cdots
+ \left[
\frac{g^2}{2}\, W^{+ \mu} W^-_\mu
+
\left( \frac{g}{2 c_w}\, Z^\mu + \frac{l s_\alpha^2}{2 c_\alpha}\, X^\mu \right)
\left( \frac{g}{2 c_w}\, Z_\mu + \frac{l s_\alpha^2}{2 c_\alpha}\, X_\mu \right)
\right] \left| n \right|^2
\no & &
+ \left(
\frac{l^2}{2}\, V^{+ \mu} V^-_\mu + \frac{l^2}{4 c_\alpha^2}\, X^\mu X_\mu
\right) \left| q \right|^2
\no & &
+ \frac{1}{2}
\left( - g W^{+ \mu} b + l V^{+ \mu} d^\ast \right)
\left( - g W^-_\mu b^\ast + l V^-_\mu d \right)
\no & &
+ \frac{1}{2}
\left( - g W^{+ \mu} d + l V^{+ \mu} b^\ast \right)
\left( - g W^-_\mu d^\ast + l V^-_\mu b \right)
\no & &
+
\left( \frac{g}{2 c_w}\, Z^\mu - \frac{l c_\alpha}{2}\, X^\mu \right)
\left( \frac{g}{2 c_w}\, Z_\mu - \frac{l c_\alpha}{2}\, X_\mu \right)
\left( \left| b \right|^2 + \left| d \right|^2 \right).
\ea
Therefore,
\ba
\mathcal{L} &=& \cdots
+ Z_l^\mu Z_{l \mu} \left[
  \left( \frac{g^2}{4 c_w^2}\, c_\psi^2
  + \frac{l^2 s_\alpha^4}{4 c_\alpha^2}\, s_\psi^2
  + \frac{g l s_\alpha^2}{2 c_w c_\alpha}\, c_\psi s_\psi \right)
  \sqrt{2} u_L \rho_L
  + \frac{l^2}{4 c_\alpha^2}\, s_\psi^2\, \sqrt{2} u_R \rho_R
  \right. \no & & \left.
  +  \left( \frac{g^2}{4 c_w^2}\, c_\psi^2
  + \frac{l^2 c_\alpha^2}{4}\, s_\psi^2
  - \frac{g l c_\alpha}{2 c_w}\, c_\psi s_\psi \right)
  \left( \sqrt{2} v_1 \rho_1 + \sqrt{2} v_2 \rho_2 \right)
  \right]
\no & &
+ \frac{W_l^{+ \mu} W_{l \mu}^-}{2} \left[
  g^2 c_\xi^2 \sqrt{2} u_L \rho_L
  + l^2 s_\xi^2 \sqrt{2} u_R \rho_R
  \right. \no & & \left.
  + \left( g^2 c_\xi^2 + l^2 s_\xi^2 \right)
  \left( \sqrt{2} v_1 \rho_1 + \sqrt{2} v_2 \rho_2 \right)
  + 2 g l c_\xi s_\xi
  \left( \sqrt{2} v_1 \rho_2 + \sqrt{2} v_2 \rho_1 \right)
  \right].
\ea
Hence,
the interactions of the scalar $S^0_5$ with pairs
of light massive gauge bosons are given by
\ba
\mathcal{L} &=& \cdots
+ \frac{Z_l^\mu Z_{l \mu} S^0_5}{2 \sqrt{2}} \left\{
  \left( \frac{g^2}{c_w^2}\, c_\psi^2
  + \frac{l^2 s_\alpha^4}{c_\alpha^2}\, s_\psi^2
  + \frac{2 g l s_\alpha^2}{c_w c_\alpha}\, c_\psi s_\psi \right)
  u_L \left( V_\rho \right)_{31}
  + \frac{l^2}{c_\alpha^2}\, s_\psi^2\, u_R \left( V_\rho \right)_{41}
  \right. \no & & \left.
  + \left( \frac{g^2}{c_w^2}\, c_\psi^2
  + l^2 c_\alpha^2 s_\psi^2
  - \frac{2 g l c_\alpha}{c_w}\, c_\psi s_\psi \right)
  \left[ v_1 \left( V_\rho \right)_{11} + v_2 \left( V_\rho \right)_{21} \right]
  \right\}
\no & &
+ \frac{W_l^{+ \mu} W_{l \mu}^- S^0_5}{\sqrt{2}} \left\{
g^2 c_\xi^2 u_L \left( V_\rho \right)_{31}
+ l^2 s_\xi^2 u_R \left( V_\rho \right)_{41}
\right. \no & & \left.
+ \left( g^2 c_\xi^2 + l^2 s_\xi^2 \right)
\left[ v_1 \left( V_\rho \right)_{11} + v_2 \left( V_\rho \right)_{21} \right]
+ 2 g l c_\xi s_\xi \left[ v_1 \left( V_\rho \right)_{21}
  + v_2 \left( V_\rho \right)_{11} \right]
\right\}.
\label{8f93r0w}
\ea
In our fits,
we identify $S^0_5$ as the observed scalar particle of mass 125\,GeV.
Equation~\eqref{8f93r0w} should therefore
be compared to the corresponding interactions
of the Higgs scalar $H$ of the SM,
which are given by
\be
\mathcal{L}_ \mathrm{SM} = \cdots
+ H\, \frac{e m_Z m_W}{\sqrt{m_Z^2 - m_W^2}}
\left( W^{+ \mu} W^-_\mu + \frac{m_Z^2}{2 m_W^2}\, Z^\mu Z_\mu \right).
\ee
We rewrite Eq.~\eqref{8f93r0w} as
\be
\mathcal{L} = \cdots
+ S^0_5\, \frac{e m_Z m_W}{\sqrt{m_Z^2 - m_W^2}}
\left( \kappa_W W^{+ \mu} W^-_\mu
+ \kappa_Z\, \frac{m_Z^2}{2 m_W^2}\, Z^\mu Z_\mu \right),
\ee
with
\bs
\label{2jvuiseoe}
\ba
\kappa_Z &=& \frac{m_W^2}{m_Z^2}\, \sqrt{\frac{m_Z^2 - m_W^2}{2
    E m_Z^2 m_W^2}} \left\{
  \frac{\left( G L + G H + L H \right) c_\psi^2
    + H^2 s_\psi^2 + 2 H \varrho_1\, c_\psi s_\psi}{L + H}\,
  u_L \left( V_\rho \right)_{31}
  \right. \no & &
  + \left( L + H \right) s_\psi^2 u_R \left( V_\rho \right)_{41}
  \no & & \left.
  + \frac{\left( G L + G H + L H \right) c_\psi^2
  + L^2 s_\psi^2 - 2 L \varrho_1\, c_\psi s_\psi}{L + H}
  \left[ v_1 \left( V_\rho \right)_{11} + v_2 \left( V_\rho \right)_{21} \right]
  \right\},
\label{nf9993} \\
\kappa_W &=& \sqrt{\frac{m_Z^2 - m_W^2}{2 E m_Z^2 m_W^2}} \left\{
  G c_\xi^2 u_L \left( V_\rho \right)_{31}
  + L s_\xi^2 u_R \left( V_\rho \right)_{41}
  + \left( G c_\xi^2 + L s_\xi^2 \right)
  \left[ v_1 \left( V_\rho \right)_{11} + v_2 \left( V_\rho \right)_{21} \right]
  \right. \no & & \left.
  + 2 g l c_\xi s_\xi \left[ v_1 \left( V_\rho \right)_{21}
    + v_2 \left( V_\rho \right)_{11} \right]
  \right\}.
\label{KW}
\ea
\es
In practice,
we always input $\psi = 0$ in Eq.~\eqref{nf9993} and,
besides,
the mixing angle $\xi$ always turns out to be extremely small in our fits.
When $\psi = \xi = 0$ one obtains,
from Eqs.~\eqref{rhos1},
\eqref{eeeee},
\eqref{nfde9re00},
and~\eqref{mn11},
\be
E = \frac{G L H}{G L + G H + L H}, \qquad
M_W = \frac{G V_L^2}{2}, \qquad
M_Z = \frac{\left( G L + G H + L H \right) V_L^2}{2}.
\ee
Then,
from Eqs.~\eqref{2jvuiseoe},
\ba
\kappa_Z = \kappa_W &=&
\frac{M_W}{V_L^2}
\sqrt{\frac{2 \left( M_Z - M_W \right)}{E M_Z M_W}}
\left[ v_1 \left( V_\rho \right)_{11} + v_2 \left( V_\rho \right)_{21}
  + u_L \left( V_\rho \right)_{31} \right]
\no &=& \frac{v_1 c_1 + v_2 s_1 c_2
  + u_L s_1 s_2 c_3}{\sqrt{V_1 + V_2 + U_L}},
\ea
where use was made of Eq.~\eqref{c93kf} in the last step.
Thus,
in our model,
with $\psi = 0$ and in the limit $\xi = 0$,
$\kappa_Z$ and $\kappa_W$ become equal and are necessarily smaller than one.
In our fits $\xi$ may deviate slightly from zero
and therefore for some points $\kappa_W$ is slightly larger than one,
but never much.

\paragraph{$\kappa_t$ and $\kappa_b$:}
The Yukawa couplings of the top and bottom quarks to the scalars include,
according to Eq.~\eqref{uf83344},
\be
\mathcal{L}_\mathrm{Yukawa} = \cdots
- \frac{y_1 \rho_1 + y_2 \rho_2}{\sqrt{2}}
\left( \bar t_L t_R + \bar t_R t_L \right)
- \frac{y_1 \rho_2 + y_2 \rho_1}{\sqrt{2}}
\left( \bar b_L b_R + \bar b_R b_L \right).
\ee
Using then Eqs.~\eqref{yuks},
\ba
\mathcal{L}_\mathrm{Yukawa} &=& \cdots
- \sum_{i=1}^4 S^0_{i+4} \left[
\frac{\left( - v_1 m_t + v_2 m_b \right) \left( V_\rho \right)_{1i}
  + \left( v_2 m_t - v_1 m_b \right)
  \left( V_\rho \right)_{2i}}{\sqrt{2} \left( V_2 - V_1 \right)}
\left( \bar t_L t_R + \bar t_R t_L \right)
\right. \no & & \left.
+ \frac{\left( - v_1 m_t + v_2 m_b \right) \left( V_\rho \right)_{2i}
  + \left( v_2 m_t - v_1 m_b \right)
  \left( V_\rho \right)_{1i}}{\sqrt{2} \left( V_2 - V_1 \right)}
\left( \bar b_L b_R + \bar b_R b_L \right) \right].
\ea
The Yukawa couplings of $S^0_5$ should be compared to the Yukawa couplings
of $H$ in the SM,
which are given by
\be
\mathcal{L}_\mathrm{SM} = \cdots
- \frac{e m_Z}{2 m_W \sqrt{m_Z^2 - m_W^2}}\,
H \left[ \left| m_t \right| \left( \bar t_L t_R + \bar t_R t_L \right)
  + \left| m_b \right| \left( \bar b_L b_R + \bar b_R b_L \right) \right].
\ee
One thus has
\bs
\label{jvuiseoe}
\ba
\kappa_t &=&
\sqrt{\frac{2 m_W^2 \left( m_Z^2 - m_W^2 \right)}{E m_Z^2 m_t^2}}\
\frac{\left( v_2 m_t - v_1 m_b \right) \left( V_\rho \right)_{21}
  + \left( v_2 m_b - v_1 m_t \right)
  \left( V_\rho \right)_{11}}{\left( V_2 - V_1 \right)},
\\
\kappa_b &=&
\sqrt{\frac{2 m_W^2 \left( m_Z^2 - m_W^2 \right)}{E m_Z^2 m_b^2}}\
\frac{\left( v_2 m_b - v_1 m_t \right) \left( V_\rho \right)_{21}
  + \left( v_2 m_t - v_1 m_b \right)
  \left( V_\rho \right)_{11}}{\left( V_2 - V_1 \right)}.
\ea
\es

\setcounter{equation}{0}
\renewcommand{\theequation}{F\arabic{equation}}

\section{Determination of the VEVs and of the gauge coupling constants}
\label{app:GLH}

\paragraph{General case:}
Part of our input is formed by $M_l$,
$M_h$,
$\bar M_l$,
$\bar M_h$,
$\xi$,
$\psi$,\footnote{In actual practice we always input $\psi = 0$,
but the method that we delineate here works for arbitrary $\psi$.}
and $E$.
Using this input we determine the VEVs and the gauge coupling constants
in the following way.
We firstly define
\be
x \equiv \frac{E}{G}, \quad
y \equiv \frac{E}{L}, \quad
z \equiv \frac{E}{H}, \qquad
a_1 \equiv E V_1, \quad
a_2 \equiv E V_2, \quad
a_3 \equiv E U_L, \quad
a_4 \equiv E U_R.
\ee
Equation~\eqref{eeeee} is equivalent to
\be
\label{glh}
x + y + z = 1.
\ee
We use the input to compute
\ba
& &
R = M_l c_\psi^2 + M_h s_\psi^2, \qquad
S = M_l s_\psi^2 + M_h c_\psi^2, \qquad
T^2 = \left( M_h - M_l \right)^2\, c_\psi^2 s_\psi^2,
\label{rrsstt} \\
& &
A = \bar M_l c_\xi^2 + \bar M_h s_\xi^2, \qquad
B = \bar M_l s_\xi^2 + \bar M_h c_\xi^2, \qquad
D^2 = \left( \bar M_h - \bar M_l \right)^2\, c_\xi^2 s_\xi^2.
\label{aabbdd}
\ea
\textit{cf.}\ Eqs.~\eqref{uf9995} and~\eqref{djfire00}.
The first Eq.~\eqref{nfde9re00} and Eq.~\eqref{mn11} imply
\ba
t \equiv \frac{A}{R} &=& \frac{G \left( L + H \right)}{G L + G H + L H}
\no &=& \left[ \frac{1}{1 - y - z}
  \left( \frac{1}{y} + \frac{1}{z} \right) \right]
\left[ \frac{1}{1 - y - z}
  \left( \frac{1}{y} + \frac{1}{z} \right) + \frac{1}{y z}
  \right]^{-1}
\no &=& \frac{z + y}{z + y + 1 - y - z}
\no &=& z + y.
\ea
Therefore,
$x$ is computed from the input as
\be
\label{xxxxx}
x = 1 - \frac{A}{R}.
\ee
Since $x = E / G$ is by definition positive,
\emph{we must enforce the condition}
\be
R \ge A
\ee
\emph{on the input}.
We then define
\be
a_s \equiv a_1 + a_2.
\ee
The first two Eqs.~\eqref{nfde9re00} and Eq.~\eqref{mn22} read,
respectively,
\bs
\label{vjfg94333}
\ba
2 x A &=& a_s + a_3, \\
2 y B &=& a_s + a_4, \\
2 y z \left( y + z \right) S &=& a_s z^2 + a_3 y^2
+ a_4 \left( y + z \right)^2.
\ea
\es
Equations~\eqref{vjfg94333} are solved as
\bs
\label{cj29932}
\ba
a_s &=& \frac{x y^2 A + y \left( y + z \right)^2 B
  - y z \left( y + z \right) S}{y \left( y + z \right)},
\\
a_3 &=& \frac{x y \left( 2 z + y \right) A - y \left( y + z \right)^2 B
  + y z \left( y + z \right) S}{y \left( y + z \right)},
\\
a_4 &=& \frac{- x y^2 A
  + y \left( y^2 - z^2 \right) B
  + y z \left( y + z \right) S}{y \left( y + z \right)}.
\ea
\es
Equation~\eqref{mn12} yields
\be
\label{ji349943}
4 T^2 = \frac{x + y + z}{x y z \left( y + z \right)^2}
\left( y a_3 - z a_s \right)^2.
\ee
Plugging Eqs.~\eqref{cj29932} into Eq.~\eqref{ji349943} one obtains
\be
4 \left( y + z \right)^2 x y z T^2
= \left( x + y + z \right)
\left[ x y A - \left( y + z \right)^2 B + z \left( y + z \right) S \right]^2.
\ee
Then,
using $x = 1 - t$ and
\be
y = t - z, \label{tvufdi}
\ee
one obtains
\be
4 t^2 \left( 1 - t \right) T^2
\left( t - z \right) z =
\left[ t^2 B + \left( t - 1 \right) \left( t - z \right) A
  - t z S \right]^2.
\hspace*{11mm}
\label{ftru88}
\ee
Equation~\eqref{ftru88} is a quadratic equation for $z$:
\ba
0 &=& 
z^2 \left\{ \left[ \left( 1 - t \right) A - t S \right]^2
+ 4 t^2 \left( 1 - t \right) T^2 \right\}
\no & &
+ 2 z \left\{ \left[ \left( 1 - t \right) A - t S \right]
\left[ t B + \left( t - 1 \right) A \right]
+ 2 t^2 \left( t - 1 \right) T^2 \right\} t
\no & &
+ \left[ t B + \left( t - 1 \right) A \right]^2 t^2.
\label{mgfogdfr0}
\ea
One solves Eq.~\eqref{mgfogdfr0},
enforcing on its solution the condition
\be
0 \le z \le t.
\ee
Equation~\eqref{tvufdi} then yields $y$.
Equations~\eqref{cj29932} produce $a_1 + a_2$,
$a_3$,
and $a_4$;
these must all turn out positive.

\paragraph{Case $\psi = 0$:}
In practice,
we input $\psi = 0$.
Then,
according to Eqs.~\eqref{rrsstt},
\be
R = M_l, \qquad S = M_h, \qquad T = 0.
\ee
Equation~\eqref{mn12} then implies
\be
H U_L = L \left( V_1 + V_2 \right),
\label{m932033}
\ee
which simplifies considerably the resolution of the system of equations.
One obtains,
\bs
\ba
G &=& \frac{E M_l}{M_l - A},
\\
L &=& \frac{E M_l \left( M_h - M_l + A \right)}{A \left( M_h - B \right)}, \label{VeVs_psi0_b}
\\
H &=& \frac{E M_l \left( M_h - M_l + A \right)}{A \left( A + B - M_l \right)}, \label{VeVs_psi0_c}
\\
V_1 + V_2 &=& \frac{2 A \left( M_h - B \right)
  \left( M_l - A \right)}{E M_l \left(M_h - M_l + A \right)},
\\
U_L &=& \frac{2 A \left( M_l - A \right)
  \left( A + B - M_l \right)}{E M_l \left( M_h - M_l + A \right)},
\\
U_R &=& \frac{2 A \left( M_h - B \right)
  \left(A + B - M_l \right)}{E M_l \left( M_h - M_l + A \right)}.
\ea
\label{VeVs_psi0}
\es
Note that one must input values of $\bar M_l$,
$\bar M_h$,
and $\xi$ such that
\be
A < M_l, \qquad M_l - A < B < M_h,
\label{mfoder0ewe9}
\ee
where $A$ and $B$ are given by the first two Eqs.~\eqref{aabbdd}.

\paragraph{$V_1$ and $V_2$:}
It remains to separate $a_1$ from $a_2$.
This we do by having recourse to the third Eq.~\eqref{nfde9re00}:
\be
a_1 a_2 = x y D^2.
\ee
Therefore,
\be
\label{uf8e9r9ee}
a_1 = \frac{1}{2} \left[ a_s - \sqrt{\left( a_s \right)^2
  - 4 x y D^2} \right],
\qquad
a_2 = \frac{1}{2} \left[ a_s + \sqrt{\left( a_s \right)^2
  - 4 x y D^2} \right].
\ee
In Eqs.~\eqref{uf8e9r9ee},
$\sqrt{\left( a_s \right)^2 - 4 x y D^2}$ is by definition positive,
\textit{viz.}\ we assume,
without loss of generality,
that $V_1 \le V_2$.

\paragraph{Proof that $\bar M_h < M_h$:}

Because of Eqs.~\eqref{nfde9re00}--\eqref{7fuf8} and~\eqref{m932033},
when $\psi = 0$
\bs
\ba
A &=& \frac{G \left( L + H \right)}{2 L}\, U_L, \\
B &=& \frac{L}{2}\, U_R + \frac{H}{2}\, U_L, \\
M_h &=& \frac{L + H}{2}\, U_R + \frac{H}{2}\, U_L
\no &=& B + \frac{H}{2}\, U_R. \label{86598549}
\ea
\es
Now,
\be
\bar M_h = \frac{1}{2} \left[ B + A + \sqrt{\left( B - A \right)^2
    + 4 D^2} \right].
\ee
Therefore,
using Eq.~\eqref{86598549},
\ba
M_h - \bar M_h &=& \frac{1}{2} \left[ B + H U_R - A
  - \sqrt{\left( B - A \right)^2 + 4 D^2} \right]
\no &\approx& \frac{1}{2} \left( H U_R
  - \frac{2 D^2}{B - A} \right)
\no &>& 0,
\ea
because both $A$ and $D$ are of order the electroweak scale squared,
\textit{viz.}\ $V_1$,
$V_2$,
and $U_L$,
while $B \sim U_R$ is much larger.
One thus concludes that the heavy neutral gauge boson
is always heavier than the heavy charged gauge boson,
at least when $\psi = 0$.

\setcounter{equation}{0}
\renewcommand{\theequation}{G\arabic{equation}}

\section{Mass matrices of the scalars}
\label{app:scalarmasses}

\paragraph{Mass matrix of the scalars:} The mass terms of the scalars
are contained in the mass matrix $M_\rho$,
which is defined through
\be
V = \cdots
+ \frac{1}{2} \left( \begin{array}{cccc} \rho_1, & \rho_2, & \rho_L, & \rho_R
\end{array} \right) M_\rho
\left( \begin{array}{c} \rho_1 \\ \rho_2 \\ \rho_L \\ \rho_R
\end{array} \right).
\ee
That matrix may be computed from the input observables as
\be
\label{computemrho}
M_\rho = V_\rho \times
\mathrm{diag} \left( \mu_5^2,\ \mu_6^2,\ \mu_7^2,\ \mu_8^2
\right) \times V_\rho^T,
\ee
where $V_\rho$ is a function of the mixing angles $\theta_i$
($i = 1, 2, \ldots, 6$)
through Eq.~\eqref{uidfr30}.
Using Eqs.~\eqref{expansion},
\eqref{pott2},
and~\eqref{mu2} one finds the matrix elements of $M_\rho$:
\bs
\label{mrmrho}
\ba
\left( M_\rho \right)_{11} &=&
\frac{m_1 v_2 - m_2 v_1}{V_1 - V_2}\, u_L u_R
+ \frac{V_2}{V_1 - V_2} \left[ \left( \lambda_{3L} - \lambda_{4L} \right) U_L
  + \left( \lambda_{3R} - \lambda_{4R} \right) U_R \right]
\no & &
+ 4 \lambda_1 V_1
+ \left( 8 \lambda_2 + 4 \lambda_3 \right) V_2
+ 8 \lambda_4 v_1 v_2,
\hspace*{7mm} \label{r11} \\
\left( M_\rho \right)_{22} &=&
\frac{m_1 v_2 - m_2 v_1}{V_1 - V_2}\, u_L u_R
+ \frac{V_1}{V_1 - V_2} \left[ \left( \lambda_{3L} - \lambda_{4L} \right) U_L
  + \left( \lambda_{3R} - \lambda_{4R} \right) U_R \right]
\no & &
+ 4 \lambda_1 V_2
+ \left( 8 \lambda_2 + 4 \lambda_3 \right) V_1
+ 8 \lambda_4 v_1 v_2,
\label{r22} \\
\left( M_\rho \right)_{12} &=&
\frac{m_1 v_1 - m_2 v_2}{V_2 - V_1}\, u_L u_R
+ \frac{v_1 v_2}{V_2 - V_1} \left[ \left( \lambda_{3L} - \lambda_{4L} \right) U_L
  + \left( \lambda_{3R} - \lambda_{4R} \right) U_R \right]
\no & &
+ \left( 4 \lambda_1 + 8 \lambda_2 + 4 \lambda_3 \right) v_1 v_2
+ 4 \lambda_4 \left( V_1 + V_2 \right),
\label{r12} \\
\left( M_\rho \right)_{13} &=&
m_2 u_R + 2 \left( \lambda_{4L} v_1 + \lambda_{5L} v_2 \right) u_L,
\label{r13} \\
\left( M_\rho \right)_{14} &=&
m_2 u_L + 2 \left( \lambda_{4R} v_1 + \lambda_{5R} v_2 \right) u_R,
\label{r14} \\
\left( M_\rho \right)_{23} &=&
m_1 u_R + 2 \left( \lambda_{3L} v_2 + \lambda_{5L} v_1 \right) u_L,
\label{r23} \\
\left( M_\rho \right)_{24} &=&
m_1 u_L + 2 \left( \lambda_{3R} v_2 + \lambda_{5R} v_1 \right) u_R,
\label{r24} \\
\left( M_\rho \right)_{33} &=& 4 \lambda_L U_L
- \left( m_1 v_2 + m_2 v_1 \right) \frac{u_R}{u_L},
\label{r33} \\
\left( M_\rho \right)_{44} &=& 4 \lambda_R U_R
- \left( m_1 v_2 + m_2 v_1 \right) \frac{u_L}{u_R},
\label{r44} \\
\left( M_\rho \right)_{34} &=&
m_1 v_2 + m_2 v_1 + 2 \lambda_{LR} u_L u_R.
\label{r34}
\ea
\es

\paragraph{Mass matrix of the pseudoscalars:}
The mass terms of the pseudoscalars
are contained in the mass matrix $M_\eta$,
which is defined through
\be
V = \cdots + \frac{1}{2}
\left( \begin{array}{cc} \eta_a, & \eta_b \end{array} \right)
M_\eta
\left( \begin{array}{c} \eta_a \\ \eta_b \end{array} \right),
\ee
\textit{cf.}\ Eq.~\eqref{Meta}.
The matrix $M_\eta$ may be computed from the input observables as
\be
M_\eta =
\left( \begin{array}{cc} c_\eta & s_\eta \\ - s_\eta & c_\eta \end{array} \right)
\left( \begin{array}{cc} M_{\eta 1} & 0 \\ 0 & M_{\eta 2} \end{array} \right)
\left( \begin{array}{cc} c_\eta & - s_\eta \\ s_\eta & c_\eta \end{array} \right),
\ee
\textit{cf.}\ Eq.~\eqref{diageta}.
Using Eqs.~\eqref{expansion},
\eqref{etaaetab},
\eqref{pott2},
and~\eqref{mu2} one finds the matrix elements of $M_\eta$:
\bs
\label{jw4838484}
\ba
\left( M_\eta \right)_{11} &=&
u_L u_R\ \frac{V_1 + V_2}{V_1 - V_2}
\left( \frac{m_1}{v_2} - \frac{m_2}{v_1} \right)
- \frac{u_L u_R \left( m_1 v_1^3 + m_2 v_2^3 \right)}{v_1 v_2
  \left( V_1 + V_2 \right)}
\no & &
+ \left( V_1 + V_2 \right) \left[ 4 \lambda_3 - 8 \lambda_2
  + \frac{\left( \lambda_{3L} - \lambda_{4L} \right) U_L
    + \left( \lambda_{3R} - \lambda_{4R} \right)
    U_R}{V_1 - V_2} \right],
\label{64a} \\
\left( M_\eta \right)_{22} &=&
- \frac{T_2}{T_1 u_L u_R} \left( m_1 v_2 + m_2 v_1 \right),
\label{64b} \\
\left( M_\eta \right)_{12} &=&
\frac{\sqrt{T_2}}{T_1} \left( m_1 v_1 - m_2 v_2 \right),
\label{64c}
\ea
\es
where $T_1$ and $T_2$ are defined in Eqs.~\eqref{T1T2}.

\paragraph{Mass matrix of the charged scalars:}
The mass terms of the charged scalars
are contained in the mass matrix $M_\varphi$,
which is defined through
\be
V = \cdots +
\left( \begin{array}{cc} \varphi_a^-, & \varphi_b^- \end{array} \right)
M_\varphi
\left( \begin{array}{c} \varphi_a^+ \\ \varphi_b^+ \end{array} \right),
\ee
\textit{cf.}\ Eq.~\eqref{mvarphi}.
The matrix $M_\varphi$ may be computed from the input observables as
\be
M_\varphi = \left( \begin{array}{cc} c_\varphi & s_\varphi \\
  - s_\varphi & c_\varphi \end{array} \right)
\left( \begin{array}{cc} M_{\varphi 1} & 0 \\ 0 & M_{\varphi 2} \end{array} \right)
\left( \begin{array}{cc}
  c_\varphi & - s_\varphi \\ s_\varphi & c_\varphi \end{array} \right),
\ee
\textit{cf.}\ Eq.~\eqref{diagvarphi}.
Using Eqs.~\eqref{expansion},
\eqref{phiab},
\eqref{pott2},
and~\eqref{mu2} one finds the matrix elements of $M_\varphi$:
\bs
\label{mphi}
\ba
\left( M_\varphi \right)_{11} &=&
\frac{\lambda_{3L} - \lambda_{4L}}{V_1 - V_2}\, K_1
+ \frac{\lambda_{3R} - \lambda_{4R}}{V_1 - V_2}\,
\frac{4 V_1 V_2 U_L U_R}{K_1}
\no & &
+ \frac{m_1 v_2 + m_2 v_1}{u_L u_R \left( V_2 - V_1 \right)^2}
\left( - U_R K_1 - U_L\, \frac{4 V_1 V_2 U_L U_R}{K_1} \right)
\no & &
+ \frac{4 \left( m_1 v_1 + m_2 v_2 \right)
  v_1 v_2 u_L u_R}{\left( V_2 - V_1 \right)^2},
\label{mvarphi11} \\
\left( M_\varphi \right)_{22} &=&
\frac{K_2}{K_1}
\left[ \left( \lambda_{3R} - \lambda_{4R} \right) \left( V_1 - V_2 \right)
- \frac{\left( m_1 v_2 + m_2 v_1 \right) u_L}{u_R} \right],
\label{mvarphi22} \\
\left( M_\varphi \right)_{12} &=&
- 2 v_1 v_2 u_L u_R \frac{\sqrt{K_2}}{K_1}
\left[ \lambda_{3R} - \lambda_{4R}
  - \frac{\left( m_1 v_2 + m_2 v_1 \right) u_L}{u_R \left( V_1 - V_2 \right)}
  \right]
\no & &
+ \frac{m_1 v_1 + m_2 v_2}{V_2 - V_1}\, \sqrt{K_2}.
\label{mvarphi12}
\ea
\es
where $K_1$ and $K_2$ are defined in Eqs.~\eqref{K1K2}.

\paragraph{A constraint:}
From Eqs.~\eqref{jw4838484} and~\eqref{mphi} one obtains
\ba
& & T_2 \left[ 2 v_1 v_2 u_L u_R \left( M_\varphi \right)_{22}
+ \sqrt{K_2} \left( V_1 - V_2 \right) \left( M_\varphi \right)_{12} \right]
\no &=& K_2 \left[ 2 v_1 v_2 u_L u_R \left( M_\eta \right)_{22}
+ \sqrt{T_2} \left( V_2 - V_1 \right) \left( M_\eta \right)_{12}
\right].
\label{constraint2}
\ea
Equation~\eqref{constraint2}
is a clear-cut prediction of our model for observable quantities
(remember that the VEVs,
and hence $K_2$ and $T_2$ too,
are functions of the input observables,
\textit{cf.}\ appendix~\ref{app:GLH}).
That prediction may be written
\ba
\label{meta2}
M_{\eta 2} &=& \left\{
T_2 \left[ 2 v_1 v_2 u_L u_R \left( M_\varphi \right)_{22}
  + \sqrt{K_2} \left( V_1 - V_2 \right) \left( M_\varphi \right)_{12} \right]
\right. \no & & \left.
 + \left[ K_2 \sqrt{T_2} \left( V_2 - V_1 \right) c_\eta s_\eta
   - 2 v_1 v_2 u_L u_R K_2 s_\eta^2 \right] M_{\eta 1}
\right\}
\no & &
\times \left[ 2 v_1 v_2 u_L u_R K_2 c_\eta^2
  + K_2 \sqrt{T_2} \left( V_2 - V_1 \right) c_\eta s_\eta \right]^{-1}.
\ea
Thus,
given $M_{\varphi 1}$,
$M_{\varphi 2}$,
$M_{\eta 1}$,
and the mixing angles $\varphi$ and $\eta$,
one may calculate $M_{\eta_2}$.

\paragraph{Determination of the parameters of the potential:}
Suppose that we already know the values of the VEVs $u_L$,
$u_R$,
$v_1$,
and $v_2$ and that we input all the scalar masses and mixing angles.
We may then
\begin{enumerate}
\item Use Eqs.~\eqref{64b} and~\eqref{64c} to determine $m_1$ and $m_2$.
\item Then use Eqs.~\eqref{r33} to determine $\lambda_L$,
  \eqref{r44} to determine $\lambda_R$,
  and~\eqref{r34} to determine $\lambda_{LR}$.
\item Next use either Eq.~\eqref{mvarphi22} or Eq.~\eqref{mvarphi12}
  to determine $\lambda_{3R} - \lambda_{4R}$,
  and then Eq.~\eqref{mvarphi11} to determine $\lambda_{3L} - \lambda_{4L}$.
\item Equations~\eqref{r13}--\eqref{r24} may now be employed
  to separately determine $\lambda_{3R}$, $\lambda_{4R}$, $\lambda_{5R}$,
  $\lambda_{3L}$, $\lambda_{4L}$, and~$\lambda_{5L}$.
\item Equation~\eqref{64a} pins down $\lambda_3 - 2 \lambda_2$.
\item Equations~\eqref{r11}--\eqref{r12} separately determine $\lambda_1$,
  $\lambda_2$,
  $\lambda_3$,
  and~$\lambda_4$.
\end{enumerate}

\setcounter{equation}{0}
\renewcommand{\theequation}{H\arabic{equation}}

\section{The ranges of the parameters of the LRM}
\label{app:results}

In this appendix we provide more information on the fits
mentioned in section~\ref{sec:numerical}.
We present scatter plots with green,
blue,
and red points following the code introduced in that section.
However,
the scatter plots in this appendix
do not dwell on the fitting of the $Z b \bar b$ vertex,
rather on the parameters of the LRM itself.
Moreover,
in the scatter plots of this appendix
we only display the points of Fig.~\ref{fig_gLgR}
that satisfy the 2$\sigma$ bounds on both $R_b$ and $A_b$,
\viz\ the points that are both in between the dashed orange lines
and in between the magenta dashed lines in the left panel of that figure.

\begin{figure}[ht]
  \begin{center}
    \includegraphics[width=0.6\textwidth]{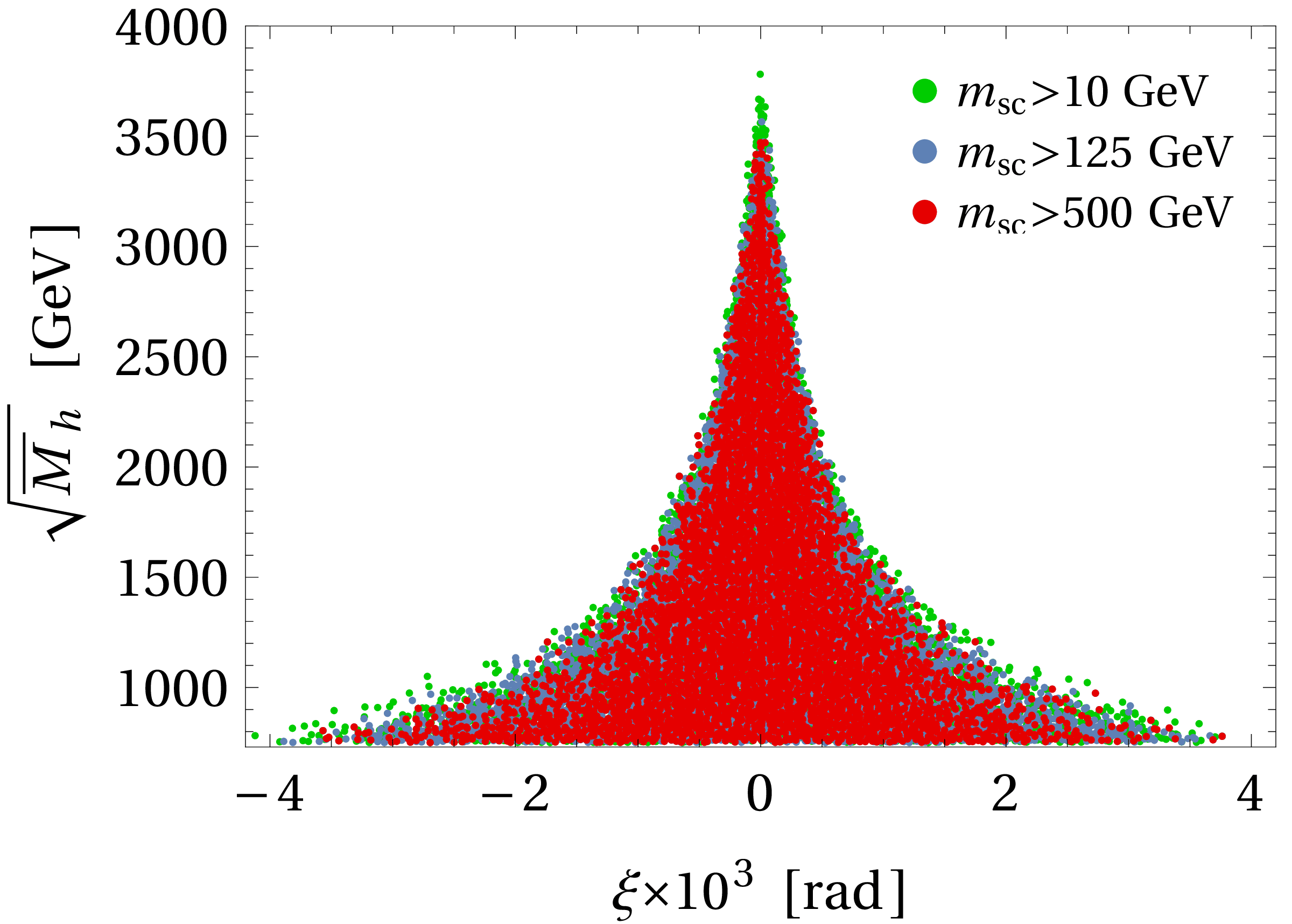}
  \end{center}
  \caption{Scatter plot of the mass of the heavy charged gauge boson
    ($\sqrt{\bar{M}_h}$)
    \vvs the mixing angle between the two charged gauge bosons
    ($\xi$),
    for the three cases where all the scalars have masses above
    10\,GeV (green points),
    125\,GeV (blue points),
    and  500\,GeV (red points).}
  \label{fig_Ml_vs_xi}
\end{figure}

In Fig.~\ref{fig_Ml_vs_xi} we plot
the mass $\sqrt{\bar{M}_h}$ of the heavy charged gauge boson
\vvs the mixing angle $\xi$ between the two charged gauge bosons.
One sees that the mixing angle lies within a narrow range  
$\left[ - 0.004,\, + 0.004 \right]$.
This happens because of the lower bound 750\,GeV
that we have imposed on $\sqrt{\bar M_h}$ and $\sqrt{M_h}$;
if we had opted for a more stringent and realistic
lower bound~\cite{TWIST:2011jfx,Civitarese:2016hbg,Babu:1993hx,
	Dekens:2014ina,Bertolini:2014sua,ValeSilva:2016dcg,Blanke:2011ry,
	Babu:2020bgz,CMS:2019gwf,ThomasArun:2021rwf,CMS:2021mux,Osland:2022ryb},
say 2 or 3\,TeV,
then $\xi$ would have to lie in an even narrower range.
The various lower bounds on the masses of the scalars
have no effect on this scatter plot,
which results almost exclusively from the procedure in Appendix~\ref{app:GLH}.
Namely,
if one chooses larger values for $\xi$,
then either one violates the first inequality~\eqref{mfoder0ewe9}
or one ends up with negative values for either $G$,
$V_1$,
or $U_L$---which must all be positive,
because they are the squares of real quantities.
Therefore,
$\xi$ must always be very small.

\begin{figure}[ht]
  \begin{center}
    \includegraphics[width=1.0\textwidth]{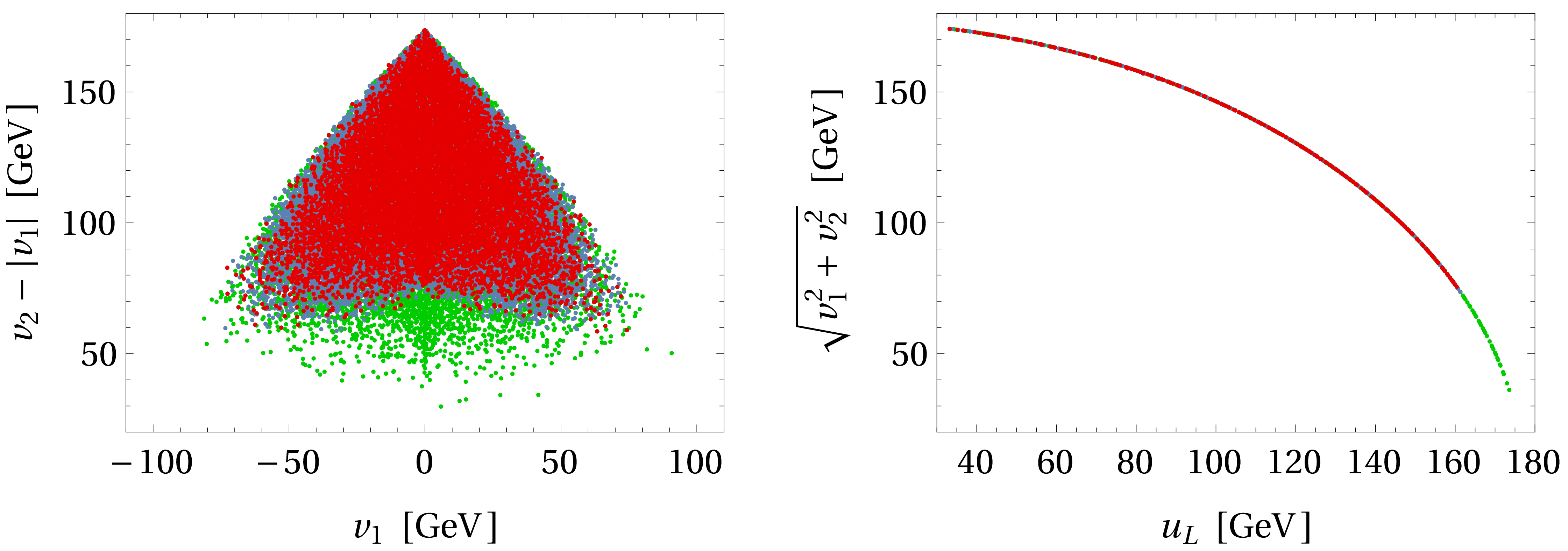}
  \end{center}
  \caption{ 
    Scatter plot of the relations among the vacuum expectation values $v_1$,
    $v_2$,
    and $u_L$ that break $SU(2)_L$.
    The points,
    and the colour code for them,
    are the same as in Fig.~\ref{fig_gLgR}.
    Left panel: $v_2 - \left| v_1 \right|$ \vvs $v_1$.
    Right panel: $\sqrt{v_1^2 +v_2^2}$ \vvs $u_L$.
    Remember that in our conventions $v_2$ and $u_L$ are always positive,
    and $v_2$ is always larger than $\left| v_1 \right|$.
  }
  \label{fig_vevs}
\end{figure}

In Fig.~\ref{fig_vevs} we display the correlations
among the vacuum expectation values.
In the left panel one sees that $\left| v_1 \right|$
is always at least 30\,GeV smaller than $v_2$.
Furthermore,
$\sqrt{v_1^2 + v_2^2} \lesssim 174$\,GeV provides an upper bound
both on $\left| v_1 \right|$ and on $v_2$.
In the right panel one observes the almost exact curve
$\sqrt{v_1^2 + v_2^2 + u_L^2} \approx 174$\,GeV
that arises from the masses of the gauge bosons $W_l$ and $Z_l$
being fixed to their observed values.

\begin{figure}[ht]
  \begin{center}
    \includegraphics[width=1.0\textwidth]{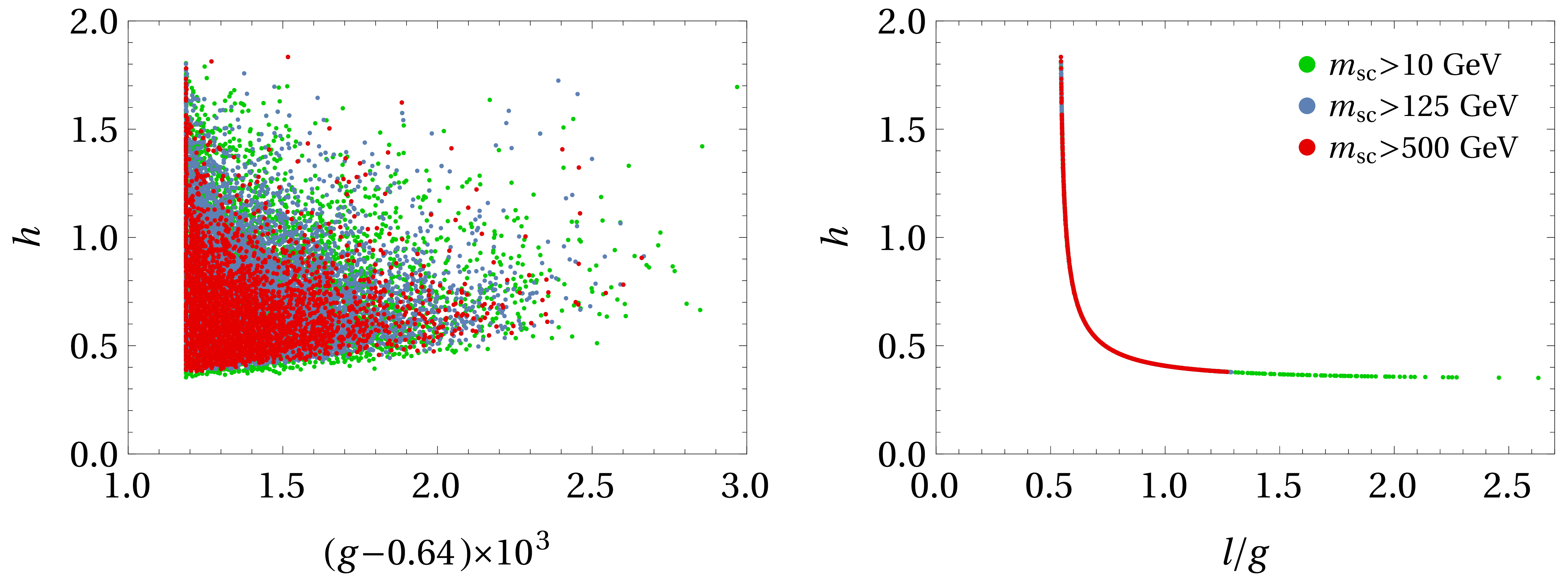}
  \end{center}
  \caption{
    Scatter plot of the gauge coupling constants $g$ of $SU(2)_L$,
    $l$ of $SU(2)_R$,
    and $h$ of $U(1)_X$.
    Both the points and the colour code used to mark them
    are the same as in Fig.~\ref{fig_gLgR}.
    Left panel: $h$ \vvs $g$.
    Right panel: $h$ \vvs $l/g$.
  }
  \label{fig_gauge}
\end{figure}

Figure~\ref{fig_gauge} displays the gauge constants $g$ of $SU(2)_L$,
$l$ of $SU(2)_R$,
and $h$ of $U(1)_X$.
One sees in the horizontal scale of the left panel
that $g$ hardly deviates from it	s SM value.
On the other hand,
$h$ may vary from $2 g / 3$ to $3 g$,
approximately.

In the right panel of Fig.~\ref{fig_gauge} one sees the counterpart,
for the gauge coupling constants,
of the relation observed in the right panel of Fig.~\ref{fig_vevs}
for the VEVs.
Indeed,
because we set $\psi = 0$,
Eq.~\eqref{m932033} forces $h / l = \left. \sqrt{v_1^2 + v_2^2} \right/ \! u_L$.
One sees that $l/g$ saturates the bound in Eq.~\eqref{nvkff0005},
but may also be much higher:
in a separate dedicated search
we have found points with $l/g \sim 6$ that satisfy
(at the price of very low-mass scalars)
the 1$\sigma$ bounds on $R_b$ and $A_b$.

\begin{figure}[ht]
  \begin{center}
    \includegraphics[width=0.6\textwidth]{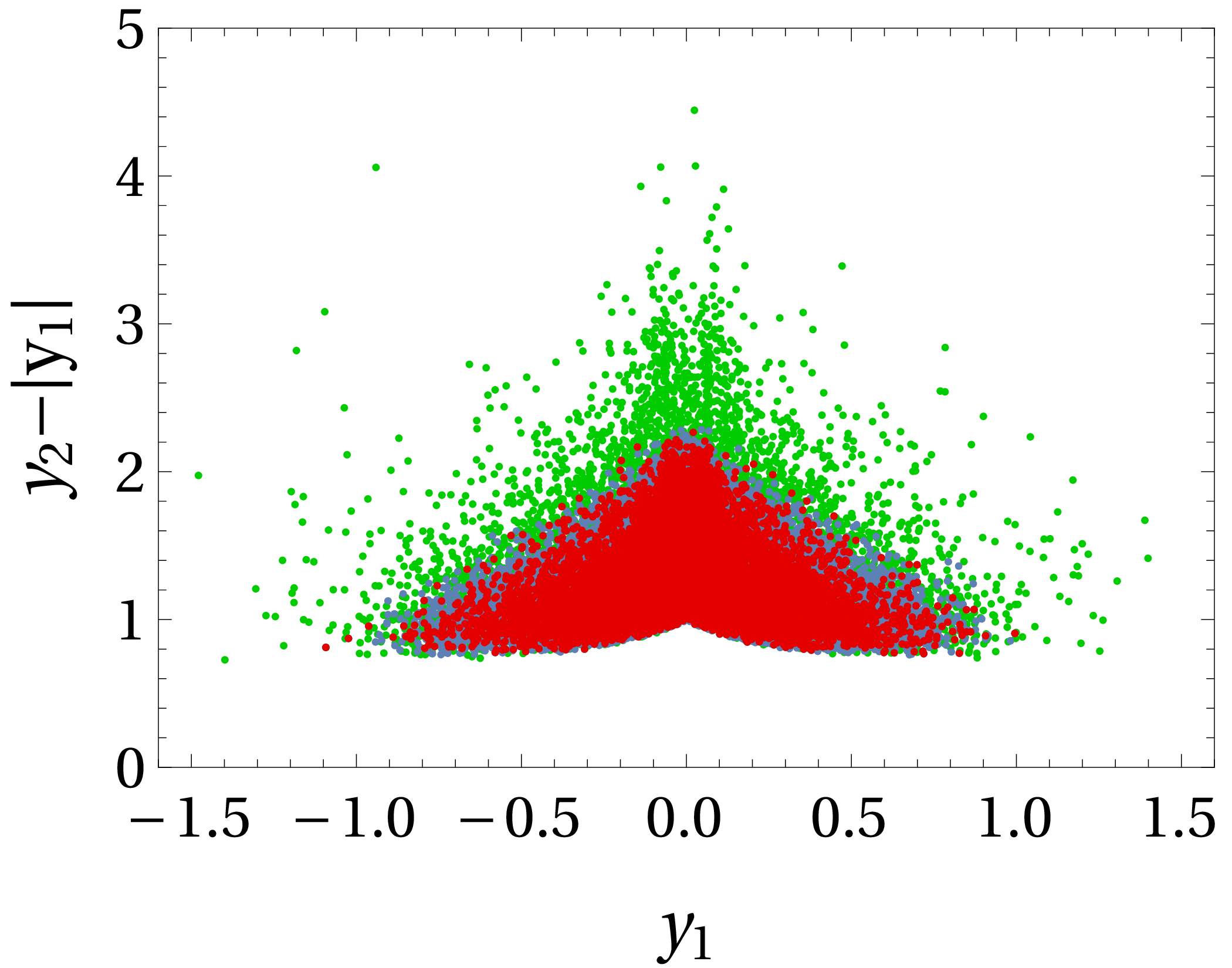}
  \end{center}
  \caption{Scatter plot of the Yukawa coupling $y_2$ and $y_1$.
    Remember that in our convention $y_2$ is always positive
    and larger than $\left| y_1 \right|$.
    The points and the colour code are the same as in Fig.~\ref{fig_gLgR}.
  }
  \label{fig_y2_vs_y1}
\end{figure}

In figure~\ref{fig_y2_vs_y1} we illustrate the Yukawa couplings $y_1$ and $y_2$.
One sees that in most cases $\left| y_1 \right|$ and $y_2$
are below $\sqrt{4 \pi} \approx 3.5$;
only a few points have them larger than 3.5,
but always quite below our relaxed perturbativity bound $4 \pi$.
One sees that
$y_2 - \left| y_1 \right|$ is always larger than~0.8,
that $y_1$ is either positive or negative with the same likelihood,
and that $\left| y_1 \right| \lesssim 1.5$.

\begin{figure}[ht]
  \begin{center}
    \includegraphics[width=0.6\textwidth]{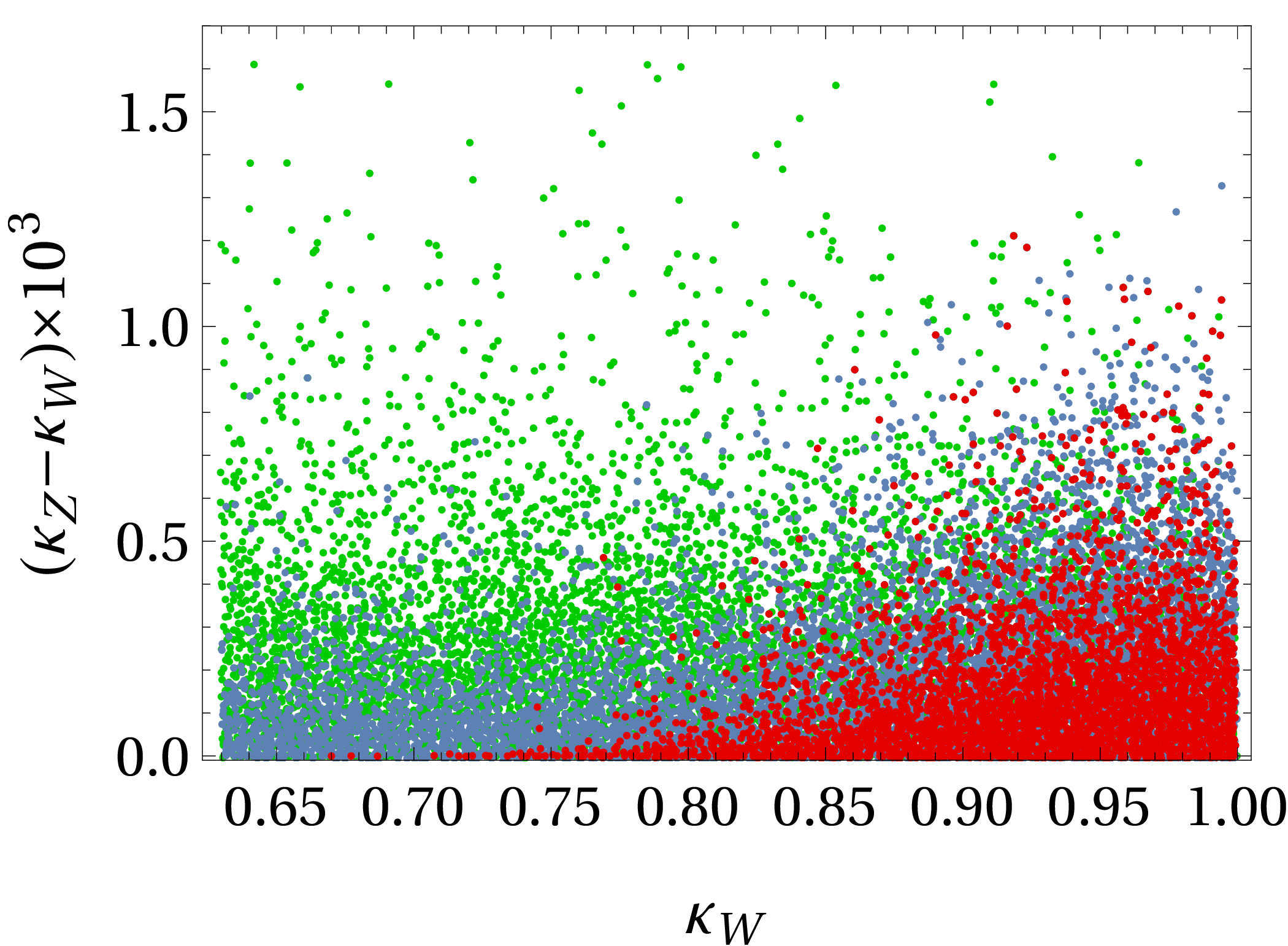}
  \end{center}
  \caption{Scatter plot of the coupling modifiers $\kappa_W$ and $\kappa_Z$,
    for the three cases where all the scalars
    have masses above 10\,GeV (green points),
    125\,GeV (blue points),
    and 500\,GeV (red points).
  }
  \label{fig_kappas}
\end{figure}

In Fig.~\ref{fig_kappas} we depict the correlation between
the coupling modifiers $\kappa_W$ and $\kappa_Z$.
One observes that in the LRM they are equal for all practical purposes
and they are always smaller than~1.
We have also examined the parameters $\kappa_t$ and $\kappa_b$;
they do not correlate with each other
and they fully cover their full ranges in conditions~\cite{CMS:2018uag}.

In table~\ref{tab:potParam}
we show the approximate ranges of some parameters of the scalar potential.
Note that in a left--right-symmetric model
one must have $\lambda_L = \lambda_R$,
$\lambda_{3L} = \lambda_{3R}$,
$\lambda_{4L} = \lambda_{4R}$,
and $\lambda_{5L} = \lambda_{5R}$;
these relations of course lead
to restrictions on the masses and mixing of the scalars.
\begin{table}[h!]\centering
   \begin{tabular}{lc}
    \hline\hline
    parameter & range \\ \hline 
    $m_1$~[GeV] & -300 to 3 \\
    $m_2$~[GeV] & -140 to 130 \\
    $\lambda_1$ & 0 to 3.6 \\
    $\lambda_2$ & -1 to 0.9 \\
    $\lambda_3$ & -3 to 3 \\
    $\lambda_4$ & -1.1 to 1.1 \\
    $\lambda_L$ & 0 to 4.2 \\
    $\lambda_R$ & 0 to 0.2 \\ 
    $\lambda_{LR}$ & -1.0 to 0.6 \\
    $\lambda_{3L}$ & -6 to 11 \\
    $\lambda_{3R}$ & -0.7 to 0.6 \\
    $\lambda_{4L}$ & -3.8 to 9.5 \\
    $\lambda_{4R}$ & -0.7 to 0.6 \\
    $\lambda_{5L}$ & -6 to 6 \\
    $\lambda_{5R}$ & -0.8 to 0.8 \\
       \hline\hline
     \end{tabular}
   \caption{\label{tab:potParam}
     Approximate ranges of the parameters of the scalar potential.
   }
\end{table}
It should be noted that the various lower bounds on the scalar masses
that we have imposed do not greatly impact
the ranges of the parameters of the potential.

\end{appendix}


\begin{thebibliography}{10}

\bibitem{PDG2022}
{\bf Particle Data Group} Collaboration, R.~L. Workman, {\it {Review of
  Particle Physics}},  {\em PTEP} {\bf 2022} (2022) 083C01.

\bibitem{Field:1997gz}
J.~H. Field, {\it {Indications for an anomalous right-handed coupling of the b
  quark from a model independent analysis of LEP and SLD data on Z decays}},
  {\em Mod. Phys. Lett. A} {\bf 13} (1998) 1937--1954,
  [\href{http://arxiv.org/abs/hep-ph/9801355}{{\tt hep-ph/9801355}}].

\bibitem{Haber:1999zh}
H.~E. Haber and H.~E. Logan, {\it {Radiative corrections to the Z b anti-b
  vertex and constraints on extended Higgs sectors}},  {\em Phys. Rev. D} {\bf
  62} (2000) 015011, [\href{http://arxiv.org/abs/hep-ph/9909335}{{\tt
  hep-ph/9909335}}].

\bibitem{freitas}
J.~Erler and A.~Freitas, {\it {Electroweak model and constraints on new
  physics}},  {\em in} ref.~\cite{PDG2022},~pp.~177~ff.

\bibitem{Jurciukonis:2021wny}
D.~Jur\v{c}iukonis and L.~Lavoura, {\it {Fitting the $ Zb\overline{b} $ vertex
  in the two-Higgs-doublet model and in the three-Higgs-doublet model}},  {\em
  JHEP} {\bf 07} (2021) 195, [\href{http://arxiv.org/abs/2103.16635}{{\tt
  arXiv:2103.16635}}].

\bibitem{Choudhury:2001hs}
D.~Choudhury, T.~M.~P. Tait, and C.~E.~M. Wagner, {\it {Beautiful mirrors and
  precision electroweak data}},  {\em Phys. Rev. D} {\bf 65} (2002) 053002,
  [\href{http://arxiv.org/abs/hep-ph/0109097}{{\tt hep-ph/0109097}}].

\bibitem{Fontes:2019fbz}
D.~Fontes, L.~Lavoura, J.~C. Rom\~ao, and J.~a.~P. Silva, {\it {One-loop
  corrections to the $Zb\bar{b}$ vertex in models with scalar doublets and
  singlets}},  {\em Nucl. Phys. B} {\bf 958} (2020) 115131,
  [\href{http://arxiv.org/abs/1910.11886}{{\tt arXiv:1910.11886}}].

\bibitem{Lee:1999kw}
K.~Y. Lee, {\it {Probing anomalous right-handed current in Z b anti-b vertex}},
   {\em Phys. Lett. B} {\bf 472} (2000) 366--372,
  [\href{http://arxiv.org/abs/hep-ph/9904435}{{\tt hep-ph/9904435}}].

\bibitem{He:2002ha}
X.-G. He and G.~Valencia, {\it {The $Z \to b \bar{b}$ decay asymmetry and
  left-right models}},  {\em Phys. Rev. D} {\bf 66} (2002) 013004,
  [\href{http://arxiv.org/abs/hep-ph/0203036}{{\tt hep-ph/0203036}}]. [Erratum:
  Phys.Rev.D 66, 079901 (2002)].

\bibitem{He:2003qv}
X.-G. He and G.~Valencia, {\it {A**b(FB) and R(b) at LEP and new right-handed
  gauge bosons}},  {\em Phys. Rev. D} {\bf 68} (2003) 033011,
  [\href{http://arxiv.org/abs/hep-ph/0304215}{{\tt hep-ph/0304215}}].

\bibitem{Liu:2017xmc}
D.~Liu, J.~Liu, C.~E.~M. Wagner, and X.-P. Wang, {\it {Bottom-quark
  Forward-Backward Asymmetry, Dark Matter and the LHC}},  {\em Phys. Rev. D}
  {\bf 97} (2018), no.~5 055021, [\href{http://arxiv.org/abs/1712.05802}{{\tt
  arXiv:1712.05802}}].

\bibitem{Dermisek:2011xu}
R.~Dermisek, S.-G. Kim, and A.~Raval, {\it {New Vector Boson Near the Z-pole
  and the Puzzle in Precision Electroweak Data}},  {\em Phys. Rev. D} {\bf 84}
  (2011) 035006, [\href{http://arxiv.org/abs/1105.0773}{{\tt
  arXiv:1105.0773}}].

\bibitem{Dermisek:2012qx}
R.~Dermisek, S.-G. Kim, and A.~Raval, {\it {Z' near the Z-pole}},  {\em Phys.
  Rev. D} {\bf 85} (2012) 075022, [\href{http://arxiv.org/abs/1201.0315}{{\tt
  arXiv:1201.0315}}].

\bibitem{Agashe:2006at}
K.~Agashe, R.~Contino, L.~Da~Rold, and A.~Pomarol, {\it {A Custodial symmetry
  for $Zb \bar b$}},  {\em Phys. Lett. B} {\bf 641} (2006) 62--66,
  [\href{http://arxiv.org/abs/hep-ph/0605341}{{\tt hep-ph/0605341}}].

\bibitem{DaRold:2010as}
L.~Da~Rold, {\it {Solving the $A_{FB}^b$ anomaly in natural composite models}},
   {\em JHEP} {\bf 02} (2011) 034, [\href{http://arxiv.org/abs/1009.2392}{{\tt
  arXiv:1009.2392}}].

\bibitem{Papavassiliou:2000pq}
J.~Papavassiliou and A.~Santamaria, {\it {Extra dimensions at the one loop
  level: Z ---\ensuremath{>} b anti-b and B anti-B mixing}},  {\em Phys. Rev.
  D} {\bf 63} (2001) 016002, [\href{http://arxiv.org/abs/hep-ph/0008151}{{\tt
  hep-ph/0008151}}].

\bibitem{Oliver:2002up}
J.~F. Oliver, J.~Papavassiliou, and A.~Santamaria, {\it {Universal extra
  dimensions and Z ---\ensuremath{>} b anti-b}},  {\em Phys. Rev. D} {\bf 67}
  (2003) 056002, [\href{http://arxiv.org/abs/hep-ph/0212391}{{\tt
  hep-ph/0212391}}].

\bibitem{Jha:2014faa}
T.~Jha and A.~Datta, {\it {$ Z\to b\overline{b} $ in non-minimal Universal
  Extra Dimensional Model}},  {\em JHEP} {\bf 03} (2015) 012,
  [\href{http://arxiv.org/abs/1410.5098}{{\tt arXiv:1410.5098}}].

\bibitem{Choudhury:2013jta}
D.~Choudhury, A.~Kundu, and P.~Saha, {\it {Z-pole observables in an effective
  theory}},  {\em Phys. Rev. D} {\bf 89} (2014), no.~1 013002,
  [\href{http://arxiv.org/abs/1305.7199}{{\tt arXiv:1305.7199}}].

\bibitem{dEnterria:2020cgt}
D.~d'Enterria and C.~Yan, {\it {Revised QCD effects on the Z $\to b\bar{b}$
  forward-backward asymmetry}},  \href{http://arxiv.org/abs/2011.00530}{{\tt
  arXiv:2011.00530}}.

\bibitem{Crivellin:2020oup}
A.~Crivellin, C.~A. Manzari, M.~Alguero, and J.~Matias, {\it {Combined
  Explanation of the Z $\to b\bar{b}$ Forward-Backward Asymmetry, the Cabibbo
  Angle Anomaly, and $\tau \to \mu \nu \nu$ and $b\to s \ell^+ \ell^-$ Data}},
  {\em Phys. Rev. Lett.} {\bf 127} (2021), no.~1 011801,
  [\href{http://arxiv.org/abs/2010.14504}{{\tt arXiv:2010.14504}}].

\bibitem{Murphy:2015cha}
C.~W. Murphy, {\it {Bottom-Quark Forward-Backward and Charge Asymmetries at
  Hadron Colliders}},  {\em Phys. Rev. D} {\bf 92} (2015), no.~5 054003,
  [\href{http://arxiv.org/abs/1504.02493}{{\tt arXiv:1504.02493}}].

\bibitem{Gori:2015nqa}
S.~Gori, J.~Gu, and L.-T. Wang, {\it {The $Zb\bar b$ couplings at future e$^+$
  e$^-$ colliders}},  {\em JHEP} {\bf 04} (2016) 062,
  [\href{http://arxiv.org/abs/1508.07010}{{\tt arXiv:1508.07010}}].

\bibitem{Yan:2021veo}
B.~Yan and C.~P. Yuan, {\it {Anomalous $Z b \bar b$ Couplings: From LEP to
  LHC}},  {\em Phys. Rev. Lett.} {\bf 127} (2021), no.~5 051801,
  [\href{http://arxiv.org/abs/2101.06261}{{\tt arXiv:2101.06261}}].

\bibitem{Dong:2022ayy}
H.~Dong, P.~Sun, B.~Yan, and C.~P. Yuan, {\it {Probing the $Z b \bar b$
  anomalous couplings via exclusive Z boson decay}},  {\em Phys. Lett. B} {\bf
  829} (2022) 137076, [\href{http://arxiv.org/abs/2201.11635}{{\tt
  arXiv:2201.11635}}].

\bibitem{Yan:2021htf}
B.~Yan, Z.~Yu, and C.~P. Yuan, {\it {The anomalous $Z b \bar b$ couplings at
  the HERA and EIC}},  {\em Phys. Lett. B} {\bf 822} (2021) 136697,
  [\href{http://arxiv.org/abs/2107.02134}{{\tt arXiv:2107.02134}}].

\bibitem{Li:2021uww}
H.~T. Li, B.~Yan, and C.~P. Yuan, {\it {Jet charge: A new tool to probe the
  anomalous $Z b \bar b$ couplings at the EIC}},  {\em Phys. Lett. B} {\bf 833}
  (2022) 137300, [\href{http://arxiv.org/abs/2112.07747}{{\tt
  arXiv:2112.07747}}].

\bibitem{Pati:1974yy}
J.~C. Pati and A.~Salam, {\it {Lepton Number as the Fourth Color}},  {\em Phys.
  Rev. D} {\bf 10} (1974) 275--289. [Erratum: Phys.Rev.D 11, 703--703 (1975)].

\bibitem{Mohapatra:1974gc}
R.~N. Mohapatra and J.~C. Pati, {\it {A Natural Left-Right Symmetry}},  {\em
  Phys. Rev. D} {\bf 11} (1975) 2558.

\bibitem{Mohapatra:1974hk}
R.~N. Mohapatra and J.~C. Pati, {\it {Left-Right Gauge Symmetry and an
  Isoconjugate Model of CP Violation}},  {\em Phys. Rev. D} {\bf 11} (1975)
  566--571.

\bibitem{Senjanovic:1975rk}
G.~Senjanovic and R.~N. Mohapatra, {\it {Exact Left-Right Symmetry and
  Spontaneous Violation of Parity}},  {\em Phys. Rev. D} {\bf 12} (1975) 1502.

\bibitem{Senjanovic:1978ev}
G.~Senjanovic, {\it {Spontaneous Breakdown of Parity in a Class of Gauge
  Theories}},  {\em Nucl. Phys. B} {\bf 153} (1979) 334--364.

\bibitem{Chang:1983fu}
D.~Chang, R.~N. Mohapatra, and M.~K. Parida, {\it {Decoupling of Parity- and
  SU(2)$_R$-Breaking Scales: A New Approach to Left-Right Symmetric Models}},
  {\em Phys. Rev. Lett.} {\bf 52} (1984) 1072.

\bibitem{Bernard:2020cyi}
V.~Bernard, S.~Descotes-Genon, and L.~Vale~Silva, {\it {Constraining the gauge
  and scalar sectors of the doublet left-right symmetric model}},  {\em JHEP}
  {\bf 09} (2020) 088, [\href{http://arxiv.org/abs/2001.00886}{{\tt
  arXiv:2001.00886}}].

\bibitem{Karmakar:2022iip}
S.~Karmakar, J.~More, A.~K. Pradhan, and S.~U. Sankar, {\it {Constraints on the
  Doublet Left-Right Symmetric Model from Higgs data}},
  \href{http://arxiv.org/abs/2211.08445}{{\tt arXiv:2211.08445}}.

\bibitem{Erler:2009jh}
J.~Erler, P.~Langacker, S.~Munir, and E.~Rojas, {\it {Improved Constraints on
  Z-prime Bosons from Electroweak Precision Data}},  {\em JHEP} {\bf 08} (2009)
  017, [\href{http://arxiv.org/abs/0906.2435}{{\tt arXiv:0906.2435}}].

\bibitem{Bobovnikov:2018fwt}
I.~D. Bobovnikov, P.~Osland, and A.~A. Pankov, {\it {Improved constraints on
  the mixing and mass of $Z'$ bosons from resonant diboson searches at the LHC
  at $\sqrt{s}=13$ TeV and predictions for Run II}},  {\em Phys. Rev. D} {\bf
  98} (2018), no.~9 095029, [\href{http://arxiv.org/abs/1809.08933}{{\tt
  arXiv:1809.08933}}].

\bibitem{Osland:2020onj}
P.~Osland, A.~A. Pankov, and I.~A. Serenkova, {\it {Updated constraints on $Z'$
  and $W'$ bosons decaying into bosonic and leptonic final states using the run
  2 ATLAS data}},  {\em Phys. Rev. D} {\bf 103} (2021), no.~5 053009,
  [\href{http://arxiv.org/abs/2012.13930}{{\tt arXiv:2012.13930}}].

\bibitem{Osland:2022ryb}
P.~Osland, A.~A. Pankov, and I.~A. Serenkova, {\it {Bounds on the mass and
  mixing of $Z^\prime$ and $W^\prime$ bosons decaying into different pairings
  of $W$, $Z$, or Higgs bosons using CMS data at the LHC}},
  \href{http://arxiv.org/abs/2206.01438}{{\tt arXiv:2206.01438}}.

\bibitem{TWIST:2011jfx}
{\bf TWIST} Collaboration, R.~Bayes et~al., {\it {Experimental Constraints on
  Left-Right Symmetric Models from Muon Decay}},  {\em Phys. Rev. Lett.} {\bf
  106} (2011) 041804.

\bibitem{Civitarese:2016hbg}
O.~Civitarese, J.~Suhonen, and K.~Zuber, {\it {Combining data from high-energy
  $pp$-reactions and neutrinoless double-beta decay: Limits on the mass of the
  right-handed boson}},  {\em Int. J. Mod. Phys. E} {\bf 25} (2016), no.~10
  1650081.

\bibitem{Babu:1993hx}
K.~S. Babu, K.~Fujikawa, and A.~Yamada, {\it {Constraints on left-right
  symmetric models from the process $b \to s \gamma$}},  {\em Phys. Lett. B}
  {\bf 333} (1994) 196--201, [\href{http://arxiv.org/abs/hep-ph/9312315}{{\tt
  hep-ph/9312315}}].

\bibitem{Dekens:2014ina}
W.~Dekens and D.~Boer, {\it {Viability of minimal left-right models with
  discrete symmetries}},  {\em Nucl. Phys. B} {\bf 889} (2014) 727--756,
  [\href{http://arxiv.org/abs/1409.4052}{{\tt arXiv:1409.4052}}].

\bibitem{Bertolini:2014sua}
S.~Bertolini, A.~Maiezza, and F.~Nesti, {\it {Present and Future K and B Meson
  Mixing Constraints on TeV Scale Left-Right Symmetry}},  {\em Phys. Rev. D}
  {\bf 89} (2014), no.~9 095028, [\href{http://arxiv.org/abs/1403.7112}{{\tt
  arXiv:1403.7112}}].

\bibitem{ValeSilva:2016dcg}
L.~Vale~Silva, {\em {Phenomenology of Left-Right Models in the quark sector}}.
\newblock PhD thesis, Saclay, 2016.
\newblock \href{http://arxiv.org/abs/1611.08187}{{\tt arXiv:1611.08187}}.

\bibitem{Blanke:2011ry}
M.~Blanke, A.~J. Buras, K.~Gemmler, and T.~Heidsieck, {\it {Delta F = 2
  observables and $B \to X_q \gamma$ decays in the Left-Right Model: Higgs
  particles striking back}},  {\em JHEP} {\bf 03} (2012) 024,
  [\href{http://arxiv.org/abs/1111.5014}{{\tt arXiv:1111.5014}}].

\bibitem{Babu:2020bgz}
K.~S. Babu and A.~Thapa, {\it {Left-Right Symmetric Model without Higgs
  Triplets}},  \href{http://arxiv.org/abs/2012.13420}{{\tt arXiv:2012.13420}}.

\bibitem{CMS:2019gwf}
{\bf CMS} Collaboration, A.~M. Sirunyan et~al., {\it {Search for high mass
  dijet resonances with a new background prediction method in proton-proton
  collisions at $\sqrt{s} =$ 13 TeV}},  {\em JHEP} {\bf 05} (2020) 033,
  [\href{http://arxiv.org/abs/1911.03947}{{\tt arXiv:1911.03947}}].

\bibitem{ThomasArun:2021rwf}
M.~Thomas~Arun, T.~Mandal, S.~Mitra, A.~Mukherjee, L.~Priya, and A.~Sampath,
  {\it {Testing left-right symmetry with an inverse seesaw mechanism at the
  LHC}},  {\em Phys. Rev. D} {\bf 105} (2022), no.~11 115007,
  [\href{http://arxiv.org/abs/2109.09585}{{\tt arXiv:2109.09585}}].

\bibitem{CMS:2021mux}
{\bf CMS} Collaboration, A.~M. Sirunyan et~al., {\it {Search for W' bosons
  decaying to a top and a bottom quark at s=13TeV in the hadronic final
  state}},  {\em Phys. Lett. B} {\bf 820} (2021) 136535,
  [\href{http://arxiv.org/abs/2104.04831}{{\tt arXiv:2104.04831}}].

\bibitem{ATLAS:2020qdt}
{\bf ATLAS} Collaboration, {\it {A combination of measurements of Higgs boson
  production and decay using up to $139$ fb$^{-1}$ of proton--proton collision
  data at $\sqrt{s}=$ 13 TeV collected with the ATLAS experiment}}, .

\bibitem{CMS:2018uag}
{\bf CMS} Collaboration, A.~M. Sirunyan et~al., {\it {Combined measurements of
  Higgs boson couplings in proton\textendash{}proton collisions at
  $\sqrt{s}=13\,\text {Te}\text {V} $}},  {\em Eur. Phys. J. C} {\bf 79}
  (2019), no.~5 421, [\href{http://arxiv.org/abs/1809.10733}{{\tt
  arXiv:1809.10733}}].

\bibitem{Fontes:2019wqh}
D.~Fontes and J.~C. Rom\~ao, {\it {FeynMaster: a plethora of Feynman tools}},
  {\em Comput. Phys. Commun.} {\bf 256} (2020) 107311,
  [\href{http://arxiv.org/abs/1909.05876}{{\tt arXiv:1909.05876}}].

\bibitem{Fontes:2021iue}
D.~Fontes and J.~C. Rom\~ao, {\it {Renormalization of the C2HDM with FeynMaster
  2}},  {\em JHEP} {\bf 06} (2021) 016,
  [\href{http://arxiv.org/abs/2103.06281}{{\tt arXiv:2103.06281}}].

\bibitem{Christensen:2008py}
N.~D. Christensen and C.~Duhr, {\it {FeynRules - Feynman rules made easy}},
  {\em Comput. Phys. Commun.} {\bf 180} (2009) 1614--1641,
  [\href{http://arxiv.org/abs/0806.4194}{{\tt arXiv:0806.4194}}].

\bibitem{Alloul:2013bka}
A.~Alloul, N.~D. Christensen, C.~Degrande, C.~Duhr, and B.~Fuks, {\it
  {FeynRules 2.0 - A complete toolbox for tree-level phenomenology}},  {\em
  Comput. Phys. Commun.} {\bf 185} (2014) 2250--2300,
  [\href{http://arxiv.org/abs/1310.1921}{{\tt arXiv:1310.1921}}].

\bibitem{Nogueira:1991ex}
P.~Nogueira, {\it {Automatic Feynman graph generation}},  {\em J. Comput.
  Phys.} {\bf 105} (1993) 279--289.

\bibitem{Mertig:1990an}
R.~Mertig, M.~Bohm, and A.~Denner, {\it {FEYN CALC: Computer algebraic
  calculation of Feynman amplitudes}},  {\em Comput. Phys. Commun.} {\bf 64}
  (1991) 345--359.

\bibitem{Shtabovenko:2016sxi}
V.~Shtabovenko, R.~Mertig, and F.~Orellana, {\it {New Developments in FeynCalc
  9.0}},  {\em Comput. Phys. Commun.} {\bf 207} (2016) 432--444,
  [\href{http://arxiv.org/abs/1601.01167}{{\tt arXiv:1601.01167}}].

\bibitem{Shtabovenko:2020gxv}
V.~Shtabovenko, R.~Mertig, and F.~Orellana, {\it {FeynCalc 9.3: New features
  and improvements}},  {\em Comput. Phys. Commun.} {\bf 256} (2020) 107478,
  [\href{http://arxiv.org/abs/2001.04407}{{\tt arXiv:2001.04407}}].

\bibitem{Fontes:2021kue}
D.~Fontes, {\em {Multi-Higgs Models: model building, phenomenology and
  renormalization}}.
\newblock PhD thesis, U. Lisbon (main), 2021.
\newblock \href{http://arxiv.org/abs/2109.08394}{{\tt arXiv:2109.08394}}.

\bibitem{Fontes:2021znm}
D.~Fontes, M.~L\"oschner, J.~C. Rom\~ao, and J.~a.~P. Silva, {\it {Leaks of CP
  violation in the real two-Higgs-doublet model}},  {\em Eur. Phys. J. C} {\bf
  81} (2021), no.~6 541, [\href{http://arxiv.org/abs/2103.05002}{{\tt
  arXiv:2103.05002}}].

\bibitem{Denner:1991kt}
A.~Denner, {\it {Techniques for calculation of electroweak radiative
  corrections at the one loop level and results for W physics at LEP-200}},
  {\em Fortsch. Phys.} {\bf 41} (1993) 307--420,
  [\href{http://arxiv.org/abs/0709.1075}{{\tt arXiv:0709.1075}}].

\bibitem{Denner:2019vbn}
A.~Denner and S.~Dittmaier, {\it {Electroweak Radiative Corrections for
  Collider Physics}},  {\em Phys. Rept.} {\bf 864} (2020) 1--163,
  [\href{http://arxiv.org/abs/1912.06823}{{\tt arXiv:1912.06823}}].

\bibitem{Khachatryan:2015qxa}
{\bf CMS} Collaboration, V.~Khachatryan et~al., {\it {Search for a charged
  Higgs boson in pp collisions at $ \sqrt{s}=8 $ TeV}},  {\em JHEP} {\bf 11}
  (2015) 018, [\href{http://arxiv.org/abs/1508.07774}{{\tt arXiv:1508.07774}}].

\bibitem{Arbey:2017gmh}
A.~Arbey, F.~Mahmoudi, O.~Stal, and T.~Stefaniak, {\it {Status of the Charged
  Higgs Boson in Two Higgs Doublet Models}},  {\em Eur. Phys. J. C} {\bf 78}
  (2018), no.~3 182, [\href{http://arxiv.org/abs/1706.07414}{{\tt
  arXiv:1706.07414}}].

\bibitem{Kannike:2021fth}
K.~Kannike, {\it {Vacuum stability conditions and potential minima for a matrix
  representation in lightcone orbit space}},  {\em Eur. Phys. J. C} {\bf 81}
  (2021), no.~10 940, [\href{http://arxiv.org/abs/2109.01671}{{\tt
  arXiv:2109.01671}}]. [Addendum: Eur.Phys.J.C 82, 247 (2022)].

\bibitem{Kannike:2012pe}
K.~Kannike, {\it {Vacuum Stability Conditions From Copositivity Criteria}},
  {\em Eur. Phys. J. C} {\bf 72} (2012) 2093,
  [\href{http://arxiv.org/abs/1205.3781}{{\tt arXiv:1205.3781}}].

\bibitem{Chauhan:2019fji}
G.~Chauhan, {\it {Vacuum Stability and Symmetry Breaking in Left-Right
  Symmetric Model}},  {\em JHEP} {\bf 12} (2019) 137,
  [\href{http://arxiv.org/abs/1907.07153}{{\tt arXiv:1907.07153}}].


\end{thebibliography}

\end{document}